\documentclass{aa}  

\usepackage{graphicx}
\usepackage{txfonts}
\usepackage{graphicx}	
\usepackage{amsmath}	
\usepackage{amssymb}	
\usepackage{multirow}

\usepackage[utf8]{inputenc}
\usepackage[export]{adjustbox}
\usepackage{wrapfig}
\usepackage{dutchcal}
\usepackage{braket}
\usepackage{grffile} 
\usepackage{xspace} 

\usepackage{pdflscape}
\usepackage{booktabs}

\usepackage{ulem}

\usepackage[framemethod=tikz]{mdframed}
\usepackage{lipsum}
\graphicspath{ {images/} }

\newcommand{\lya}{\mbox{${\rm Ly}\alpha$}\xspace}
\newcommand{\Lya}{Ly$\alpha$\xspace}

\newcommand{\flareon}{\texttt{FLaREON}\xspace}
\newcommand{\lyart}{\texttt{LyaRT}\xspace}
\newcommand{\zelda}{\texttt{zELDA}\xspace}

\newcommand{\igmz}{\texttt{IGM+z}\xspace}
\newcommand{\igm}{\texttt{IGM-z}\xspace}
\newcommand{\noigm}{\texttt{NoIGM}\xspace}
\newcommand{\recz}{\texttt{REC+z}\xspace}

\newcommand{\kms}{\,\ifmmode{\mathrm{km}\,\mathrm{s}^{-1}}\else km\,s${}^{-1}$\fi\xspace}
\newcommand{\vexp}{$V_{\rm exp}$\xspace}
\newcommand{\nh}{$N_{\rm H}$\xspace}
\newcommand{\ta}{$\tau_a$\xspace}
\newcommand{\ew}{$EW_{\rm in}$\xspace}
\newcommand{\w}{$W_{\rm in}$\xspace}

\newcommand\fA{$f_{\rm esc}^{4\AA}$\xspace }
\newcommand\fAo{$f_{\rm esc}^{1\AA}$\xspace }
\newcommand\fAt{$f_{\rm esc}^{2\AA}$\xspace }
\newcommand{\mfa}{$\langle f_{\rm esc}^{4\AA} \rangle$\xspace}
\newcommand{\fa}{$f_{\rm esc}^{4\AA}$\xspace}

\newcommand{\mone}{\texttt{Mock1}\xspace}
\newcommand{\mtwo}{\texttt{Mock2}\xspace}
\newcommand{\mthree}{\texttt{Mock3}\xspace}
\newcommand{\mfour}{\texttt{Mock4}\xspace}

\newcommand{\mevo}{\texttt{Mock\_Evo}\xspace}
\newcommand{\mfix}{\texttt{Mock\_fix}\xspace}

\newcommand{\zp}{\texttt{ZP22}\xspace}

\newcommand{\dlt}{$\Delta \lambda _{\rm True}$\xspace}

\newcommand{\wg}{$W_{\rm g}$\xspace}
\newcommand{\dl}{$\Delta \lambda_{\rm Pix}$\xspace}
\newcommand{\sn}{$S/N_p$\xspace}

\definecolor{mycolor}{rgb}{0.122, 0.435, 0.698}
\definecolor{dscolor}{rgb}{0.000, 0.435, 0.698}

\newmdenv[innerlinewidth=1.0pt, roundcorner=6pt,linecolor=mycolor,innerleftmargin=6pt,
innerrightmargin=6pt,innertopmargin=4pt,innerbottommargin=2pt]{mybox}

\newmdenv[innerlinewidth=1.0pt, roundcorner=6pt,linecolor=red,innerleftmargin=6pt,
innerrightmargin=6pt,innertopmargin=4pt,innerbottommargin=2pt]{cbbox}

\begin{document} 

   \title{zELDA II: reconstruction of galactic Lyman-alpha spectra
attenuated by the intergalactic medium using neural networks }

   \subtitle{}


   \titlerunning{ Reconstruction of galactic Ly$\alpha$ spectra }
    
   \author{
    Siddhartha Gurung-López\inst{1,2}\thanks{E-mail: gurung.lopez@gmail.com}
    \and Chris Byrohl \inst{3}
    \and Max Gronke \inst{4}
    \and Daniele Spinoso \inst{5}
    \and Alberto Torralba\inst{1,2}
    \and Alberto Fernández-Soto \inst{7}
    \and Pablo Arnalte-Mur \inst{1,2}
    \and Vicent J. Martínez\inst{\ref{1}, \ref{2}, \ref{6}}
          }

   \institute{
    Observatori Astron\`omic de la Universitat de Val\`encia, Ed. Instituts d’Investigaci\'o, Parc Cient\'ific. C/ Catedr\'atico Jos\'e Beltr\'an, n2, 46980 Paterna, Valencia, Spain\label{1}
    \and Departament d’Astronomia i Astrof\'isica, Universitat de Val\`encia, 46100 Burjassot, Spain\label{2}
    \and Universität Heidelberg, Institut für Theoretische Astrophysik, ZAH, Albert-Ueberle-Str. 2, 69120 Heidelberg, Germany \label{3} 
    \and Max Planck Institute for Astrophysics, Karl-Schwarzschild-Str. 1, 85748 Garching, Germany \label{4}
    \and Department of Astronomy, Physics Building, Tsinghua University, 100084 Beijing, China \label{5}
    \and Unidad Asociada ``Grupo de Astrof\'{\i}sica Extragal\'actica y Cosmolog\'{\i}a'', IFCA-CSIC/Universitat de Val\`encia, Val\`encia, Spain\label{6}
    \and Instituto de F\'{\i}sica de Cantabria (CSIC-UC), Avda. Los Castros s/n, 39005 Santander, Spain\label{7}
    }

   \date{Received ???; accepted ???}

 
  \abstract
   {The observed  Lyman-Alpha (\lya) line profile is a convolution of the complex \lya radiative transfer taking place in the interstellar, circumgalactic and intergalactic medium (ISM, CGM, and IGM, respectively). Discerning the different components of the \lya line is crucial in order to use it as a probe of galaxy formation or the evolution of the IGM.  }
   {We present the second version of \zelda  (redshift Estimator for Line profiles of Distant Lyman-Alpha emitters), an open-source Python module focused on modeling and fitting observed \lya line profiles. This new version of \zelda  focuses on disentangling the galactic from the IGM effects.   }
   {We build realistic \lya line profiles that include the ISM and IGM contributions, by combining the Monte Carlo radiative transfer simulations for the so called `shell model' (ISM) and IGM transmission curves generated from IllustrisTNG100. We use these mock line profiles to train different artificial neural networks. These use as input the observed spectrum and output the outflow parameters of the best fitting `shell model' along with the redshift and \lya emission IGM escape fraction of the source.   }
   {We measure the accuracy of \zelda on mock \lya line profiles. We find that \zelda is capable of reconstructing the ISM emerging \lya line profile with high accuracy (Kolmogórov-Smirnov<0.1)  for 95\% of the cases for HST/COS-like observations and 80\% for MUSE-WIDE-like.   \zelda is able to measure the IGM transmission with the typical uncertainties below 10\% for HST/COS and MUSE-WIDE data. }
   {This work represents a step forward in the high-precision reconstruction of IGM attenuated \lya line profiles. \zelda allows the disentanglement of the galactic and IGM contribution shaping the \lya line shape, and thus allows us to use \lya as a tool to study galaxy and ISM evolution. }


   \maketitle
%
\section{Introduction}\label{sec:intro}

\begin{figure*} 
        \begin{center}
            \includegraphics[width=6.2in]{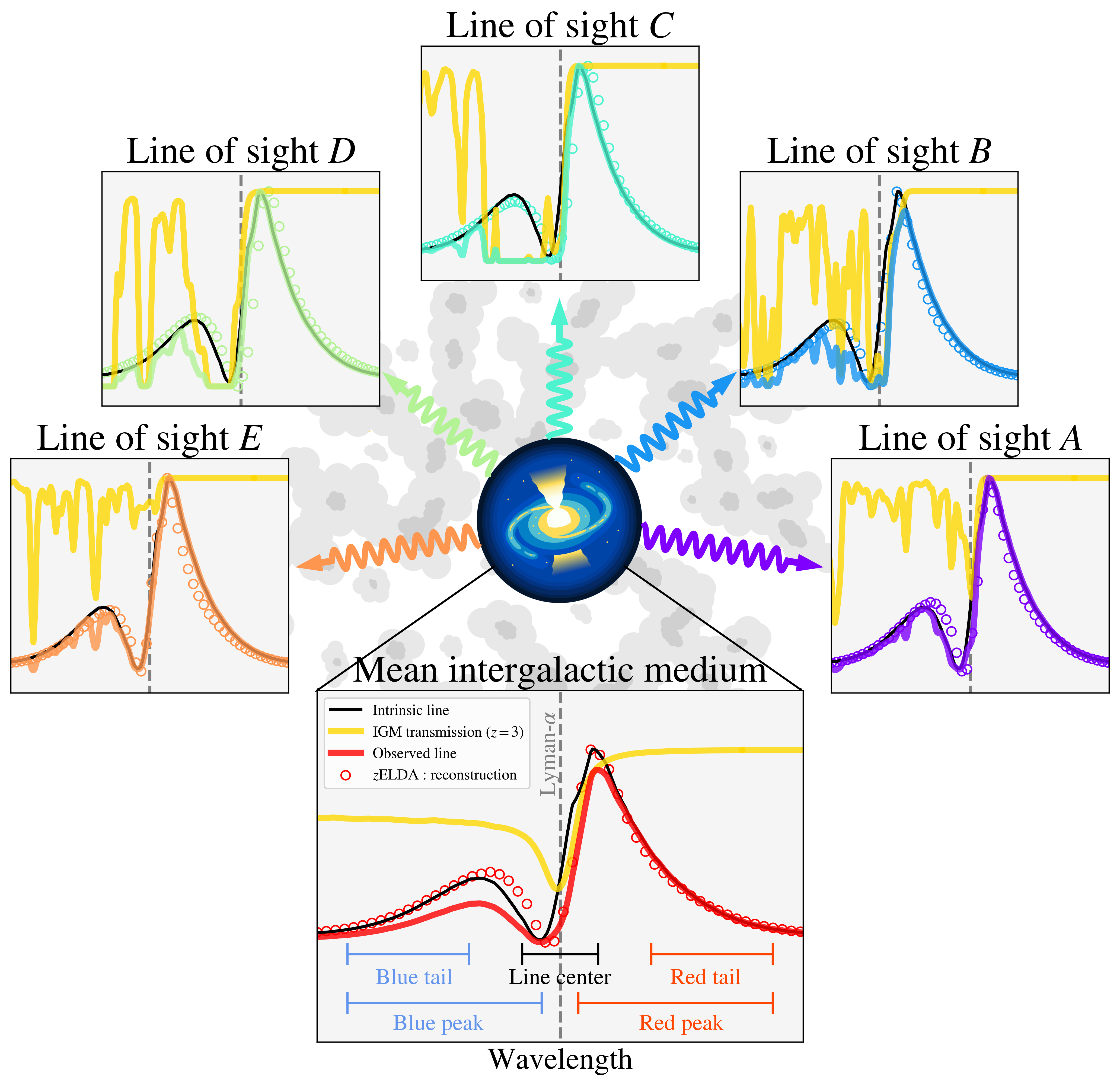}%
        \end{center}
        \caption{ Illustration of the impact of different line of sights in the same intrinsic spectrum. In the middle bottom panel we show the intrinsic spectrum escaping the source (black) convolved with the mean IGM transmission at $z=3.0$ (yellow). The color line shows the convolution of the intrinsic spectrum and IGM transmission. The colored circles show \zelda reconstruction using the \igm model (discussed later). In the other top 5 panels, individual IGM transmission through different line of sights are used at $z=3.0$.  }
        \label{fig:Main_illustration}
        \end{figure*}

Due to the large abundance of hydrogen in the Universe, the Lyman-$\alpha$ (\Lya) emission line is of imperious importance in astrophysics. \Lya photons are produced when an electron decays from the first excited level to the ground energy level in hydrogen. The \Lya line is very luminous in extragalactic sources and thus, is used to identify galaxies throughout the evolution of the Universe
\citep[for a review, see,][]{OuchiObservationsLymana2020}.

In the recent years many experiments have expanded the number of known \Lya emitting galaxies (LAEs), such as HETDEX ($\sim 0.8$ million \Lya emitting galaxies at $1.9<z<3.5$; \citealp{hill08,Farrow_2021,Weiss_2021}),  SILVERRUSH ($\sim 2,000$ at $6<z<7$; \citealp{Ouchi2018a,Kakuma_2019}), \textit{MUSE WIDE} ($\sim 500$ at $3\lesssim z\lesssim 6$; \citealp{Herenz2017,Urrutia2019A&A...624A.141U,Caruana_2020}) or the J-PLUS \citep[$\sim 14,500$ at $2\lesssim z\lesssim 3.3$;][]{spinoso_2020}, 67 at $2<z<3.75$ in miniJPAS/J-NEP \citep{Torralba_2023} and  in PAUS  \citep[591 at $2.7<z<5.3$ ][]{Torralba_PAUS_2024}. Meanwhile,  {\it The Prime Focus Spectrograph Galaxy Evolution Survey} \citep{PFS_2022} will increase our knowledge about LAEs from z$\sim2$ up to the epoch of reionization at $z\sim7$.



The \lya line constitutes an unique tracer of the composition and kinematics of cold gas. 
This is because of the resonant nature of \Lya, which implies that \Lya photons are absorbed and reemitted by neutral hydrogen atoms in a short time scale ($\sim 10^{-8}\,$s). \Lya emitting galaxies typically exhibit a hydrogen column density is $N_{\rm HI}\sim 10^{17}-10^{20}\,\mathrm{cm}^{-2}$  \citep{Gronke_2016}, and the scattering cross section at the center of the line is $\sigma\sim 6\times 10^{-14}\,\mathrm{cm}^{2}$ (assuming gas at $T\sim 10^4\,$K). This causes the \lya photons to experience thousands of scattering events before leaving the galaxy, in which the frequency of the photon changes, mostly due to Doppler boosting. In general, this modifies the shape of the \lya line profile emerging from the interstellar medium.

The \lya photons furthermore interact on larger scales after leaving their emitting galaxy. In the circumgalactic medium (CGM), the diffuse gas bound to the galaxy's hosting halo, and the clumpy intergalactic medium (IGM)  \citep{zheng11,laursen11,Behrens_2019,Byrohl_2019,Byrohl2020,GurungLopez_2020a}. A fraction of the \lya photons interact with neutral hydrogen and are scattered outside the line of sight. This further modify the \lya line profile shape that reaches the observer.

In fact, some theoretical works show that the measured clustering of LAEs could be potentially influenced by the radiative transfer of \lya. In the first place, the observability of LAEs can potentially depend on IGM the large-scale properties such as the density, velocity with respect \lya sources and their gradients \citep{zheng11,Behrens_2019,GurungLopez_2020a}.  In second place, the determination of the redshift from the profile of the \lya line profile is complex \citep{steidel10,Rudie_2012,Verhamme:2018aa,GurungLopez_2019b,Byrohl_2019,Runnholm_2020} and might introduce further distortions in the measured clustering \citep{gurung_2020b}.


The profile of the \lya line is affected at the same time by the ISM and IGM. This makes challenging the study of each of these mediums independently of the rest of the observed \lya line profile. There are works that analyze possible correlation between galaxy properties and the features of the \lya line profile \citep[e.g.]{Hayes_2023} at $z<0.5$, where the IGM is mostly transparent to \lya. However, in order to conduct the same study at high redshift, it would be ideal to have access to the \lya line profile emerging from the ISM, without the influence of the IGM. At the same time, disentangling between the contributions of the ISM and the IGM would provide the IGM selection function on LAEs. This could clarify whether \lya visibility depends on the large-scale properties of the IGM. Therefore, splitting the \lya line profile by the contributions of ISM, IGM  will be key in future works based on \lya emission.

The \lya radiative transfer process is nontrivial and only few analytical solutions exist in relatively simple gas geometries \citep[e.g.,][]{neufeld90,dijkstra06}. Due to the complexity of solving the radiative transfer equations analytically, typically Monte Carlo radiative transfer codes are being employed. \lya Monte-Carlo radiative transfer, although typically computationally expensive, allows for a large flexibility in gas properties and specially in gas geometry, allowing from the relatively simple `shell model' (ISM/CGM), a moving spherical shell which surrounds a \lya emitting source \citep{ahn03}, to the intricate ISM, CGM and IGM gas distributions in cosmological simulations \citep{Byrohl_2019}.

The  `shell-model' has been very successful in reproducing the shape of the observed \lya line profiles across the Universe \citep[e.g.,][]{Verhamme_2007,Schaerer2011,Gronke2017,gurung_lopez_2022}. The radiative transfer process in the clumpy and intricate ISM and inner CGM is intrinsically different from that taking place in the smoother and colder IGM. In principle, the low redshift ($z<0.5$) observed \lya line profiles should be dominated by the radiative transfer in the ISM. However, at high redshift, the \lya line profile should be affected by the radiative transfer in the ISM, CGM and IGM.


The use of mean IGM transmission curves might work on stacked line profiles, but for individual sources it is a very limited technique given the huge diversity of the IGM at the same redshift and even for the same source \cite{Byrohl2020}. This is illustrated in Fig.~\ref{fig:Main_illustration} where the same \lya line profile emerging from the ISM (black) is convolved with the mean IGM transmission at $z=3$ (bottom panel) and five line of sights ({\it A} to {\it E}) and their individual IGM transmission curves (yellow). The line profile emerging from the IGM is shown as solid colored lines, while the reconstruction of the code presented in this work is displayed as colored empty circles. In addition, a sketch of the parts of the \lya line profile can be found in the bottom panel. While applying the mean IGM transmission modulates the \lya line profile smoothly, the IGM features in the individual line of sight (LoS) are sharper. This is especially noticeable in LoS {\it B} and {\it D}. In addition, the IGM topography is quite diverse, even for the same source. While there will be LoS with almost no neutral hydrogen (LoS {\it A}), other will be very optically thick (LoS {\it C}). Thus, the IGM emerging \lya line profile depends on the individual LoS and could exhibit a huge variety even if the ISM emerging line profile is assumed to be the same. This also shows the limitation of reconstructing IGM attenuated \lya line profiles using the mean IGM transmission at the source redshift.  


In this work, we present the second version of \zelda \citep[based on][]{gurung_lopez_2022}, an open source Python package based on \lyart \citep{orsi12} and \flareon  \citep{GurungLopez_2019b} . \zelda has two main scientific motivations i) modeling \lya line profiles and escape fraction using the shell model for cosmological simulations and \citep[as in][]{orsi14,GurungLopez_2019a,gurung_2020b,gurung_lopez_2021}, ii) fitting observed \lya line profiles to the shell model. In the first version of \zelda we focused on modeling the \lya spectrum affected only by the ISM. \zelda was able to nicely fit observed \lya line profiles at $z<0.5$. In the version presented here, we focus on fitting \lya line profiles affected by ISM and IGM. For this, we make use of machine learning algorithms in which the input is basically the observed spectrum, convoluted with IGM and ISM. In Fig.~\ref{fig:Main_illustration} we show six examples of the reconstructed \lya line profile using \zelda of the same intrinsic line profile traveling through five different lines of sights (open circles). 

\zelda\ is publicly available and ready to use\footnote{\url{https://github.com/sidgl/zELDA_II}}. \zelda\ contains all the scripts necessary to reproduce all the results presented in this work. Documentation and several tutorials on how to use \zelda\ are also available\footnote{\url{https://zelda-ii.readthedocs.io/index.html}}.

This work is organized as follows. In Sect.~\ref{sec:simulations} we describe the data sets used to model the observed \lya line profiles. In Sect.~\ref{sec:methodoogy} we detail the pipeline to reconstruct the ISM emerging \lya line profiles from the observed line profile. First, we test the accuracy of our methodology in mock \lya spectrum in Sect.~\ref{sec:results_mock}.  Finally, we draw our conclusions in Sect.~\ref{sec:conclusions}.


Through this work, we show \lya line profiles and IGM transmission curves in $\Delta \lambda _0$, i.e., the rest frame difference to the \lya wavelength. We also provide redshift accuracy in the same units. This quantity can be expressed in velocity units as $\Delta v = c \Delta\lambda_0/\lambda_{\rm Ly\alpha}\sim(247km/s)\times \Delta\lambda / 1$\AA{}, where $c$ is the speed of light and $\lambda_{\rm Ly\alpha}\approx 1215.67$\AA{}. 

Another convention that we use through this work is the notion of `\lya  IGM escape fraction'. We refer to the \lya IGM escape fraction of a source as the ratio between intrinsic and observed \lya photons along the line of sight. We note that the \lya photons are not, in general, destroyed by dust grains in the IGM, and hence this ratio is also referred to as the `transmission fraction' in the literature. Instead, they are scattered out of the line of sight. Thus, although for the observer the IGM causes absorption features, globally, the missing photons escape the IGM in another direction.

\section{Simulating Lyman alpha line profiles}\label{sec:simulations}

In this section, we detail the data sets used to produce mock \lya line profiles that include the ISM, CGM and IGM. In Sect.~\ref{ssec:model_ISM} we show the 'shell model' simulation that are used, based on the first \zelda version. Meanwhile, in Sect.~\ref{ssec:model_IGM} we detail the CGM/IGM transmission curves based on \cite{Byrohl2020}.

\subsection{Radiative transfer in the interstellar medium}\label{ssec:model_ISM}

As in \cite{gurung_lopez_2022} (hereafter \zp), \zelda uses a set of precomputed \lya line profiles. These lines are computed using the Monte Carlo radiative transfer code \lyart \citep{orsi12}, which performs the entire radiative transfer computation photon by photon. This set of lines would contain the ISM radiative transfer component, while lacking the IGM influence.

The regular grid of \lya line profiles used in this work is the same as that described in \zp. Basically, it consists in a 5 dimensional parameter grid with 3,132,000 nodes. The 5 parameters are those of the thin shell model in \lyart. These are, the outflow expansion velocity \vexp, the neutral hydrogen column density \nh, the dust optical depth \ta, the intrinsic equivalent width \ew and the line width \w of the \lya emission before entering into the thin shell. The ranges of these parameters covered by the regular grid are \vexp$\in [0,1000]km/s$, \nh$\in[10^{17},10^{21.5}]cm^{-2}$, \ta$\in[0.0001,0.0]$, \ew$\in[0.1,1000]\AA$ and \w$\in[0.01,6]\AA$. For further information on the grid specifications, we refer the reader to \zp.

\lya line profiles within the boundaries of \zelda's grid are computed by 5D lineal interpolation between nodes, as described in \zp. This leads to a typical accuracy of 0.04 in the Kolmogorov-Smirnov estimator (KS), which is, the maximum difference between cumulative distributions, in the prediction of \lya line profiles from the thin shell model. 

\begin{figure*} 
        \includegraphics[width=7.2in]{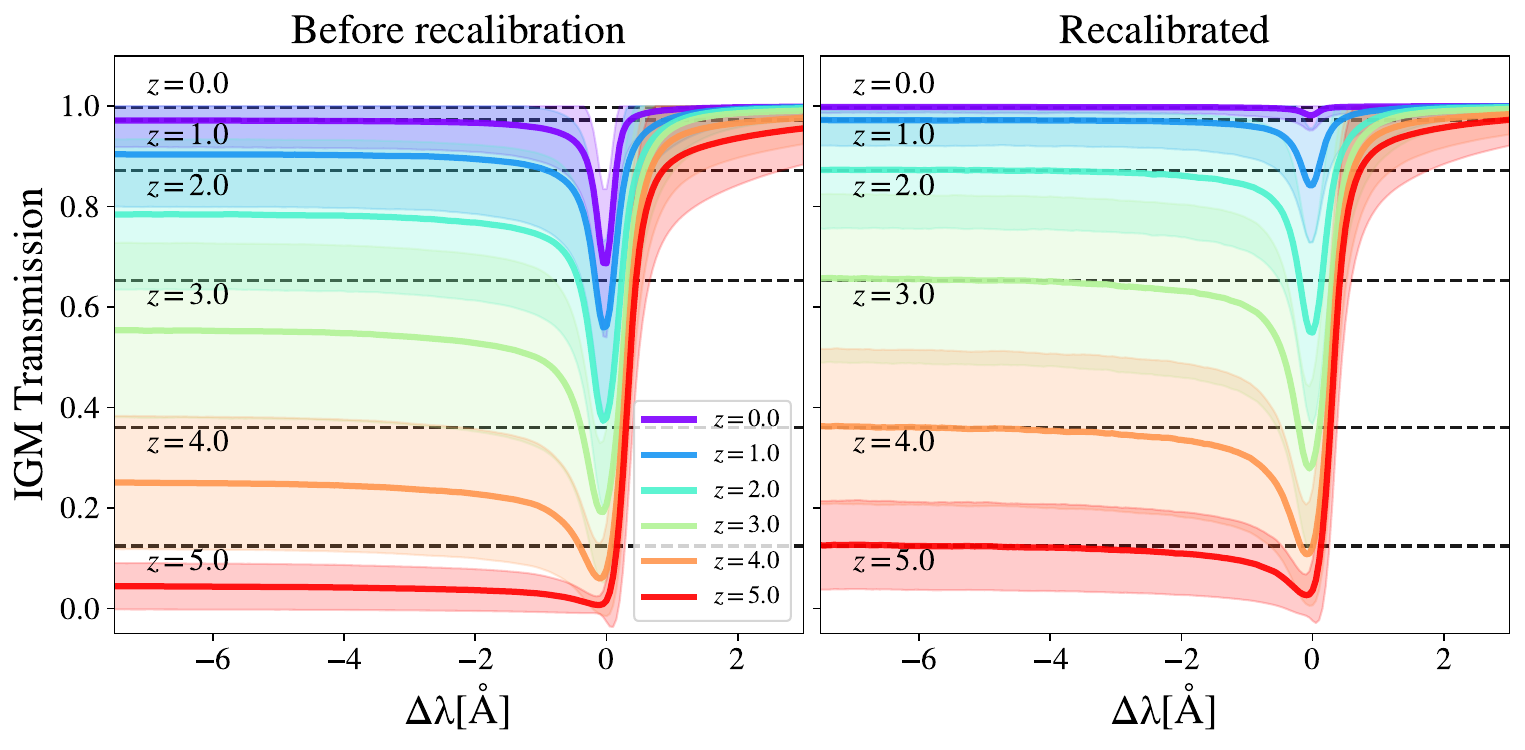}%
        \caption{ Mean IGM transmission curves without recalibration (left) and after recalibration (right). Each color shows a different redshift snapshot. The horizontal black dashed show the mean IGM transmission given by \citep{Faucher-Giguere08} at $z=0$,1,2,3,4,5 from top to bottom. }
        \label{fig:IGM_T_mean}
        \end{figure*}

\subsection{Radiative transfer in the intergalactic medium}\label{ssec:model_IGM}

In order to include radiative transfer in the intergalactic and circumgalactic medium, we make use of the transmission curves from \cite{Byrohl2020}. These were calculated in six snapshots of the IllustrisTNG100 simulation \citep{Naiman_2018,Nelson_2019,Marinacci_2018,Pillepich_2018,Springel_2018}, at redshift 0.0, 1.0, 2.0, 3.0, 4.0 and 5.0. The \lya radiative transfer was computed using a modified version of the code ILTIS  \citep{Behrens_2019,Byrohl2020,Byrohl_21}.The radiative transfer analysis was performed for every halo with mass greater than $5\times10^9 M_{\odot}$ in 1000 different line of sight. For more information, see \cite{Byrohl2020}. 



The mean IGM transmission curves of \cite{Byrohl2020} are shown in the left panel of Fig.~\ref{fig:IGM_T_mean}. These show the typical structure found in the literature, i.e., a transmission close to unity redward \lya, a well of absorption at \lya and a plateau at bluer wavelengths than \lya that decreases with redshift \citep{laursen11,gurung_lopez_2022}. The shaded regions mark the 25 and 75 percentiles. As shown by \cite{Byrohl2020}, the scatter around the mean is significant, as there is a lot of variability in the individual line of sights. 

As mentioned above, the IGM transmission curves from \cite{Byrohl2020} are given in discrete redshift bins (0.0, 1.0, 2.0, 3.0, 4.0 and 5.0). However, for training our artificial neural networks we require a continuous redshift sampling. In order to obtain an IGM transmission curve at a given $z_t$ we proceed as follows. First, we obtain the target mean optical depth $\tau_t$ at $z_t$ from \cite{Faucher-Giguere08}. Next, we recalibrate the snapshot closest to $z_t$ so that its mean optical depth matches $\tau_t$. For this we use the wavelength range from -8\AA{} to -6\AA{} from \lya. Finally, we draw a random IGM transmission curve.

The right panel of Fig.~\ref{fig:IGM_T_mean} shows the mean IGM transmission curves after the recalibration. The black horizontal lines show the mean IGM transmission by \cite{Faucher-Giguere08}. Before recalibration, we find that there is up to a 10\% difference between -8 to -6 \AA from \lya. After recalibration, by construction, both mean IGM transmissions match perfectly.

Examples of the large diversity of IGM transmission curves at different redshifts are shown throughout this work. Individual IGM transmission curves are shown as solid yellow lines in Fig.~\ref{fig:Main_illustration}.

\begin{figure*} 
        \includegraphics[width=7.2in]{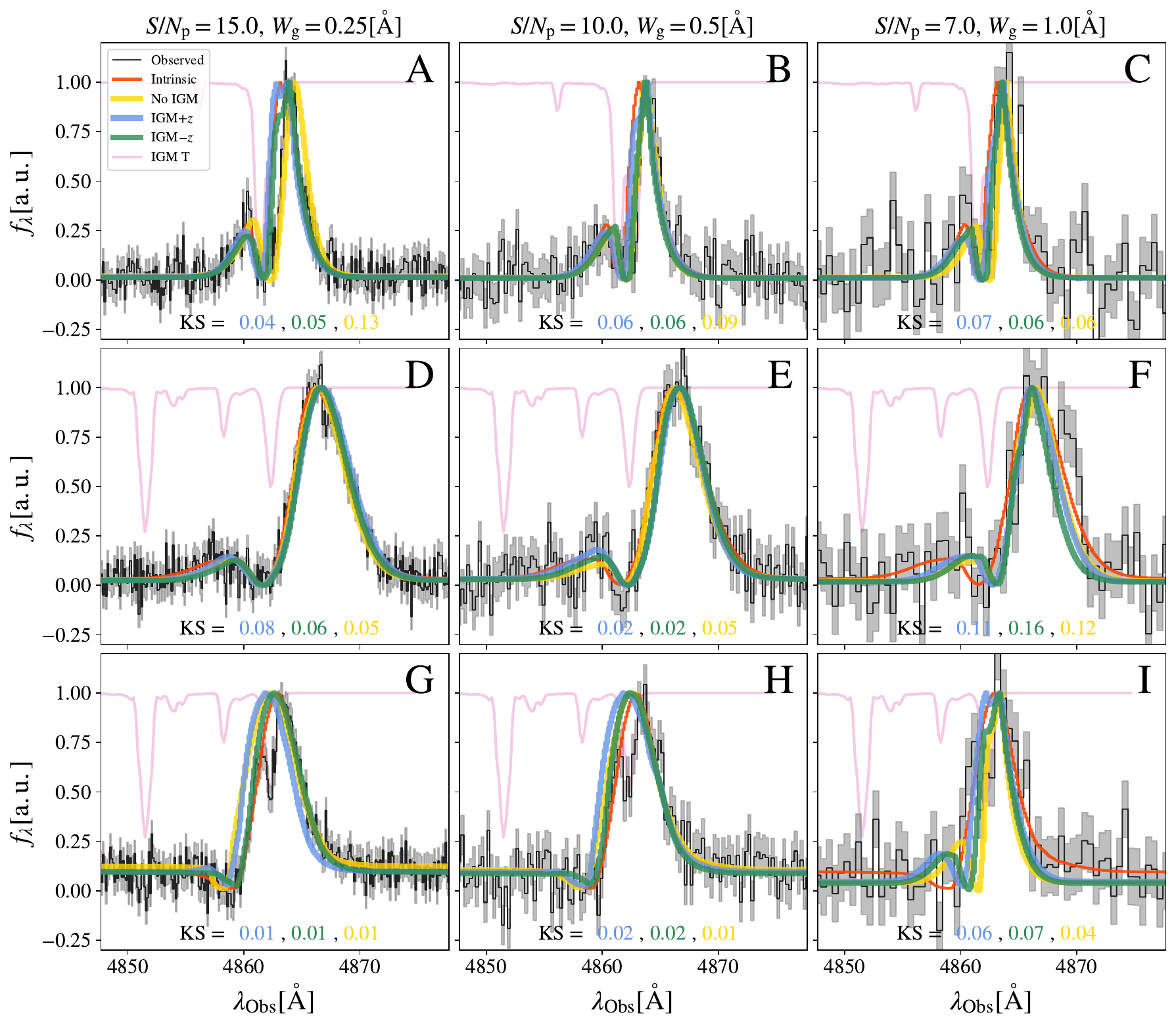}%
        \caption{ Example of line profile reconstruction at different line profile qualities and using our different models. The true \lya line before passing through the IGM is displayed in red. The IGM transmission curve is shown in pink. The true \lya line profile is fixed in each row. The observed line profile, after IGM absorption and mocking observation conditions, is shown in black. The observation conditions are fixed in each column as $W_{\rm g}=0.25\AA$, $S/N_{\rm p}=15.0$, $W_{\rm g}=0.5\AA$, $S/N_{\rm p}=10.0$, $W_{\rm g}=1.0\AA$, $S/N_{\rm p}=7.0$, from left to right respectively. The \zelda prediction for the models \igmz, \igm and \noigm are displayed in blue, green and yellow respectively. In the bottom of each panel the KS between the true \lya line profile before the IGM absorption and \zelda prediction is displayed in different colors matching the model used.  }
        \label{fig:EXAMPLES_quality}
        \end{figure*}

\begin{figure*} 
        \includegraphics[width=3.6in]{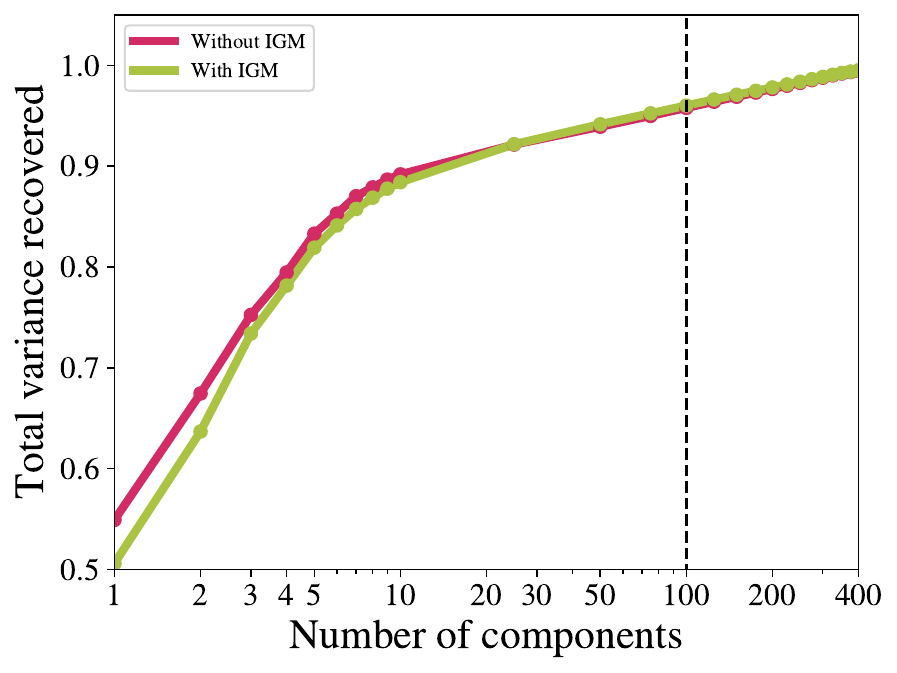}%
        \includegraphics[width=3.6in]{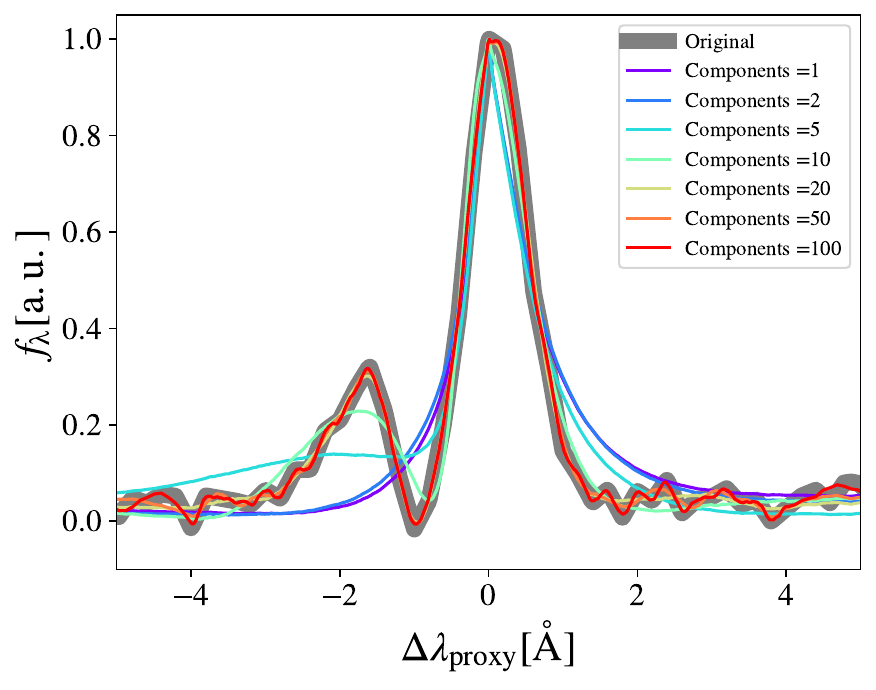}
        \caption{ Left: Total variance recovered as a function of the number of principle components. In red (green), \lya line profiles spamming \zelda's grid without (with) IGM absorption.  The dashed black line mark the number of principal components used for the input of the artificial neural networks. Right: example of the PCA decomposition in an IGM clean shell model line profile. The line profiles are shown in the proxy rest frame. The original mock line profile is shown in grey. Meanwhile, the reconstructed line profiles using the $N$ first principal components as displayed in the legend.   }
        \label{fig:PCA_n_feature}
        \end{figure*}

\subsection{ Mocking observed Lyman-$\alpha$ line profiles}\label{ssec:model_mocks}

\lya line profiles predicted by \zelda using the \lyart grid of line profiles and the IGM transmission curves of \cite{Byrohl2020} are ideal, both in terms of spectral resolution and signal-to-noise ratio. In contrast, measurements of \lya line profiles present limitations in the spectral resolution, spectral binning and signal to noise.

In order to produce a mock line profile of a given set of \{ \vexp, \nh, \ta, \ew , \w \} and through a given IGM line of sight, we follow the next procedure. First, we produce the {\it intrinsic} \lya line profile escaping the galaxy from the \lyart grid as described above. Second, the spectrum after traveling through the IGM is obtained by the convolution of the ideal thin shell \lya line profile with the chosen IGM transmission curve. Third, we downgrade the quality of the \lya\ line profile to match the desired observation configuration as in \zp.

The quality of a \lya line profile is set by three parameters: i) the signal to noise ratio of the peak of the \lya line, \sn, ii) the wavelength resolution element in the observed frame, \wg and iii) the pixel size in the observed frame, \dl. To downgrade the quality of the \lya line profile, first we convolve the ideal spectrum with a Gaussian kernel of FWHM=\wg. Next, we pixelate the \lya line profile following, 
\begin{equation}
            \label{eq:pixelization}
            \displaystyle
            f_{\lambda , \rm  pix }^{\rm Ly\alpha}(\lambda_{\rm pix})  =  
            {
            \displaystyle
            {\int ^{\lambda_{\rm pix}+\Delta \lambda_{\rm pix}/2} 
            _{\lambda_{\rm pix}-\Delta \lambda_{\rm pix}/2}
            {f_{\lambda}^{\rm Ly\alpha} (\lambda) \;  d\lambda }} 
            \over 
            {\Delta \lambda_{\rm pix} } 
            }
            .
            \end{equation} 
The intensity of the maximum of the line profile is computed and Gaussian white noise is added to the spectrum with an amplitude fixed by \sn. Note that in the training set \dl and \wg are independent variables. However, in order to show the results of \zelda we fixed \dl=\wg/2 across all the plots and tables to reduce from three dimensions \{\sn,\wg,\dl\} to only two \{\sn,\wg\}. 


In Fig.~\ref{fig:EXAMPLES_quality} we show three spectrum quality configurations progressing from best to worst from left to right. In particular, the left column uses \sn=15.0 and \wg=0.25\AA{}, in the middle \sn=10.0 and \wg=0.5\AA{} and in the right \sn=7.0 and \wg=1.0\AA{}. \lya profiles are fixed at redshift 3.0. The intrinsic \lya line profile emerging from the galaxy (modeled with the \lyart grid) is shown in red, while a randomly chosen IGM transmission is shown in pink. The mock observed \lya line is shown in black. The other colored lines are \zelda's reconstructions of the observed \lya line profile that will be discussed later in Sect.~\ref{sec:methodoogy}. Basically, each row shows the same outflow configuration \{ \vexp, \nh, \ta, \ew , \w \}, listed in Tab.~\ref{tab:quality_params}, in Appendix ~\ref{sssec:results_mock_parameters_parameters}, through the same line of sight. In the left column (cases {\it A}, {\it D} and {\it G}), the observed line profile closely follows the convolution of IGM and the intrinsic line profile. In particular, the pixel size is small in comparison with the size of the \lya peaks. Also, most of the pixels with the peaks are above the noise.  Meanwhile, the quality of the middle columns is worse. For example, the blue peak in {\it D} was clearly visible in the observations, while it is more difficult to see in {\it E}. Finally, in the right most column shows  \lya spectrum heavily affected by noise and a relatively low spectral resolution. While case {\it F} is relative well recovered given its width, the strong absorption feature present in {\it G} is mostly removed from {\it I}.

\section{Reconstructing attenuated Lyman-$\alpha$ emission lines}\label{sec:methodoogy}

In this section, we explain our methodology to reconstruct \lya line profiles attenuated by the intergalactic medium, as well as to estimate the \lya IGM escape fraction. 

We present three models based on artificial neural networks. The idea of obtaining the redshift of a source using artificial neural networks in the \lya line profile was initially explored in \cite{gurung_lopez_2021}. These models are referred to as \igmz, \igm and \noigm and have a different input and training sample. Basically, \igmz includes the redshift of the source as input and it is trained with a realistic IGM transmission curve redshift evolution. The input of the \igm model does not include the redshift of the source and it is trained with redshift-randomized IGM transmission curves. Finally, the input of the \noigm model includes the redshift of the source, but no IGM transmission curve is applied to the line profiles of the training set. 

This section is structured as follows. First, we detail the ANN input in Sect.~\ref{ssec:methodoogy_ANN}, while the training sets are discussed in Sect.~\ref{ssec:methodoogy_training}. Then the ANN output is described in Sect.~\ref{ssec:methodoogy_output} along with the ANNs architecture in ~\ref{ssec:methodoogy_architecture}. We present a feature importance analysis in ~\ref{ssec:methodoogy_feature}. Finally, the parameter estimation is shown in ~\ref{ssec:methodoogy_estimation}.

\subsection{ Input of artificial neural networks.}\label{ssec:methodoogy_ANN}

The input in the three presented models (\igmz, \igm and \noigm) follows the same philosophy as those introduced in \zp. Basically, the input consists of the line profile and its observational quality. The main difference between the models of this work and those presented in \zp is how the \lya line profile is provided to the artificial neural networks (ANN). 

\subsubsection{Line profile treatment}\label{sssec:methodoogy_ANN_lineprof}

In the \igmz, \igm and NoIGM models, we treat the observed line as follows:
\begin{enumerate}
        \item The wavelength position of the global maximum of the observed line profile $\lambda_{\rm max}$ is used as a proxy for the true \lya\ wavelength $\lambda_{\rm True}$. Thus, the proxy redshift is $z_{\rm  max}= \lambda _{\rm max} / \lambda_{\rm  Ly\alpha}-1$.
        
        \item The observed line profile is moved to the proxy rest frame, $f_{\lambda, \rm max }^{\rm Ly\alpha}$. In particular, we convert the array where $f_{\lambda}^{\rm Ly\alpha}$ is evaluated in the observed frame, $\lambda^{\rm Obs}_{\rm Arr}$ to the rest frame wavelength as if $\lambda_{\rm True} = \lambda_{\rm max}$, i.e., $\lambda^{\rm 0}_{\rm Arr} = \lambda^{\rm Obs}_{\rm Arr}/(1+z_{\rm max}) $. 
        
        \item The line profile is normalized by its maximum $f_{\lambda}^{\rm Ly\alpha}(\lambda_{\rm max})$. 
        
        \item The normalized line profile $f_{\lambda, \rm max }^{\rm Ly\alpha}$ is rebinned into 600 bins from $\lambda_{\rm Ly\alpha}-12.0$\AA{} to $\lambda_{\rm Ly\alpha}+12.0$\AA{} by linear interpolation between the values of $f_{\lambda, \rm max }^{\rm Ly\alpha}$ evaluated in $\lambda^{\rm 0}_{\rm Arr}$.  
        \item  The line profile is decomposed using a principal component analysis (PCA). The first 100 principal components are used, as detailed below.  
        
    \end{enumerate}

Steps 1 to 4 are almost identical to those in \zp with only minor changes in the wavelength range used. The PCA analysis is a new addition to \zelda.

In Fig.~\ref{fig:PCA_n_feature} we show the total explained variance as a function of the number of principal components. We present two PCA models, one for lines without IGM (red, used for the \noigm  model) and another for lines with IGM absorption (green, used for the \igmz and \igm models ). We find that both PCA models exhibit the same behavior. The total explained variance grows rapidly with the first $\sim 7$ components up to $\sim$90\%. After a knee, the total recovered variance grows slowly, reaching $\sim$95\% at 100 and $\sim$98\% at 400 principal components. Below 25 principal components, the total explained variance for in the lines without IGM is greater than those including the IGM for a fixed value of principal components. This seems reasonable since the IGM absorption would add complexity to the observed \lya line profiles. From 25 principal components onward, the total variance in both models is the same. 

We used the first 100 principal components in the models presented in this work. We tested that using the first 200 and 400 principal components did not increase the precision in reconstructing the intrinsic \lya line profiles.  

The right panel of Fig.~\ref{fig:PCA_n_feature} shows an example of the PCA decomposition of a shell model line profile that is unobscured by the IGM (grey). The different colored lines show the PCA decomposition using only the fist 1, 2, 5, 10, 20, 50 and 100 components (from purple to red). The first 2 components focus on the red peak. The blue peak is progressively recovered as we increase the components from 3 to 20. The first 20 principal components give an accurate, although smoothed, version of the original line profile. Meanwhile, from the 20th up to 100th components, small features are captured, including the noise pattern. Although not displayed here, the IGM featured are encapsulated from the 20th to 100th principal components.

\begin{figure*} 
        \includegraphics[width=7.2in]{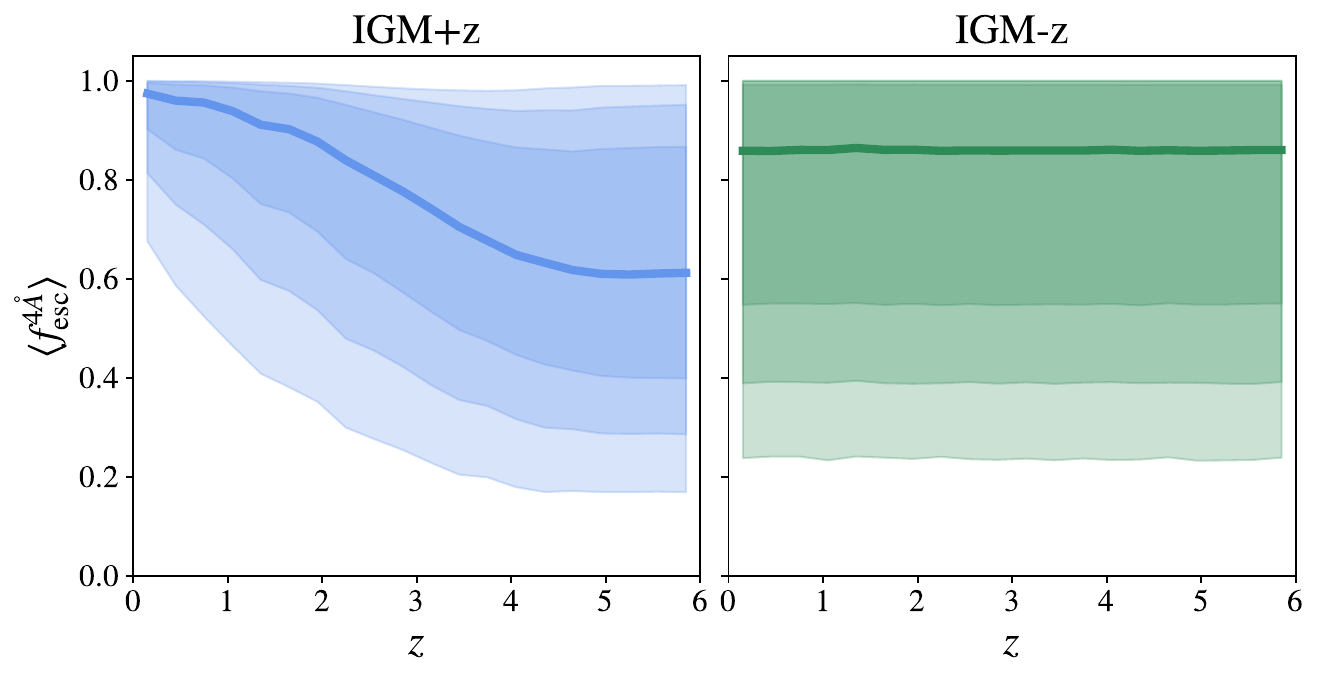}%
        \caption{ \fa used for the training in the \igm (left) and \igmz (right) models. The solid thick line marks the median \fa. Meanwhile, the shaded regions show the scatter between the 16 and 84 percentiles (darkest), 5 and 95 percentiles (medium dark), 1 and 99 percentiles (clearest).   }
        \label{f_mean_training}
        \end{figure*}

\subsubsection{ Line profile quality and redshift}\label{sssec:methodoogy_ANN_quality}

As in \zp we include the line profile quality as features in the input of the artificial neural networks. In particular, we include the wavelength element of resolution and the pixel size, both in the observed frame. This is the same in all three models.

Then, in \zp we included the redshift of the sources through the proxy $z_{\rm max}$. In this work, we do the same in \igmz and \noigm. Meanwhile, the redshift is excluded from \igm. On one hand, the motivation behind \igmz is to produce a model with the same IGM distribution as in \cite{Byrohl2020}. On the other hand, the goal behind \igm  is to provide a model as unbiased as possible by the redshift-dependent quantities and that matches \cite{Faucher-Giguere08} mean IGM transmission.

\subsubsection{Total input}\label{sssec:methodoogy_ANN_input}

The input for each model is slightly different. The models \igmz and \noigm use 103 features:
\begin{itemize}
    \item{ {\texttt Input}  = [ $ \dots \;  100 \; {\rm PCA} \;  \dots , W_{\rm g} \; , \Delta\lambda_{\rm pix} \; , z_{\rm max} $], }
\end{itemize}
However, the PCA model used for \igmz includes the IGM absorption, while the PCA model used for \noigm does not.

Then, the input for the \igm model is the same as before but excluding $z_{\rm max}$, i.e., 102 features:
\begin{itemize}
    \item{ {\texttt Input}  = [ $ \dots \;  100 \; {\rm PCA} \;  \dots , W_{\rm g} \; , \Delta\lambda_{\rm pix} \;  $], }
\end{itemize}
where the PCA model is the same as in \igmz and includes the IGM absorption.\\

\subsection{ Training sets of artificial neural networks.  }\label{ssec:methodoogy_training}


The training sets for \igmz, \igm and \noigm are different from each other. Nevertheless, they share the same number of \lya line profiles ($4.5\times 10^{6}$) and the outflow parameters \{ \vexp, \nh, \ta, \ew , \w \} are equally homogeneously randomly drown from the space covered by the \lyart grid. The mock \lya line profiles cover uniformly from \w=0.01\AA{} to 4.0\AA, \dl=0.01\AA to 2.0\AA and from \sn=5.0 to \sn=15.0.  We tested that for this training set size, our artificial neural networks have converged. In addition, for the three models, redshifts from 0 to 6 were homogeneously sampled. Notice that the IGM transmission curves from \cite{Byrohl2020} are computed up to the snapshot at $z=5$. Therefore, the predictions given by \igmz and \igm at $z>5$ should be taken with caution. The particularities of the training sets, therefore, depend on how the IGM is treated:

\begin{itemize}
    \item \igmz: This training set includes line profiles with the IGM absorption. In this model, we use the uncalibrated IGM transmission curves. In particular, the IGM transmission curve to each \lya line profile uses the actual redshift of the source. Therefore, there is an evolution in the IGM where \lya line profiles at higher redshift are more attenuated. 
    \item \igm: This training set also includes line profiles with the IGM absorption. However, in contrast to \igmz, we use the recalibrated transmission curves, which are drawn randomly irrespective of the source redshift. Thus, the IGM absorption distribution is constant with redshift.  There  will be \lya basically unabsorbed  (typical of $z=0$) and greatly absorbed (typical of high $z$) at all redshifts.
    \item \noigm: This training set does not use IGM transmission curves. 
\end{itemize}

In Fig.~\ref{f_mean_training} we compare the distribution of \fa (IGM \lya escape fraction $\pm$2\AA{} around \lya) as a function of redshift in the training sets for the \igmz (left) and \igm (right).  The model \igmz exhibits an evolution in the median \fa as a function of the redshift, given by the evolution of the opacity of IGM. At $z<1$ the median \fa is close to 1.0 while at $z>4.0$ it stalls at 0.6. In particular, there is no evolution from $z=5.0$ to $z=6.0$ as the last snapshot with IGM transmission curves is at $z=5.0$. Meanwhile, at $z<0.5$ the training set for \igmz shows less scatter than at high redshift, exhibiting more than 98\% of the sample with \fa>0.6. Then, at higher redshift the scatter increases and at $z=4$ as 98\% of the sample is between 1.0 and 0.2. Then, focusing on \igm, as the redshift of the IGM transmission curve is randomized, there is no evolution in the distribution of \fa as a redshift function. The \igm model is conceived as a 'redshift unbiased' model. 

The \igmz and \igm models have the same global motivation: reconstructing an IGM attenuated \lya line profile and obtaining the IGM escape fraction. \igmz and \igm are complementary to each other. In machine learning, the input features and the training set are very important for the output of the neural networks. If a training set exhibits some particular biases, the output can potentially show the same biases. In principle, \igmz uses in the input the proxy redshift of the source and the evolution of the IGM transmission curves with redshift of the IllustrisTNG100 simulation  \cite{Byrohl2020}. Furthermore, in Appendix \ref{ssec:methodoogy_feature} we performed a feature importance analysis on \igmz, \igm and \noigm, finding that the input proxy redshift had a strong influence in the determination of the outflow parameters and especially in \fa. Thus, the output of \igmz can potentially be biased towards the IGM redshift evolution in \cite{Byrohl2020}. For these reasons, we developed \igm, which does not include the proxy redshift in the input and has no IGM redshift evolution in the training set. Thus, \igm should be a redshift unbiased model.  In the following sections, we compare the results obtained by \igmz and \igm, finding that both perform really similarly with only small differences at $z<1$. Finally, we remark that if a redshift-dependent evolution (e.g. on \vexp or \fa) if found by both, \igmz and \igm, this would give robustness to the result. 


\subsection{ Output of artificial neural networks }\label{ssec:methodoogy_output}

The three models \igmz, \igm and \noigm have a similar output, almost like the ANN in \zp. As in \zp, there are five output variables associated to the outflow configuration \{ \vexp, \nh, \ta, \ew , \w \}. In order to estimate the redshift of the source, another output is the difference between the wavelength set as \lya and the true \lya wavelength in the proxy rest frame, $\Delta \lambda_{\rm True}$. The true \lya wavelength in the observed frame, $\lambda^{\rm Obs}_{\rm True}$, can be recovered as
    
    \begin{equation}
        \Delta \lambda_{\rm True} = \lambda_{\rm True}^0 - \lambda _{\rm Ly\alpha} = \lambda _{\rm Ly\alpha} \left( \frac{\lambda^{\rm Obs}_{\rm True}}{\lambda_{\rm max}} - 1 \right),
    \end{equation}
    Then, the redshift of the source is  $z=\lambda^{\rm Obs}_{\rm True}/\lambda_{\rm Ly\alpha}-1$. For further details see  \cite{gurung_lopez_2022}.

In the case of  \igmz, \igm we included additional variables. In each of the \lya line profiles used for the training set, we measure the fraction of photons that escape the IGM in wavelength intervals centered around \lya. We refer to these variables as $f_{\rm esc}^{x\AA}$ where $x$ is the width of the wavelength window in rest frame used. For example, \fA is the \lya IGM escape fraction in $\lambda _{\rm Ly\alpha}\pm2\AA$ in rest frame.  \zelda's \igmz, \igm models include wavelength windows from 1\AA{} to 10\AA{} . We find that for a wavelength window of 4\AA{} the \lya IGM escape fraction converge. Increasing the window size does not increase $f_{\rm esc}^{x\AA}$. Also, up to a wavelength window of 4\AA{}  the escape fraction increases with window size. This is reasonable since the mean IGM transmission curves show a drop close to the center of the \lya line, then stabilize to the cosmic mean IGM transmission. 
\subsection{ Architecture of artificial neural networks. }\label{ssec:methodoogy_architecture}

For each output property, we trained an independent ANN. We tested different configurations. We found that for  \{ \vexp, \nh, \ta, \ew , \w \} and $f_{\rm esc}^{x\AA}$ the best configuration was a three-layer ANN with sizes (103 , 53 , 25). Meanwhile, for $\Delta \lambda_{\rm True}$ we found that a nine layer ANN had the best accuracy with sizes (103,90,80,70,60,50,40,30,20).

We have performed a feature importance analysis in Appendix ~\ref{ssec:methodoogy_feature}. Overall, we find that the region $\pm 4\AA{}$ around the \lya alpha wavelength contains the most important information for most predicted shell parameters. 

\subsection{ Redshift, outflow and IGM escape fraction estimation }\label{ssec:methodoogy_estimation}

The shell properties \vexp, \nh, \ta, \ew , \w and  the IGM escape fractions $f_{\rm esc}^{x\AA}$ and $\Delta \lambda_{\rm True}$ are obtained as the median of the distribution of outputs resulting from the ANN using as input 1000 perturbations of the original observed \lya line profile by its noise pattern \citep[as in][]{gurung_lopez_2022}. The percentiles 16 and 84 are used as the 1$\sigma$ uncertainty of these properties. In \zp we demonstrated that this methodology achieves better accuracy in contrast to directly using the ANN output of a single realization. Additionally, making multiple iterations provides the uncertainty for the measurement. This is further discussed in Appendix \ref{ap:uncertainty_estimation}.

\section{Results on mock Lyman-$\alpha$ line profiles}\label{sec:results_mock}
In this section, we show the results of the three ANN models presented in this work on mock observed \lya line profiles. First, we characterize the accuracy in reconstructing the line profiles in Sect.~\ref{ssec:results_mock_parameters}. Then we show the performance of \zelda in recovering the evolution of the IGM escape fraction through time in Sect.~\ref{ssec:results_mock_fesc}. Finally, analyze \zelda's capability to reconstruct the intrinsic stack spectrum emerging from galaxies and before the IGM radiative transfer in Sect.~\ref{ssec:results_mock_stack}.

\begin{figure*} 
            \includegraphics[width=7.00in]{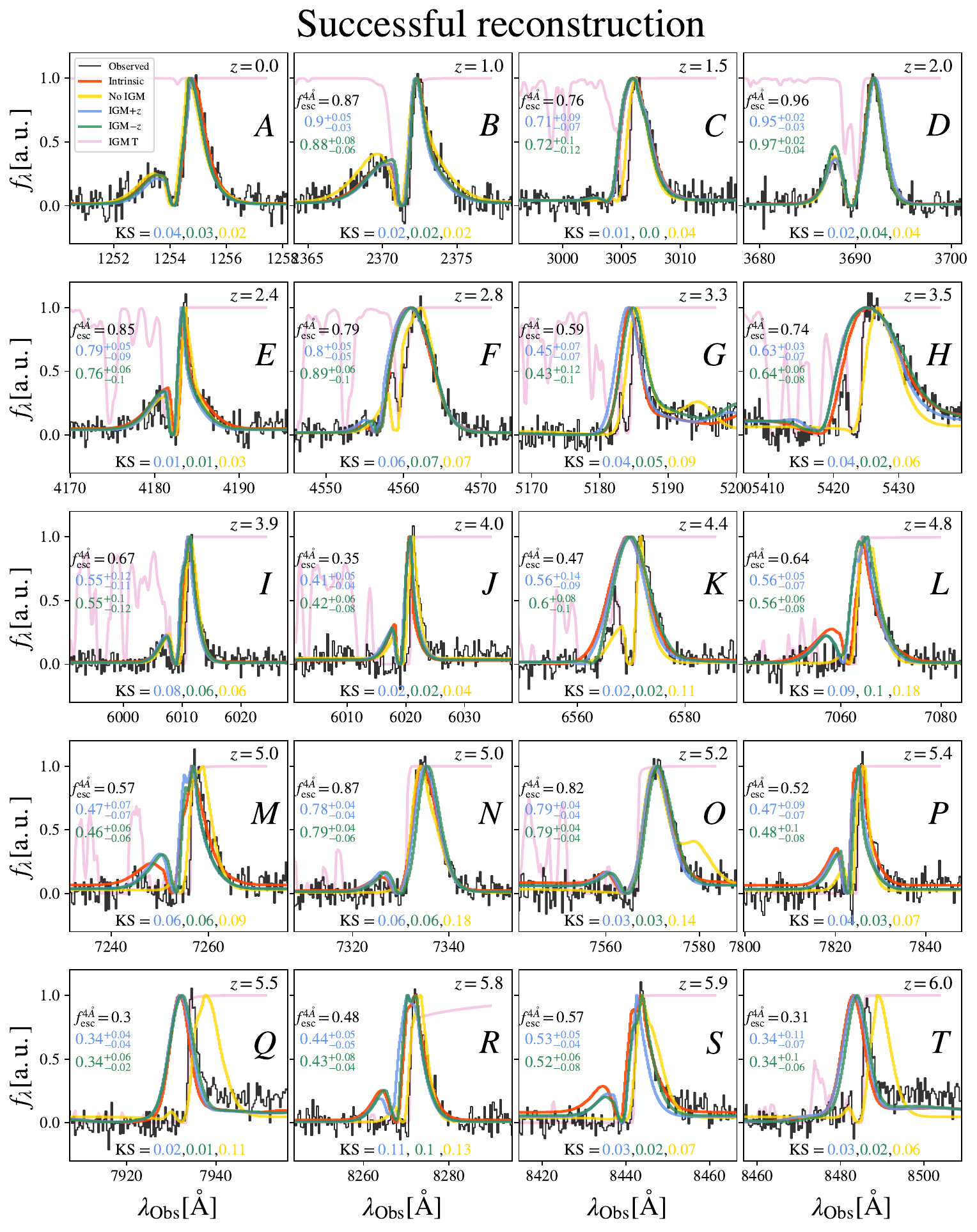}%
        \caption{ Examples of line profile successful reconstruction at different redshift. The redshift of the mock line profile is in the top right corner of each subpanel. The \lya line after the ISM and before traveling through the IGM is shown in red. The IGM transmission curve is shown in pink. The observed line profile, after IGM absorption and mocking observation conditions, is shown in black. \sn=15.0 is fixed for all the line profiles. Meanwhile, \wg is $0.1(1+z)$ so that the resolution element is constant in rest frame. \zelda's prediction for the models \igmz, \igm and \noigm are displayed in blue, green and yellow, respectively. In each panel the true \fa is displayed in black while \zelda's predictions are shown in color text matching the model used with their uncertainties below it. In the bottom of each panel the KS between the true \lya line profile before the IGM absorption and \zelda prediction is displayed in different colors matching the model used. }
        \label{fig:mock_example_successful_reconstruction}
        \end{figure*}
\begin{figure*} 
        \includegraphics[width=7.2in]{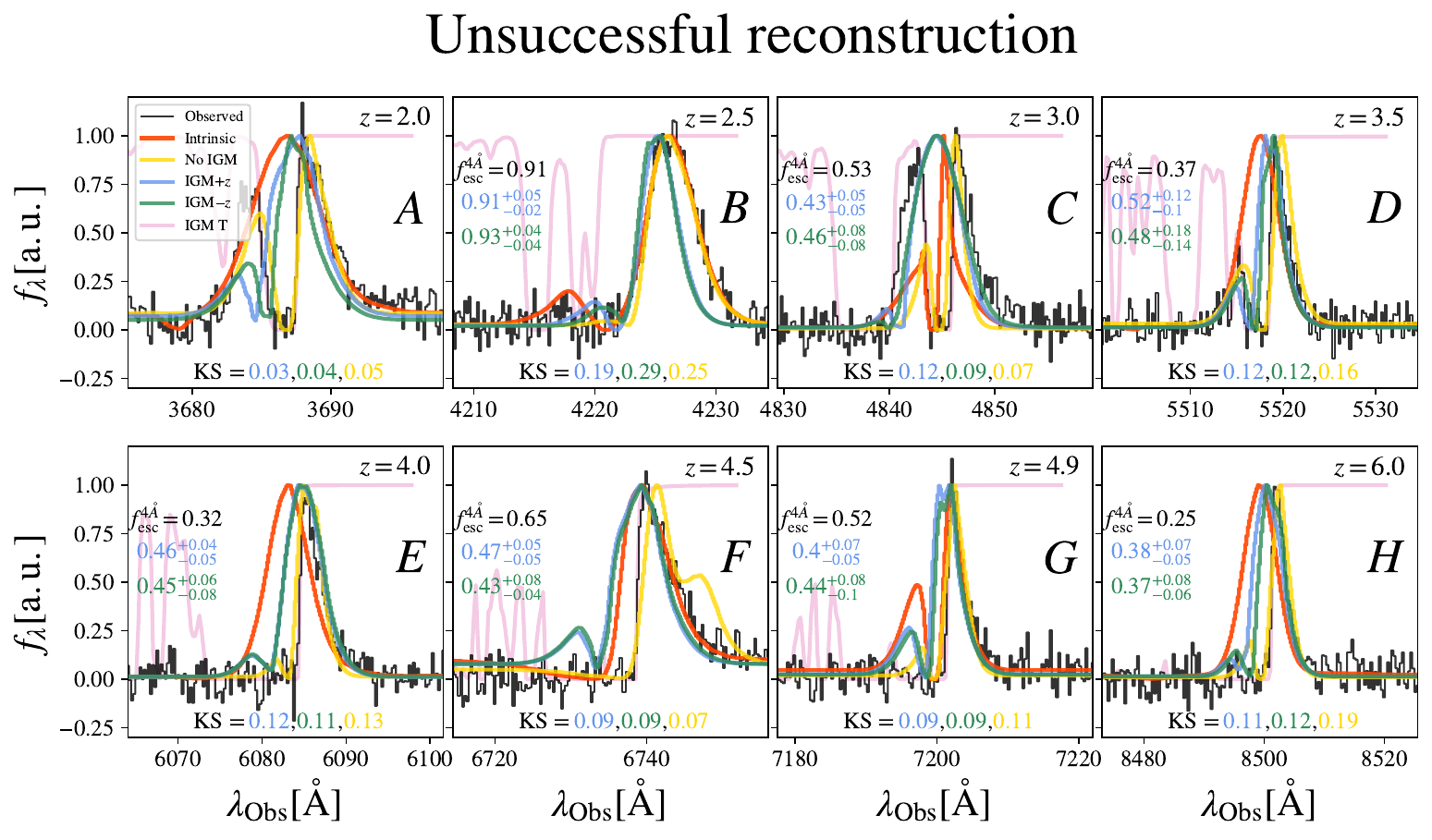}%
        \caption{ Same as Fig.~\ref{fig:mock_example_successful_reconstruction} but for unsuccessful reconstructions. }
        \label{fig:mock_example_unsuccessful_reconstruction}
        \end{figure*}

\subsection{ Accuracy of the ANN models }\label{ssec:results_mock_parameters}

Here we characterize the accuracy of \igmz, \igm and \noigm for recovering the redshift, outflow parameters and \lya IGM escape fraction. First, we show some individual examples. Further details on the accuracy of each parameter are given in Appendix \ref{sssec:results_mock_parameters_parameters}. Moreover, we show the precision in the recovered line profile shape in Appendix \ref{sssec:results_mock_parameters_line}.


In Fig.~\ref{fig:mock_example_successful_reconstruction} we display 20 individual mock line profiles that were successfully reconstructed by \zelda's \igmz, \igm models. In each panel, the intrinsic \lya line profile leaving the galaxy is shown in red, while the IGM transmission curve is shown in pink. The mock observed \lya line profiles used to build the input for the ANN models are shown in black. The quality of the mock line profiles is \sn=15.0, \wg is $0.1(1+z)$ so that the resolution element is constant in rest frame, and \dl=\wg/2. \zelda's prediction using \igmz , \igm and \noigm are displayed in blue, green and yellow, respectively. The true \fa is shown in black, while the predictions of \igmz , \igm are given in top blue and bottom green, respectively, with their 1$\sigma$ uncertainty. Kolmogorov-Smirnov is shown for \igmz , \igm and \noigm, from left to right, between the predicted line and the intrinsic (red).

\lya line profiles in Fig.~\ref{fig:mock_example_successful_reconstruction} are typically reconstructed by \igmz , \igm with KS<0.1 . Both, the red peaks and the blue peak are properly recovered at the same time. For example, in cases {\it E} and {\it I} the observed \lya line profiles still show some hints of the existence of a blue peak previously to the IGM. \igmz , \igm and reconstruct quite well the intrinsic blue peak. However, despite the general good reconstruction of the blue peaks, sometimes \igmz or \igm underpredict the blue peaks (case {\it S} ). This tends to happen when the IGM absorption is so strong that most of the blue peak information is erased from the observed spectrum. However, it is remarkable that even in some scenarios of heavy IGM attenuation, both the blue and red peaks are properly reconstructed (cases {\it J}, {\it N}, {\it O}, {\it P} and {\it R}). Moreover, in some extreme cases where half or more of the line is obscured (cases {\it P}, {\it Q}, {\it S} and {\it T}), the \lya line profiles are reconstructed with typical KS<0.8. 

Meanwhile, \noigm works relatively well at low redshift (cases {\it A}, {\it B}, {\it C} ). However, \noigm fails to recover the intrinsic \lya line profiles of the heavily attenuated observed \lya line profiles. In fact, the red peaks are relatively well fitted, while the blue peaks are poorly reconstructed (cases {\it L}, {\it M}, {\it P}, {\it R}, {\it S}). There are some cases in which \noigm fails to reconstruct the red peak of the line (cases {\it Q}, {\it T}) as well. 

The examples in Fig.~\ref{fig:mock_example_successful_reconstruction} demonstrate two key aspects of the \lya line profile reconstruction. First, Fig.~\ref{fig:mock_example_successful_reconstruction} shows that in some cases,  the IGM attenuation can reshape a thin shell \lya line profile into another \lya line profile similar to another thin shell  configuration. This is made evident by comparing the \noigm output to the observed lines in cases {\it K}, {\it L}, {\it M}, {\it N}, {\it R} and {\it S}. Second, notice that especially in examples {\it K}, {\it L}  and {\it N} the red peak of the observed line profile is well fitted by the three models, including \noigm.  \noigm gives a different prediction for the blue peak than \igmz and \igm. This shows that the observed red peak degenerates with the intrinsic blue peak. These two facts result in a, perhaps inevitable, confusion in the reconstructed shell parameters, redshift and \lya IGM escape fraction for some observed \lya line profiles. 

Fig.~\ref{fig:mock_example_unsuccessful_reconstruction} shows some cases where the line profile reconstruction by \igmz , \igm could be considered deficient or improvable (KS>0.1). The color code follows Fig.~\ref{fig:mock_example_successful_reconstruction}. In general, we find that \igmz , \igm  success or fail in the same line profiles. Only in a few cases \igmz or \igm accurate recover the intrinsic line (e.g. KS=0.04) and the other IGM model gives an inaccurate reconstruction (e.g. KS=0.2). \igmz , \igm tend to give worse predictions when the observed line after IGM absorption resembles that of a wrong outflow model (here, especially cases {\it A} and {\it D}). Another source of inaccuracy is the complete destruction of the blue side information, leading to an over/underprediction of the \lya blue peak (cases {\it F} and {\it G}). In general, we also find that the \noigm model does not recover the correct intrinsic line profile when \igmz and \igm do. 

We study the ratio between successful and unsuccessful reconstructions. For simplicity, we use $KS=0.1$ as a threshold to distinguish between successful and unsuccessful reconstructions. We find that the fraction of sources with $KS<0.1$ (properly recovered) depends on the quality of the \lya line profile. The better the spectral quality, the higher the fraction of sources with $KS<0.1$. Taking into account \igmz and \igm, the fraction for sources with $KS<0.1$ is greater than 90\% for many of the quality configurations explored. We find that a good fraction of the line profiles are properly recovered even at relatively bad spectral quality. In particular,  we find that for \sn>7.5 and \fa>0.5 the  $Q(KS=0.1)$>70\% typically, even at \wg=4.0\AA. More details can be found at  Appendix \ref{sssec:results_mock_parameters_line}.

\subsection{ Reconstructing the IGM escape fraction evolution }\label{ssec:results_mock_fesc}

In this section, we explore \zelda's capability to measure the \fa evolution through cosmic time. For this purpose, we develop mock samples of \lya line profiles with different IGM escape fraction redshift dependences and analyze \zelda's performance on these mocks.

\subsubsection{ Mean \lya IGM escape fraction redshift dependence reconstruction. }\label{sssec:results_mock_fesc_mocks}

\begin{figure*} 
        \includegraphics[width=7.2in]{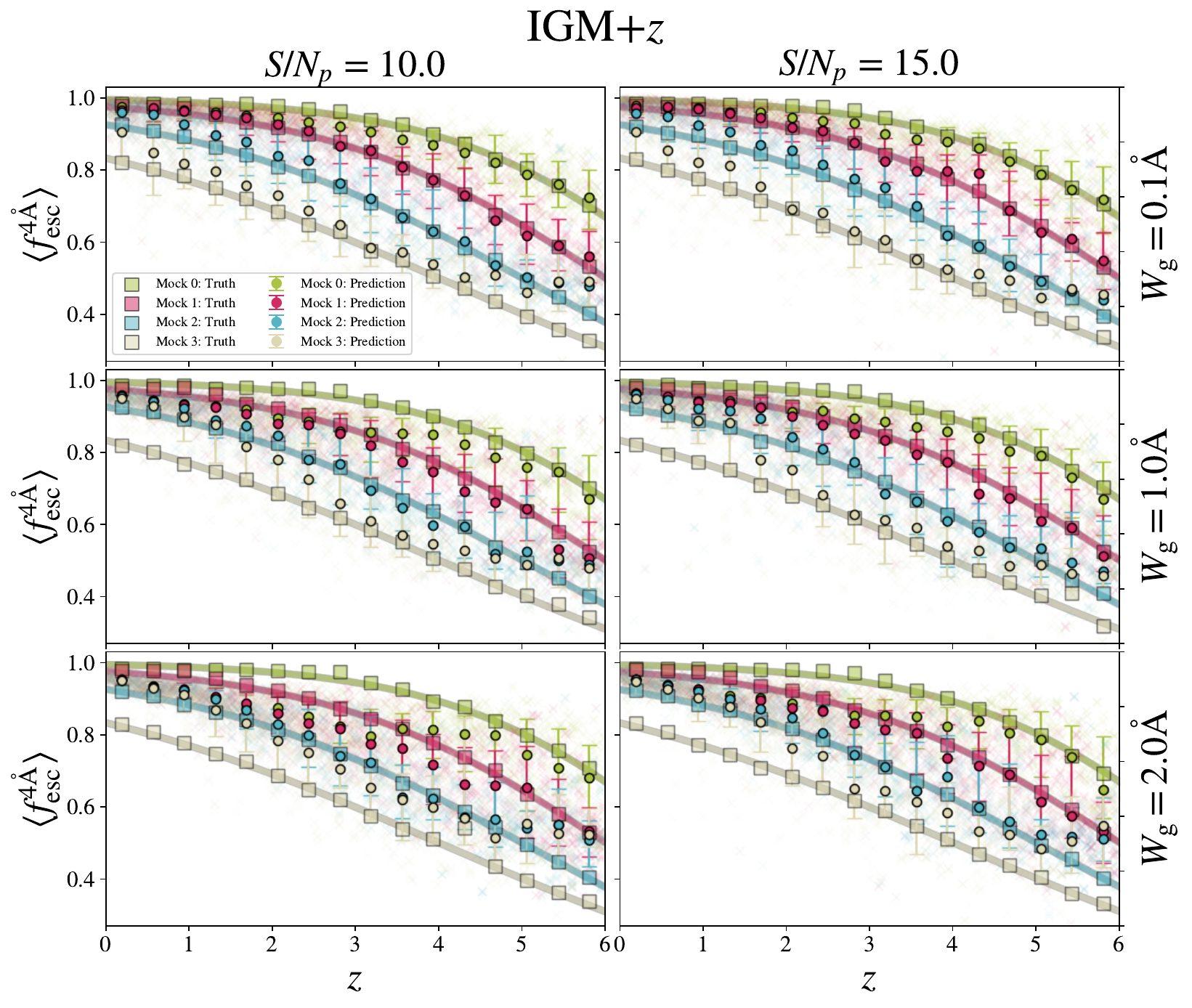}%
        \caption{ \zelda's prediction on the mean \fa for different mock \lya line profiles using the \igmz model. Each panel shows a different observation quality. The left and right column display \sn=10.0 and \sn=15.0, respectively. Each row has a constant \wg. In particular, \wg=0.1\AA, 1.0\AA\ and 4.0\AA, from top to bottom. Four mean \fa $z$ evolution scenarios are considered, which are shown in solid lines (green, red, blue and grey). The true mean \fa in the mocks is shown in colored squares. \zelda's predictions for individual \lya line profiles are marked as crosses. The mean \fa from \zelda's prediction is shown in circles with it's uncertainty.   }
        \label{fig:Mock_f_esc_IGM+z}
        \end{figure*}
\begin{figure*} 
        \includegraphics[width=7.2in]{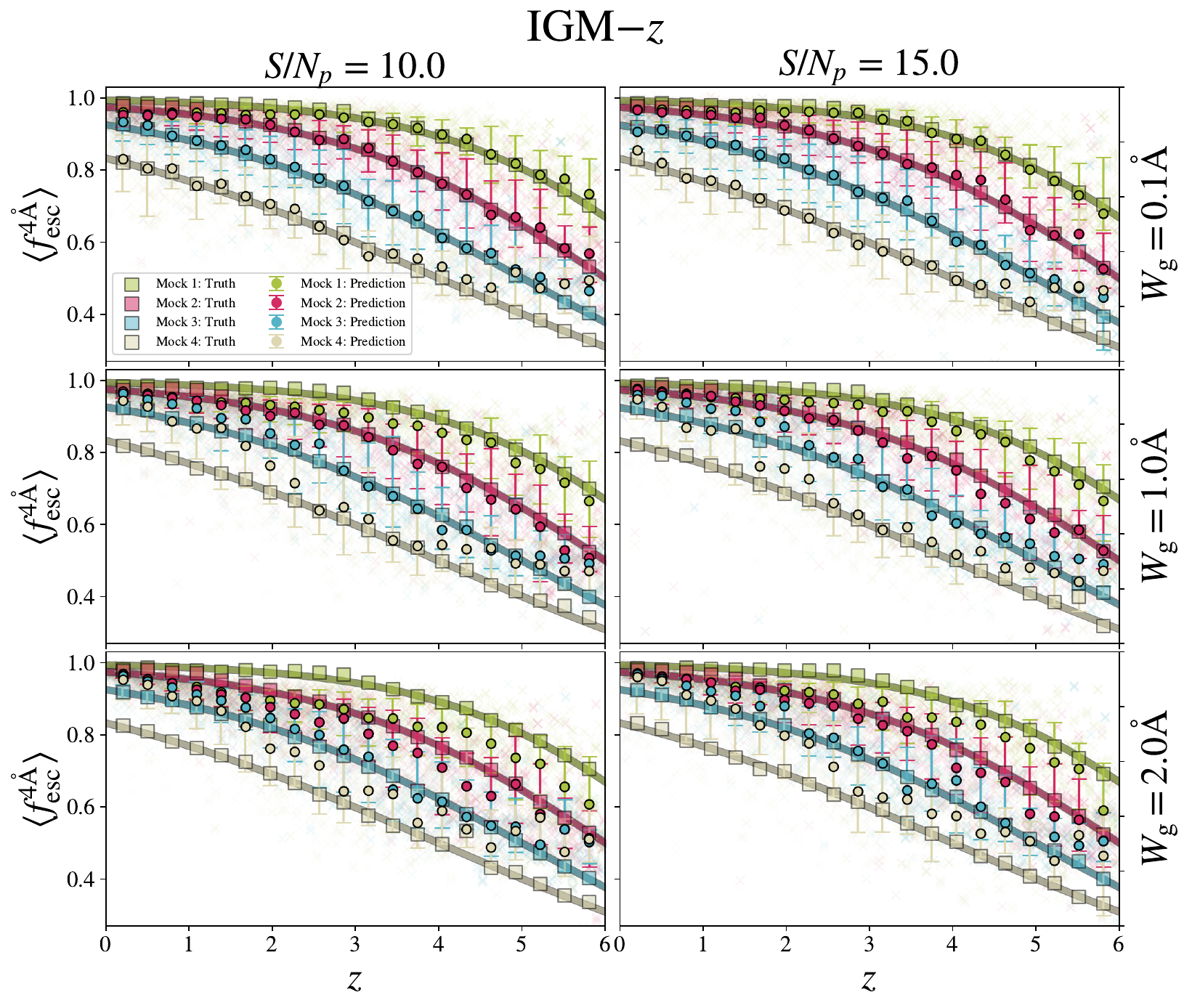}%
        \caption{ Same as Fig.~\ref{fig:Mock_f_esc_IGM+z} but for the \igmz model. }
        \label{fig:Mock_f_esc_IGM-z}
        \end{figure*}     

In this section, we study the capabilities of \zelda to reconstruct the redshift evolution of the mean \fa. For this goal, we develop \lya line profile mock samples with different mean \fa evolution. Later, we study the accuracy of \igmz and \igm in these mocks. 

We parameterize the mean \fa redshift dependence as a Fermi-Dirac  distribution, i.e.,

\begin{equation}
    \displaystyle \langle f_{\rm esc}^{4\AA} \rangle = \displaystyle\frac{1}{ e^{ b  (z-a) } + 1 },
\end{equation}
where $a$ and $b$ are two free parameters. The Fermi-Dirac distribution was chosen to reproduce the expected average evolution from an opaque (f=0.0) to transparent IGM (f=1.0) with an asymptotic behavior on both ends.

For a given combination of parameters $a$ and $b$ we generate a set of 500 \lya line profiles homogeneously distributed from $z=0$ to 6. Each \lya line profile at redshift $z_1$ is assigned an IGM transmission curve at a random redshift $z_2$. Next, the outflow line profile (intrinsic) is convolved with the chosen IGM transmission curve and \fa is measured. If the computed \fa is within 10\% of the $\langle f_{\rm esc}^{4\AA} \rangle(z_1)$, then this \lya line profile is accepted as valid. However, if this condition is not met, a new IGM transmission curve at another random redshift $z_3$ is assigned until the condition is fulfilled. By construction, all \lya line profiles will exhibit \fa values close to the Fermi-Dirac with parameters $a$ and $b$.

We have produced \lya line profile mocks for four $a$ and $b$ combinations that populate the $\langle f_{\rm esc}^{4\AA} \rangle$-$z$ space. In Figs. \ref{fig:Mock_f_esc_IGM+z}  and  \ref{fig:Mock_f_esc_IGM-z} we show  four $\{a,b\}$ combinations, which are \mone $\{7.0,0.7\}$ (green), \mtwo $\{6.0,0.6\}$ (red), \mthree $\{5.0,0.5\}$ (blue) and \mfour $\{4.0,0.4\}$ (grey). For each $a$ and $b$ combination, we produce mocks with six spectral quality configurations. For \sn=10.0 (left column) and \sn=15.0 (right column) three values of \wg are run, 0.1\AA{}, 1.0\AA{} and 2.0\AA{} from top to bottom. The colored solid lines show the parametric \mfa and the colored squares the actual $\langle f_{\rm esc}^{4\AA} \rangle$ in that redshift bin. 

In Fig.~\ref{fig:Mock_f_esc_IGM+z} we show the \igmz prediction for the mock samples. The colored crosses show the \fa predictions line per line. The colored circles mark \igmz's estimations of \mfa in the same redshift bins as for the true values. Meanwhile, the results for \igm are shown in Fig.~\ref{fig:Mock_f_esc_IGM-z}. In general, we find that both \igmz and \igm recover the \mfa evolution with redshift. For both models, the precision of the reconstructed \mfa evolution changes with the spectral quality. For line with higher resolution and \sn the recovered \mfa follows well the true \mfa evolution. However, when the spectral quality decreases, some biases appear in the \mfa estimate.  We also find that \igmz and \igm show a limitation for low \fa values ($\sim 0.4$). For example, at \sn=15.0, \wg=0.1\AA{}, \mfa stales at $\sim 0.45$ at redshift $\sim 5$ in \mfour while it should fall down to 0.3.  Also, for the worst spectral quality configuration (\wg=2.0\AA) we find that, while the general \mfa redshift evolution is recovered, \mfa is slightly biased for \mone and \mfour. In particular, the results in \mone are  up to a $\sim10\%$ under predicted, while those for \mfour are up to a  $\sim10\%$ over predicted. Meanwhile, the results for \mtwo and \mthree for \wg=2.0\AA{} is quite unbiased and most individual measurements are 1$\sigma$ compatible with the true values. 

We find that while the \sn of the line has an impact on the accuracy of the recovered \mfa, \wg has a greater influence. This is also found when determining \fa in individual line profiles. As shown in Fig.~\ref{fig:acuracy_IGM-z}, focusing in the \fa$\in[0.95,1.0]$ regime, for \wg=0.1\AA{}, changing \sn from 15.0 to 5.0 produces a drop in \fa accuracy from 0.03 to 0.04. However, for \sn=10.0, the \fa accuracy at \wg=0.25\AA{} is 0.03, while at \wg=2.0\AA{} is 0.1. This becomes even more apparent in the \fa$\in[0.65,0.8]$, where there is no clear \fa accuracy dependence on \sn, while it gets worse for larger \wg values.  

In general, Fig.~\ref{fig:Mock_f_esc_IGM+z} shows that the \igmz model provide an accurate \mfa estimation for the  explored mocks at \wg=0.1\AA{}, as most of the measurements are 1$\sigma$ compatible with the true values. Moreover, \igmz provides a relatively unbiased and accurate prediction between redshifts 2 and 5 for the four mocks presented and for all spectral quality configurations. However, we find that the \igmz model is biased at $z<1$ at every explored spectral quality (e.g. \mthree and \mfour). \igmz tends to over predict \mfa and gives values close to unity at this redshift range. This bias becomes stronger as the spectral quality decreases. For example, at $z=1$ and \wg=0.1\AA, \mfa is over predicted a $\sim$10\%.  Nevertheless, the general trend (\mfa decreases with $z$) is recovered in all spectral quality configurations and redshift bins. 

Focusing in \igm (Fig.~\ref{fig:Mock_f_esc_IGM-z}), \mfa is nicely recovered for \wg=0.1\AA{} and \sn=10.0 and \sn=15.0 for the  explored evolution cases. Also, we find that \igm is less biased toward \mfa=1.0 than \igmz at low redshift. For example at \wg=0.1\AA{}, the \mfa evolution in \mfour is recovered almost perfectly with no apparent bias at $z$<1. For \wg=1.0\AA{} , \mfa is a 10\% overestimated for \mfour at $z$<1. Moreover, unlike for \igmz, at $z<1$ the rank order in \mfa is recovered properly. Individual measurements of \mone are greater than those of \mtwo, which are greater than those of \mthree, which are above those of \mfour. Furthermore, at $z$>1 we find the same trends as in \igmz.

\begin{figure*} 
        \begin{center}
        \includegraphics[width=7.2in]{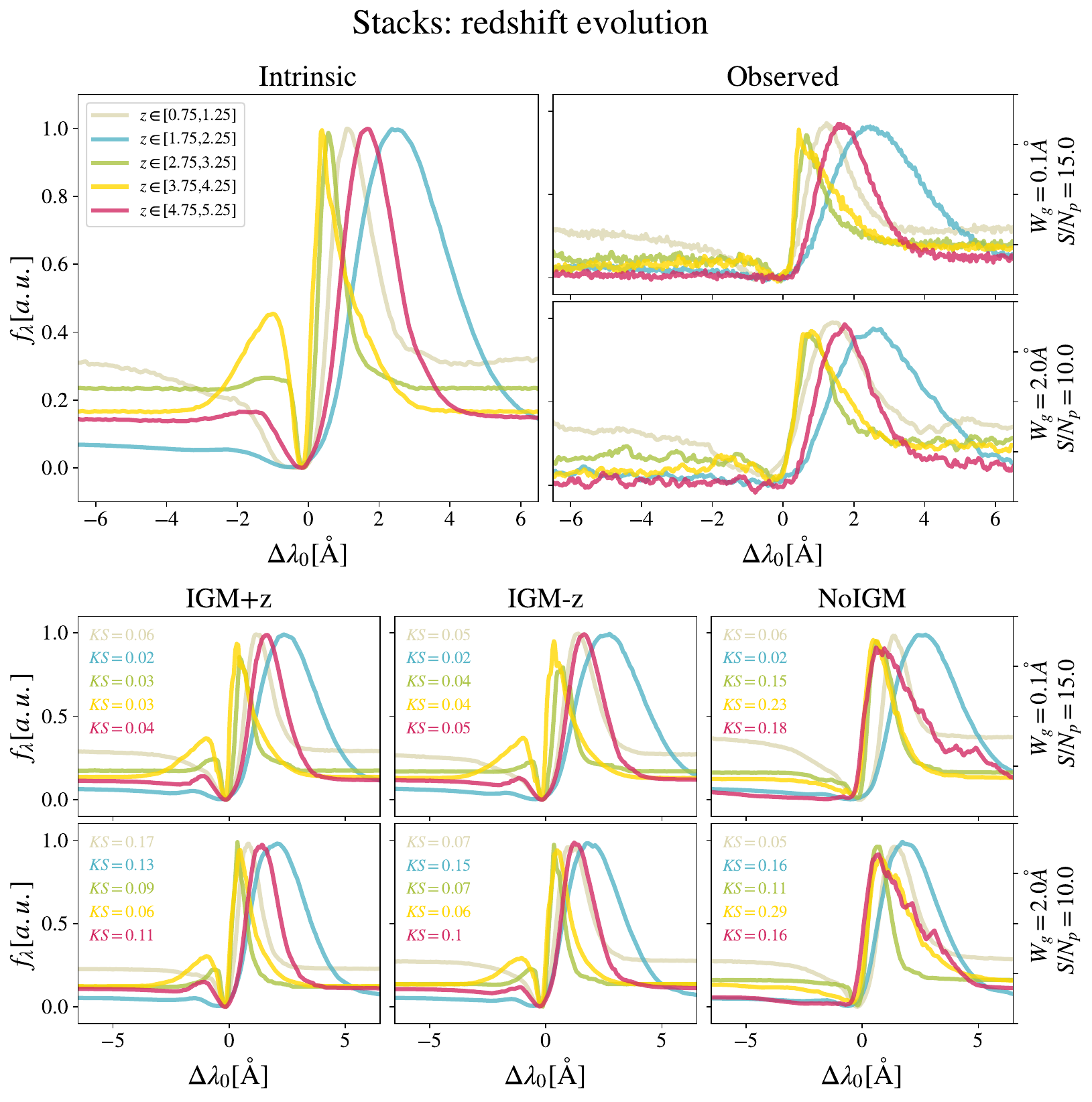}%
        \caption{ Stacked line profile reconstruction example in mock \lya  line profiles using the \lya line profiles with redshift dependence.  The stacked line profile is shown the redshift intervals [0.75,1.25] (grey), [1.75,2.25] (blue), [2.75,3.25] (green), [3.75,4.25] (yellow) and [4.75,5.25] (red). The \lya stacked line profiles using the \lya line profiles before applying the IGM absorption is displayed in the top left panel. The two top right panels display the \lya stacked line profile after applying the IGM absorption and mocking observation quality similar to HST (${\rm W}_g=0.1\AA$ and $S/N_p=15.0$, top) and MUSE (${\rm W}_g=2.0\AA$ and $S/N_p=10.0$, bottom). The six bottom panels show the reconstructed stacked \lya line profiles. Each bottom column make use of a different ANN model: \igmz, \igm and \noigm from left to right. The KS between the stacked \lya line profile before the IGM (left column) and that of the reconstructed \lya line profiles is displayed in colored text matching the redshift bin.   }
        \label{fig:Mock_lines_stack_EVO}   
        \end{center}
        \end{figure*}     
\begin{figure*} 
        \begin{center}
        \includegraphics[width=7.2in]{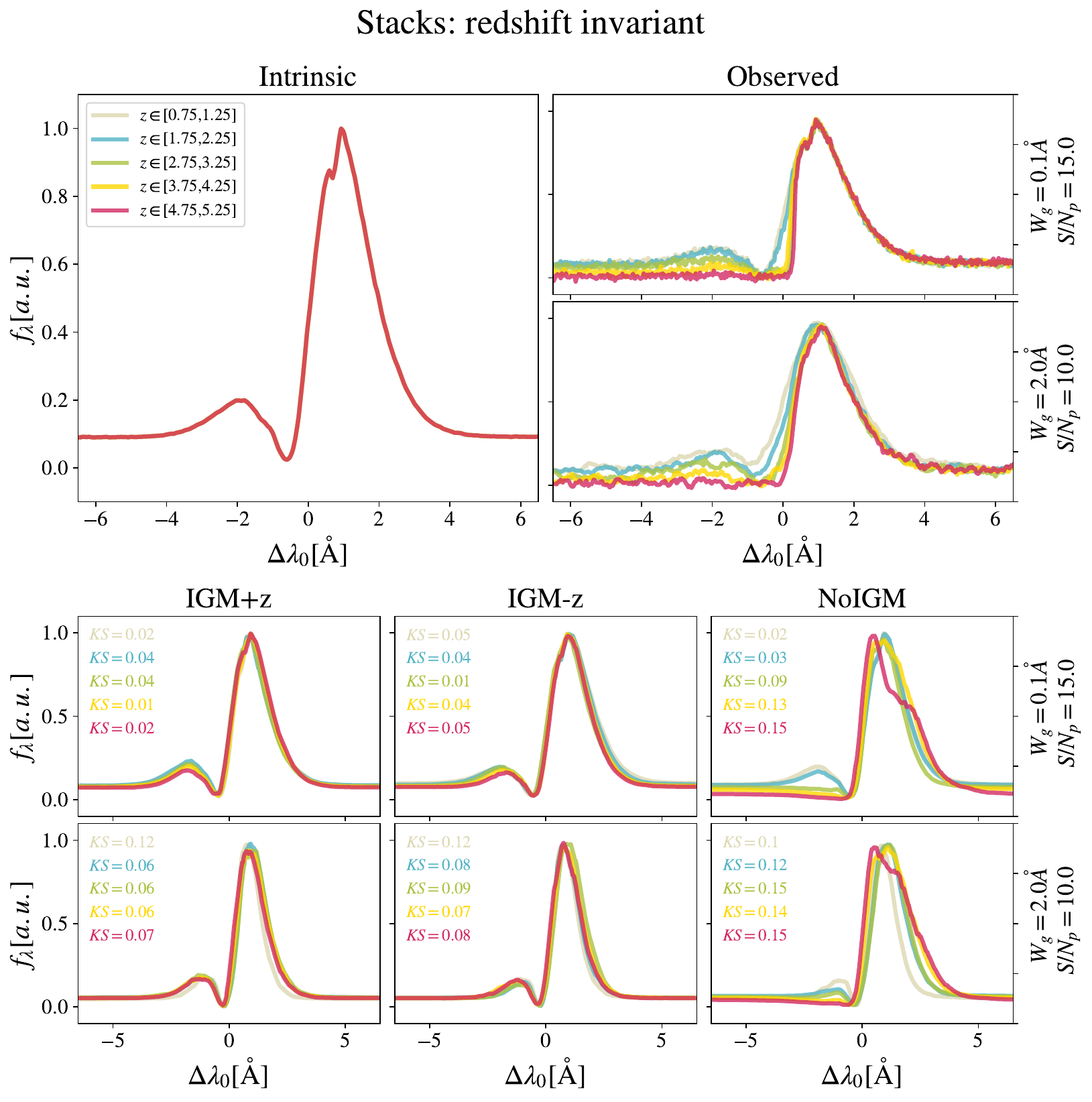}%
        \caption{ Same as Fig.~\ref{fig:Mock_lines_stack_EVO} but for the redshift invariant \lya line profiles. }
        \label{fig:Mock_lines_stack_STA}
        \end{center}
        \end{figure*}     

\begin{table}
\caption{ Outflow parameters in the redshift nodes used for the \lya line profile stacked mocks. }
\begin{center}
\begin{tabular}{cccccc}
$z$ & \vexp & $\log$ \nh & \ew & \w & \ta \\
    &  [\kms]  &      [$\rm cm^{-2}$]              & [\AA{}] & [\AA{}] &  \\ \hline
 0.0   &  200.0 & 19.3 & 18.0 & 2.0 & 0.01  \\
 1.0   &  125.0 & 19.9 & 12.0 & 0.7 & 1.0  \\
 2.0   &  190.0 & 20.7 & 50.0 & 1.6 & 0.02  \\
 3.0   &  40.0  & 19.1 & 1.0 & 0.25 & 0.0004 \\
 4.0   &  70.0  & 18.2 & 12. & 1.6 & 0.5  \\
 5.0   &  70.0  & 20.3 & 9.0 & 1.6 & 0.0001  \\
 6.0   &  170.0 & 19.3 & 300 & 1.75 & 0.0004 
\end{tabular}
\end{center}
\label{tab:stacks_z_evo}
\end{table}

\subsection{ Stacked line profile reconstruction. }\label{ssec:results_mock_stack}

In this section, we explore \zelda's capability to reconstruct the stack of intrinsic \lya line profiles. Our focus here is to study whether \zelda is able to measure the evolution or non-evolution of the intrinsic stacked \lya line profile. For this goal, we used mock samples of the \lya line profiles. Meanwhile, a detailed analysis for individual spectral recovery can be found in Appendix \ref{sssec:results_mock_parameters_line}.

Given a sample of \lya line profiles with a given redshift distribution, we follow the next procedure. First, we run \zelda (\igmz, \igm or \noigm) in order to obtain the redshift of the \lya line profile by running our ANN.  Next, we move all observed \lya line profiles to a common wavelength array in the rest frame using the redshift provided by \zelda. Every \lya line profile is normalized so that the their maximum reaches unity. Finally,  the stacked \lya line profile is computed as the median flux in each wavelength bin. 

We present two types of samples of mock \lya line profile. In the first one, \mevo, the  intrinsic \lya line profiles change with redshift. In particular, for a given \lya line profile at redshift $z_t$, the outflow properties are linearly interpolated from the $z$ nodes listed in Tab.\ref{tab:stacks_z_evo}.  For illustration, in each node, the outflow parameter combination was chosen so that the intrinsic line profile changed dramatically from node to node. The top left panel of Fig.~\ref{fig:Mock_lines_stack_EVO} shows the stacked \lya line profile at redshift between 0.75 and 1.25 (grey), between 1.75 and 2.25 (blue), between 2.75 and 3.25 (green), between 3.75 and 4.25 (yellow) and between 4.75 and 5.25 (red). 

In the second mock sample, \mfix, all the line profiles use the same outflow parameters:\vexp=200.0 \kms , \nh=19.3 $\rm cm^{-2}$, \ew 20.0 \AA{}, \w 2.0 \AA{} and \ta=0.001. These were chosen so that the intrinsic line profile resembled the observed \lya line profiles stack at low redshift \citep{Hayes_2023}. The top left panel of Fig.~\ref{fig:Mock_lines_stack_STA} shows this outflow line profile.

The \mevo observed stacked \lya line profile after including the IGM is shown in the top right panels of Fig.~\ref{fig:Mock_lines_stack_EVO} for two spectral quality combinations: \sn=15 and \wg=0.1\AA (top) and  \sn=10 and \wg=2.0\AA (bottom). For both spectral quality configurations, the higher the redshift, the more attenuated the blue side of \lya. Meanwhile, the red side of \lya remains unabsorbed. This is a direct consequence of the redshift dependence and shape of the mean IGM transmission curves (shown in Fig.~\ref{fig:IGM_T_mean}). For instance, the intrinsic stacked line profile at $z\sim1$ remains mostly unchanged after applying the IGM absorption. Furthermore, the stacked line profile at $z\sim5$ intrinsically exhibits a relatively strong blue continuum and a faint blue peak. These features are erased from the observed stack spectrum as a result of the IGM absorption. In the same way, the strong blue peak in the intrinsic stack at $z\sim4$ is also erased.

At the same time, the \mfix observed stacked \lya line profile after including the IGM (top right panels of Fig.~\ref{fig:Mock_lines_stack_STA}) shows the same trends as that of \mevo. We find that the higher the redshift, the more absorbed is the blue side of \lya. In particular, at $z\sim4$ the blue peak is erased. In addition, spectral quality has an impact on the observed stack spectrum of \lya. As \wg is fixed in the observed frame, the resolution element in rest frame is smaller for sources at higher redshift. This causes that, for \sn=10.0, \wg=2.0\AA{}, the \lya stack spectrum at $z\sim1$ is wider than at $z\sim5$.

We present the results for the stacked \lya line profile reconstruction in Fig.~\ref{fig:Mock_lines_stack_EVO} for the redshift-dependent \lya line profile (\mevo). In order to compute the reconstructed intrinsic \lya line profiles we run \zelda in each individual observed line profile. Next, we follow the same procedure for computing the stacked line profile described earlier.  The stacked \lya line profile reconstructed by \igmz, \igm and \noigm are shown in the bottom columns from left to right. The top row shows \sn=15.0, \wg=0.1\AA{} while the bottom shows \sn=10.0, \wg=2.0\AA{}. The accuracy of the reconstruction is quantified by the KS estimator between the intrinsic stack and the reconstructed one, shown within each panel.  

In general, the three ANN models assess that there is a clear evolution in the \lya stacked line profile. We find that, in general, \igmz and \igm manage to reconstruct the intrinsic stacked \lya line profile accurately with typical values below $KS=0.1$. In particular, the stack is more accurate for the \sn=15.0, \wg=0.1\AA{} configuration with typical $KS<0.06$. For example, both \igmz and \igm recover the blue peaks visible in the intrinsic stack at $z\sim3,4$ and 5. Even the small hint of a blue peak present in the $z\sim2$ bin is recovered. However, the reconstruction in the blue peaks is not perfect. For example, in the $z\sim4$ stack, the reconstructed blue peak is slightly under predicted. It is also noticeable that the amplitude of the continuum blue side \lya is relatively well recovered.

The \noigm model recovers well the intrinsic \lya line profile at $z\sim1$, with $KS=0.05$. However, at $z>1$, the stack is poorly constrained with typical $KS$ values greater than 0.1. For example, the \noigm ANN provides reconstructed \lya line profiles that exhibit the same blue continuum as the observed stacked line profile. In addition, the blue peaks of the high-redshift bins were not recovered.  

In Fig.~\ref{fig:Mock_lines_stack_STA} we show \zelda's prediction for the static \lya line profile sample (\mfix). In general, we find the same trends as in \mevo. Both the \igmz and \igm models recover accurately the non-evolution of the stacked \lya line profile. The stacked line profile is better reconstructed for better spectral quality. Typically $KS<0.10$ for \sn=10.0, \wg=2.0\AA{} and $KS<0.05$ for \sn=15.0, \wg=0.1\AA{}. The blue peak is reconstructed at all redshifts, although with slightly different amplitudes. Meanwhile, \noigm does not manage to recover the non-evolution. \noigm predicts correctly the stacked line profile at $z\sim1$ with $KS=0.02$ for \sn=15.0, \wg=0.1\AA{}. However, at $z>1.0$ does not recover the existence of the blue peak.


\section{Summary and conclusions}\label{sec:conclusions}

The observed \lya line profile is shaped by the complex radiative transfer taking place inside the galaxies in the interstellar medium, in the circumgalactic medium after escaping the galaxy and in the intergalactic medium. In this work, we have presented the second version of the open source code \zelda . \zelda's second version focuses on disentangling between the ISM and IGM contributions to the \lya line profile using artificial neural networks. \zelda can be found at \url{https://github.com/sidgl/zELDA_II} along with installation and usage tutorials at \url{https://zelda-ii.readthedocs.io/index.html}.

Our training sets contain mock \lya line profiles with ISM and IGM attenuation mimicking a wide range of observed spectral quality configurations.  The ISM contributions come directly from the first version of \zelda \zp, which counted with grid of precomputed   'shell model' line profiles using a \lya Monte Carlo radiative transfer \citep[\lyart][]{orsi12}. Meanwhile, the IGM attenuation comes from the \lya transmission curves published by \cite{Byrohl2020}. These are obtained by running a Monte Carlo \lya radiative transfer code \citep[a modified version of ILTIS,][]{Behrens_2019} in the IllustrisTNG100 simulation \citep{Nelson_2019}, at six snapshots between redshifts of 0.0 and 5.0.

We have presented three ANN models. All include the first 100 components of the PCA decomposition of the observed \lya line profile and spectral quality. First, \igmz, includes a proxy redshift of the sources and the IGM transmission lines assigned to the sources in the training are at the redshift of the source. Meanwhile, \igm does not include the redshift of the source in the input and sources are assigned random IGM transmission curves. Finally, for comparison, \noigm, which includes a proxy redshift of the source in the input but the \lya line profiles in the training set lack the IGM contribution.  

\zelda's performance on mock \lya line profile can be summarized as follows:

\begin{itemize}
    \item  We have tested our ANN models in mock \lya line profiles. We find that \igmz and \igm manage to reconstruct the shape of the \lya line profile emerging from the ISM. The accuracy of the reconstruction depends on the spectral quality of the observed line profile. For example, for the typical spectral quality of \lya line profiles obtained by  the Cosmic Origins Spectrograph \citep[{\it COS}][]{Green_2012} on board the {\it Hubble Space Telescope}, 95\% of the \lya line profiles should be recovered with a Kolmogórov-Smirnov estimator below 0.1. Meanwhile, for data with the spectral quality of the  MUSE-WIDE survey \cite{Urrutia2019A&A...624A.141U,Herenz2017}, typically, 81\% of the \lya line profiles are reconstructed with a KS<0.1. 

    \item  Additionally, we have tested \zelda's capabilities to reconstruct the stacked line profile of the line emerging from the ISM from the IGM attenuated line profiles. We find that the \lya stacked line profile can be recovered with KS<0.7 for HST and MUSE-like data by both \igmz and \igm.  Moreover, the precision in the stack reconstruction enables us to detect evolution or non-evolution in the ISM stacked line profile. 

    \item Interestingly \zelda is capable of predicting the IGM \lya escape fraction, \fa, with an uncertainty of $\sim 0.03$ in HST-like data and $\sim 0.12$ for MUSE-like data. In fact, we found that \igmz and \igm are able to detect evolution in \fa from redshift 2.0 onward for MUSE-like data. Meanwhile, \igmz provides \fa values biased towards 1 at z<1.0. In contrast, \igm seems unbiased in this redshift range. 
\end{itemize}

This work advances our modeling of the \Lya emission line and fitting by incorporating the IGM attenuation along the line of sight. \zelda's current version presents some limitations, like recovering \fa at \fa<0.5 or reconstructing the shape of \lya line profiles with \fa<0.4. Nevertheless, this work demonstrates that disentangling the ISM from the IGM a contributions is possible at the level of individual \lya line profiles. This opens multiple scientific paths.  Some examples of the many applications that \zelda can have in observed data are given below.

\begin{itemize}
    \item The exploration of what shapes the ISM emerging \lya line profiles in high-redshift galaxies. This could be done by correlating  the inferred 'shell model' parameters with the luminosity of their spectral features or galaxy properties like mass, neutral hydrogen column density and so on. 
    \item The current \lya luminosity functions (LF) are measured with the observed \lya luminosity, which is attenuated by the IGM. \zelda provides a \fa source-by-source. Therefore, \zelda can provide the \lya LF for star forming sources before the IGM attenuation. This could shed light on the cosmic star formation history at high redshift. 
    \item There is still some debate on whether the IGM large scale properties affect the visibility of \lya. A spectroscopic survey covering a large enough area and using \zelda could directly measure if there is an excess of clustering signal in \fa with respect that of LAEs. 
\end{itemize}



\begin{acknowledgements}

The authors acknowledge the financial support from the MICIU with funding from the European Union NextGenerationEU and Generalitat Valenciana in the call Programa de Planes Complementarios de I+D+i (PRTR 2022) Project (VAL-JPAS), reference ASFAE/2022/025.
This work is part of the research Project PID2023-149420NB-I00 funded by MICIU/AEI/10.13039/501100011033 and by ERDF/EU.
This work is also supported by the project of excellence PROMETEO CIPROM/2023/21 of the Conselleria de Educación, Universidades y Empleo (Generalitat Valenciana).
MG thanks the Max Planck Society for support through the Max Planck Research Group.
DS acknowledges the support by the Tsinghua Shui Mu Scholarship, funding of the National Key R\&D Program of China (grant no. 2023YFA1605600), the science research grants from the China Manned Space Project with no. CMS-CSST2021-A05, and the Tsinghua University Initiative Scientific Research Program (no. 20223080023).



This research made use of matplotlib, a Python library for publication quality graphics \citep{Hunter:2007}, NumPy \citep{harris2020array} and SciPy \citep{Virtanen_2020}.

\end{acknowledgements}


%
%

\bibliographystyle{aa}
\bibliography{ref}

\begin{appendix} 

\section{ Feature importance analysis. }\label{ssec:methodoogy_feature}

\begin{figure*} 
        \includegraphics[width=7.1in]{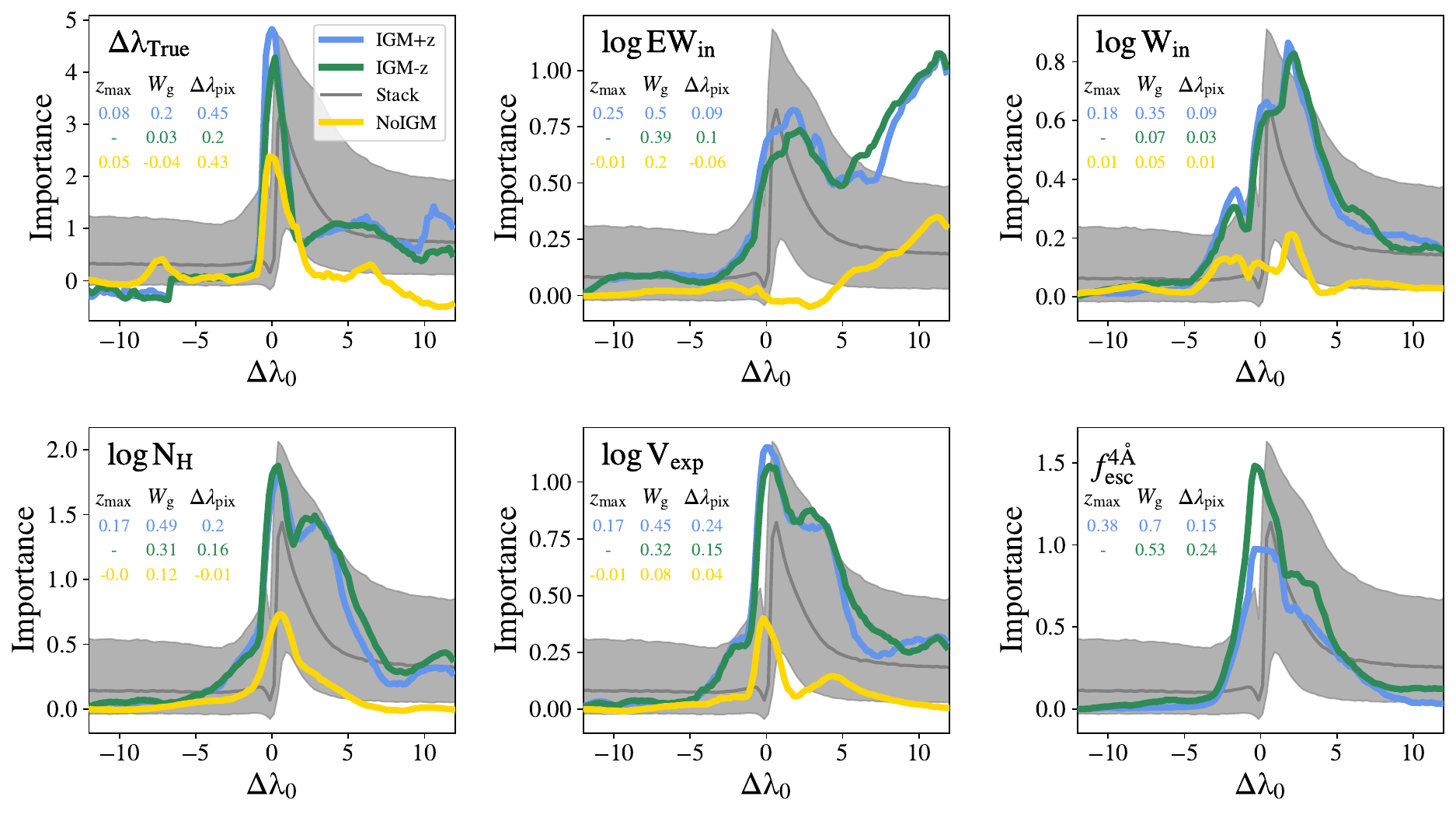}%
        \caption{ Feature importance analysis for the \igmz (blue), \igm (green) and \noigm (yellow). For comparison, the stacked line profiles with its 1$\sigma$ scatter is shown in grey. Each subplot shows the importance determining a different property. In top row, \dlt, \ew, \w from left to right. In the bottom row, \nh, \vexp, \fa from left to right. In each subplot we also show the $z_{\rm max}$, \wg and \dl importance for each model in its matching color in the small table. }
        \label{fig:feature_importance}
        \end{figure*}

The feature importance analysis of the spectral features is made by shuffling the fluxes in wavelength bins of 1\AA{} in the rest frame of the sources, as in \zp. Next, we convert the altered spectrum with the PCA model  and pass it to the ANN along with the actual  \wg, \dl and $z_{\rm max}$ values. Then, in order to compute the importance of \wg, \dl and $z_{\rm max}$, we shuffled these properties one by one. The importance is computed as $I=\sigma_{\rm original}/\sigma_{\rm shuffled}-1$, where $\sigma_{\rm original}$ is the accuracy of the ANN predicting a given output variable and $\sigma_{\rm shuffled}$ is that but using the perturbed input. In general, $\sigma_{\rm shuffled}$<$\sigma_{\rm Original}$, since the input without perturbation contains more information. Therefore, $I>0$, in general. 

In Fig.~\ref{fig:feature_importance} we show the feature importance analysis of $\Delta \lambda_{\rm True}$, \vexp, \nh, \ew , \w  and \fA for the \igmz model with \wg=0.25\AA, \dl=0.125\AA{} and \sn=15.0. The general trends found here are also present for other quality configurations and \igm and \noigm. We find the same general trends as in \zp. For $\Delta \lambda_{\rm True}$, \vexp, \nh , \w  and \fA the regions closer than 5\AA{} contains the most information. This tends to be skewed redwards \lya in the three models. However, the models including IGM, \igmz and \igm, give more importance to redder wavelengths than \noigm. This shows that \igmz and \igm\ 'trust' more the red side of \lya. This makes sense considering that the blue side of \lya could be heavily influence by the IGM. Still, some information in the blue side is being used in order to estimate the output. The ANN constraining \ew gives the most importance to the region +5\AA{} from \lya, which should be dominated by the continuum of the source. In particular, the ANN constraining  $\Delta \lambda_{\rm True}$ exhibits the narrowest importance peak around \lya ($\pm1\AA{}$). Finally, focusing in \fA, we find that in the \igmz model, the importance of $z_{\rm max}$ is 0.38, which shows that it plays a significant role in determining \fA. Actually, \fA is variable for which  $z_{\rm max}$ is the most informative.    

\section{ Individual parameters uncertainty estimation}\label{ap:uncertainty_estimation}

\begin{figure*} 
        \includegraphics[width=2.4in]{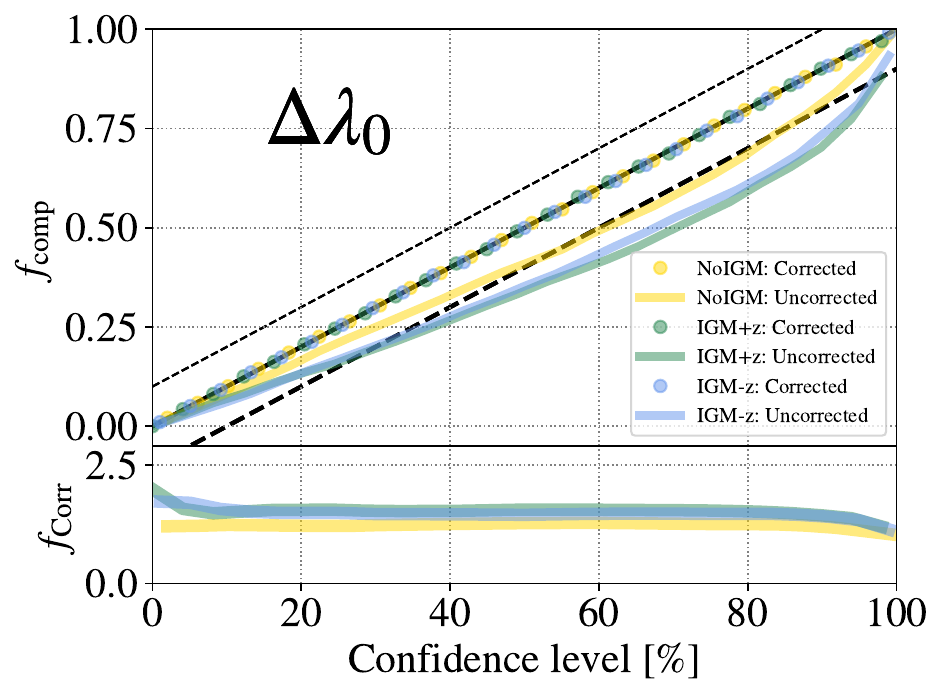}%
        \includegraphics[width=2.4in]{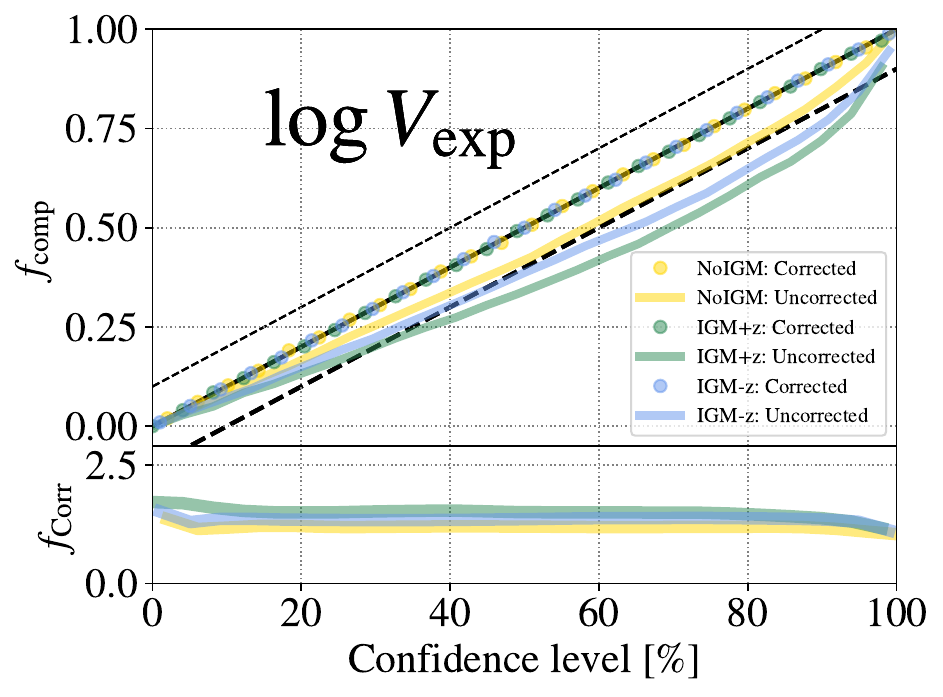}%
        \includegraphics[width=2.4in]{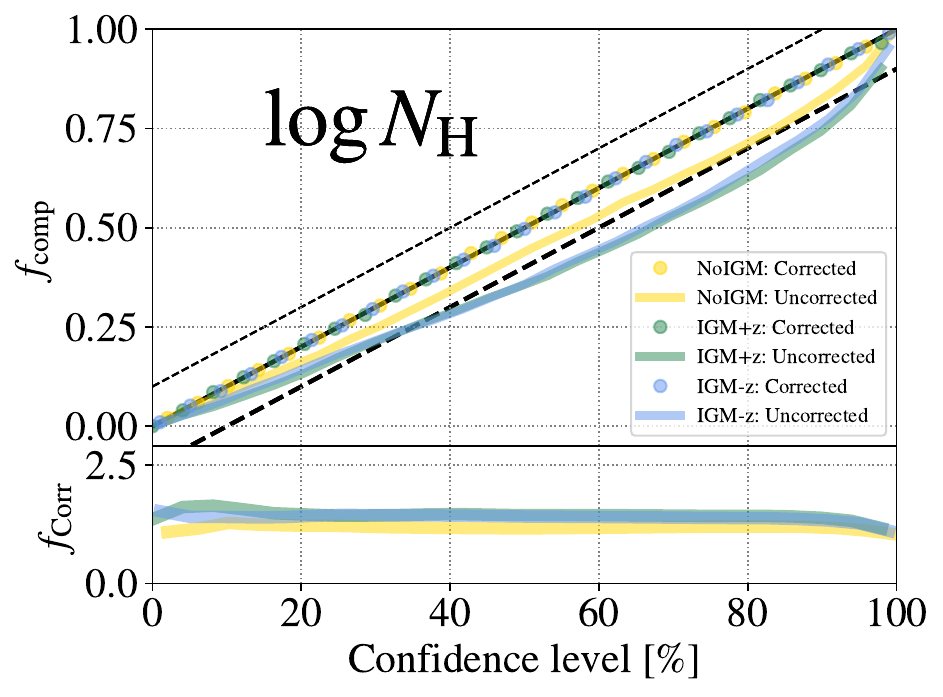}
        \includegraphics[width=2.4in]{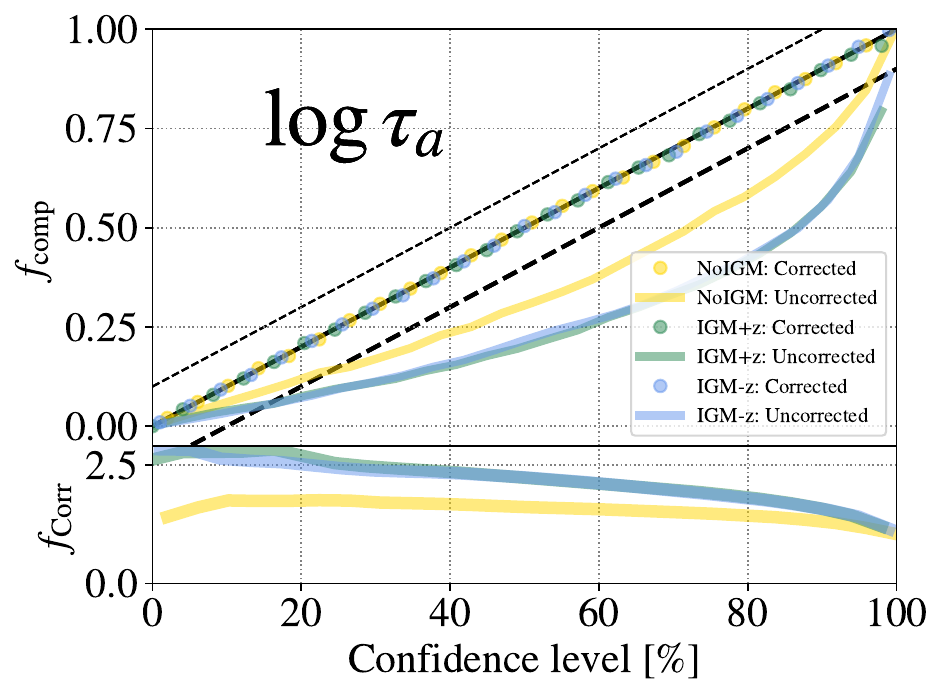}%
        \includegraphics[width=2.4in]{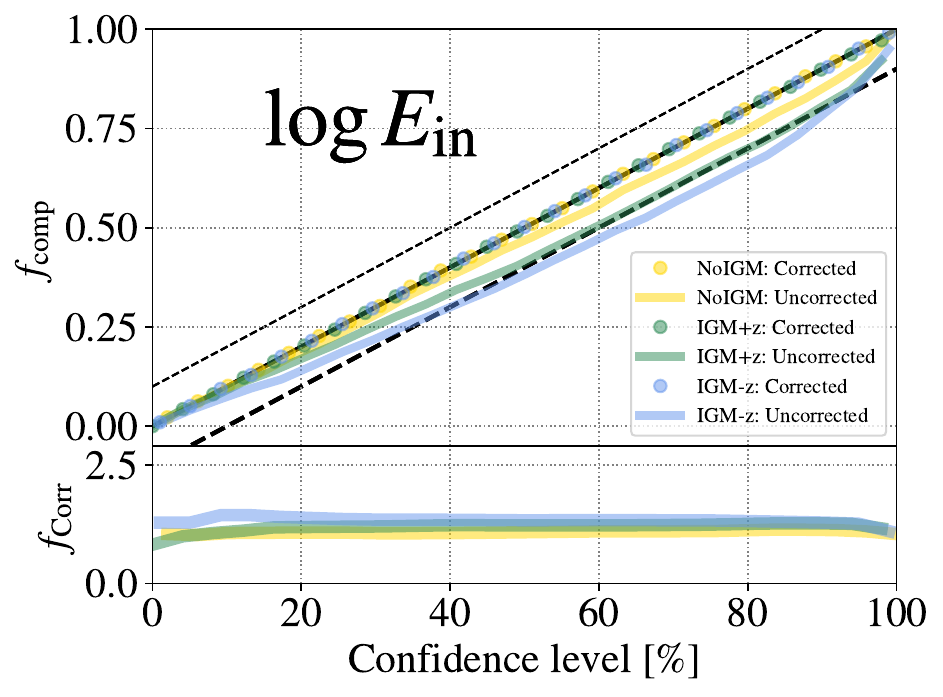}%
        \includegraphics[width=2.4in]{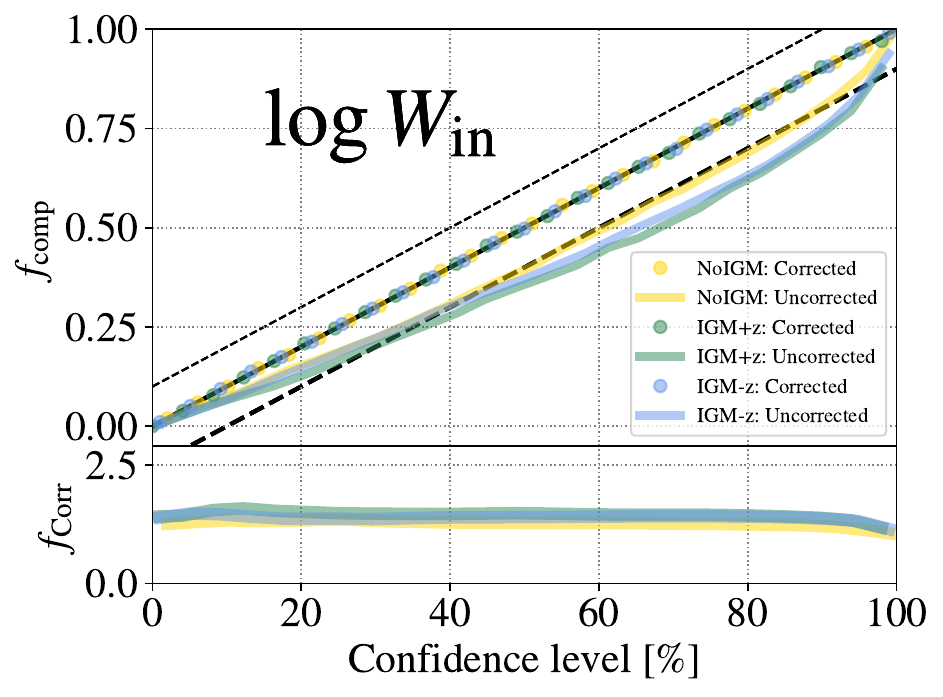}
        \includegraphics[width=2.4in]{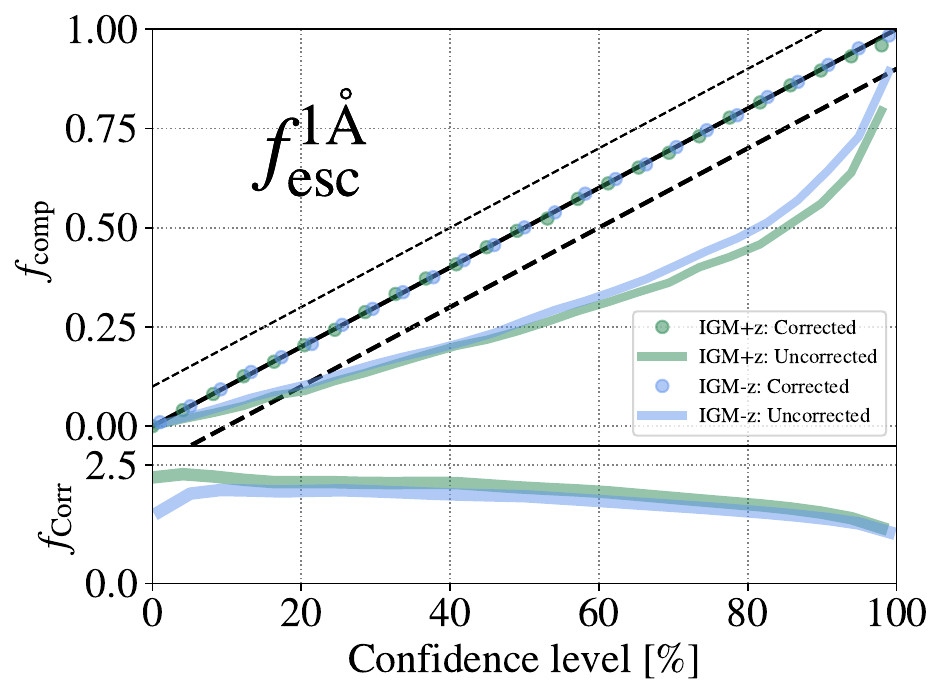}%
        \includegraphics[width=2.4in]{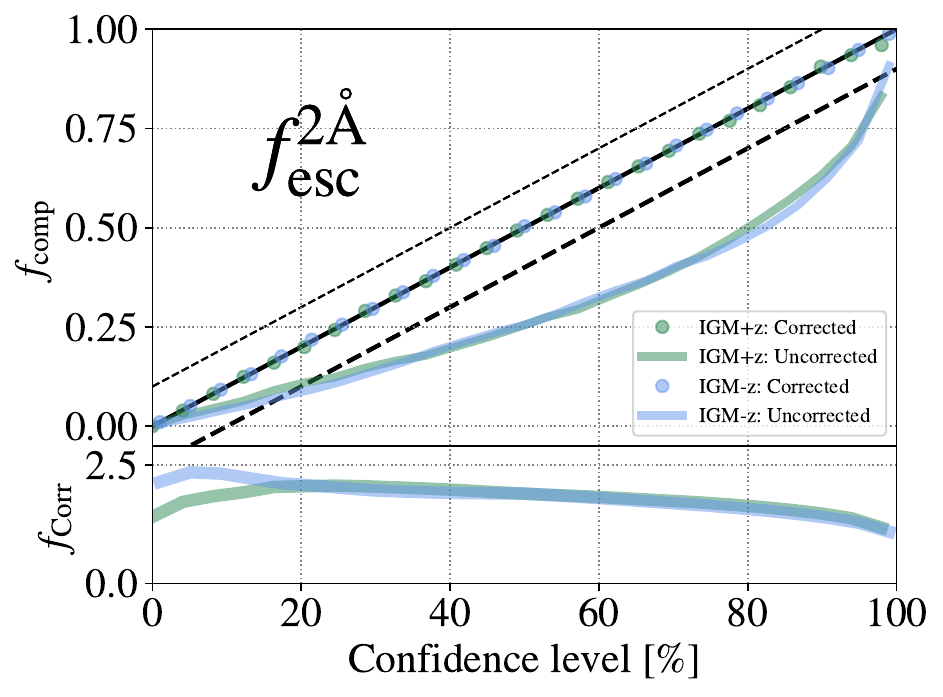}%
        \includegraphics[width=2.4in]{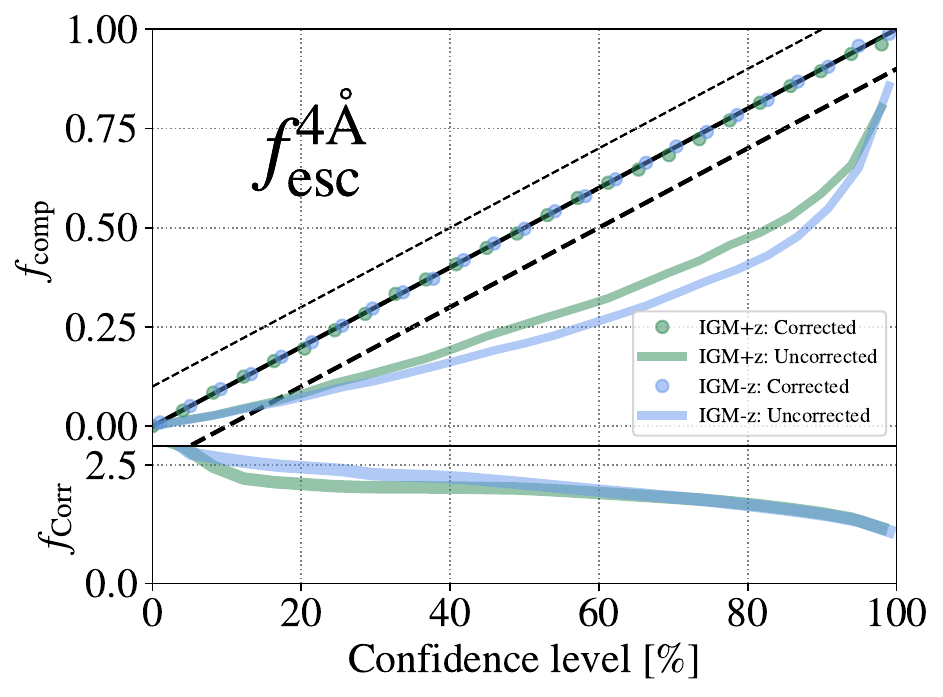}
        \caption{ Top: Comparison between the fraction of cases that a measurement is in agreement with the intrinsic true value, $f_{\rm comp}$, as a function of the confidence level. Bottom: Applied correction factor to uncertainty. The uncorrected $f_{\rm comp}$ is shown in colored solid lines, which \igmz in green, \igm in blue and \noigm in yellow. $f_{\rm comp}$ after the correction is shown in colored dots. In each panel a different output property is shown. In the first row $z_{\rm max}$, \vexp and \nh from left to right. In the second row \ta, \ew , \w , from left to right. In the third row \fAo , \fAt, and \fa from left to right. The diagonal black dashed lines show the 1:1$\pm10\%$ relation. Computed using $10^4$ line profiles with quality \wg=0.5\AA{} , \dl=0.1\AA{} and \sn = 10. }
        \label{fig:uncertainty_recalibration}
        \end{figure*}

In \zp we found that the uncertainty computed through this methodology had less than a 10\% error, and therefore needed no correction. However, we find that the models presented in this work tend to underestimate the uncertainty in the output parameters. In the top panels of Fig.~\ref{fig:uncertainty_recalibration} we show the comparison between the fraction of cases that a measurement is in compatible with the intrinsic true value, $f_{\rm comp}$, as a function of the confidence level. Each subplot shows the indicated outflow property. Ideally this relation should follow perfectly a 1:1 trend. Focusing in \noigm (yellow) we find that before correction (solid line) the uncertainty is underestimated no more than 10\%, except for \ta, as it was the case in \zp, which did not include IGM. We find that the uncertainty in the outflow parameters can be underestimated up a 20\% in \igmz and \igm. This bias in the uncertainty estimation is produced by the IGM inclusion. Some IGM transmission curve can produce observed \lya line really close to unabsorbed thin shell spectrum. In some of these cases the outflow parameters predicted correspond to the 'fake' thin shell spectrum rather than the intrinsic one. This causes than in smaller fraction of cases the true outflow parameters and the predicted ones are compatible at a given confidence level than in \noigm. We correct our uncertainty estimation so that $f_{\rm comp}$ and the confidence level follow a 1:1 relation. The correction factors depend on the confidence level and are shown in the bottom panels. $f_{\rm comp}$ as a function of the confidence level after the correction is shown in dots. 

\section{ Accuracy of individual parameters }\label{sssec:results_mock_parameters_parameters}

\begingroup
\setlength{\tabcolsep}{4.5pt}
\renewcommand{\arraystretch}{1.6}
\begin{table*}
\caption{ Parameters associated with the line profiles displayed in Fig.~\ref{fig:EXAMPLES_quality}. \vexp is given in \kms, \nh in $cm^{-2}$. \ew and \w are given in \AA{}. }
\begin{tabular}{ccccccccccccc}
 & & \multicolumn{3}{c}{Source A} & & \multicolumn{3}{c}{Source B} & & \multicolumn{3}{c}{Source C} \\ \cline{3-5} \cline{7-9} \cline{11-13}
Prop. &  True & \igmz & \igm & No IGM & & \igmz & \igm & No IGM & & \igmz & \igm & No IGM \\ \hline
$\log V_{\rm exp}$ & $1.9$ & $1.85_{0.23}^{0.17}$ & $1.81_{0.39}^{0.29}$ & $1.4_{0.11}^{0.11}$ & & $1.78_{0.27}^{0.19}$ & $1.62_{0.2}^{0.16}$ & $1.61_{0.1}^{0.14}$ & & $1.75_{0.21}^{0.19}$ & $1.62_{0.16}^{0.15}$ & $1.52_{0.21}^{0.17}$\\
$\log N_{\rm H}$ & $17.5$ & $17.91_{0.24}^{0.16}$ & $18.11_{0.32}^{0.26}$ & $19.01_{0.16}^{0.28}$ & & $18.05_{0.27}^{0.2}$ & $18.08_{0.32}^{0.26}$ & $18.46_{0.36}^{0.4}$ & & $18.26_{0.23}^{0.19}$ & $18.18_{0.38}^{0.26}$ & $18.29_{0.36}^{0.32}$\\
$\log \tau_{\rm a}$ & $-1.7$ & $-2.05_{0.17}^{0.13}$ & $-2.12_{0.31}^{0.23}$ & $-1.98_{0.69}^{0.47}$ & & $-2.16_{0.2}^{0.28}$ & $-2.24_{0.27}^{0.23}$ & $-2.2_{0.32}^{0.28}$ & & $-2.13_{0.2}^{0.2}$ & $-2.09_{0.23}^{0.27}$ & $-2.14_{0.21}^{0.22}$\\
$\log EW_{\rm in }$ & $1.7$ & $1.78_{0.15}^{0.19}$ & $1.77_{0.18}^{0.19}$ & $1.58_{0.13}^{0.12}$ & & $1.89_{0.12}^{0.19}$ & $1.82_{0.18}^{0.24}$ & $1.73_{0.19}^{0.17}$ & & $1.82_{0.13}^{0.19}$ & $1.75_{0.22}^{0.25}$ & $1.64_{0.22}^{0.22}$\\
$\log W_{\rm in }$ & $-0.6$ & $-0.62_{0.1}^{0.07}$ & $-0.6_{0.07}^{0.07}$ & $-0.61_{0.07}^{0.05}$ & & $-0.56_{0.14}^{0.11}$ & $-0.61_{0.13}^{0.13}$ & $-0.64_{0.1}^{0.09}$ & & $-0.59_{0.18}^{0.13}$ & $-0.64_{0.16}^{0.12}$ & $-0.72_{0.16}^{0.12}$\\
$f_{\rm esc}^{4\AA}$ & $0.75$ & $0.81_{0.07}^{0.09}$ & $0.82_{0.12}^{0.17}$ & - & & $0.76_{0.07}^{0.1}$ & $0.84_{0.14}^{0.16}$ & - & & $0.73_{0.09}^{0.12}$ & $0.7_{0.17}^{0.19}$ & -\\ \\
 & & \multicolumn{3}{c}{Source D} & & \multicolumn{3}{c}{Source E} & & \multicolumn{3}{c}{Source F} \\ \cline{3-5} \cline{7-9} \cline{11-13}
Prop. &  True & \igmz & \igm & No IGM & & \igmz & \igm & No IGM & & \igmz & \igm & No IGM \\ \hline
$\log V_{\rm exp}$ & $2.0$ & $1.77_{0.17}^{0.17}$ & $1.86_{0.12}^{0.1}$ & $1.86_{0.13}^{0.14}$ & & $1.7_{0.26}^{0.23}$ & $1.87_{0.24}^{0.27}$ & $2.06_{0.13}^{0.24}$ & & $1.93_{0.19}^{0.23}$ & $1.94_{0.24}^{0.22}$ & $2.05_{0.28}^{0.27}$\\
$\log N_{\rm H}$ & $19.5$ & $19.95_{0.21}^{0.3}$ & $19.77_{0.33}^{0.31}$ & $19.74_{0.24}^{0.26}$ & & $19.74_{0.34}^{0.4}$ & $19.56_{0.38}^{0.51}$ & $19.27_{0.3}^{0.31}$ & & $19.16_{0.66}^{0.7}$ & $18.86_{0.68}^{0.67}$ & $18.96_{0.55}^{0.6}$\\
$\log \tau_{\rm a}$ & $-3.75$ & $-2.52_{0.42}^{0.24}$ & $-2.28_{0.61}^{0.44}$ & $-2.02_{0.86}^{0.68}$ & & $-2.26_{0.32}^{0.32}$ & $-2.33_{0.41}^{0.39}$ & $-2.33_{0.36}^{0.27}$ & & $-2.07_{0.26}^{0.28}$ & $-2.11_{0.34}^{0.36}$ & $-2.23_{0.29}^{0.36}$\\
$\log EW_{\rm in }$ & $1.6$ & $1.73_{0.12}^{0.14}$ & $1.74_{0.13}^{0.1}$ & $1.71_{0.1}^{0.09}$ & & $1.58_{0.15}^{0.12}$ & $1.62_{0.15}^{0.14}$ & $1.58_{0.14}^{0.11}$ & & $1.81_{0.18}^{0.21}$ & $1.77_{0.2}^{0.24}$ & $1.72_{0.21}^{0.21}$\\
$\log W_{\rm in }$ & $-0.1$ & $-0.22_{0.12}^{0.15}$ & $-0.29_{0.16}^{0.15}$ & $-0.23_{0.14}^{0.17}$ & & $-0.17_{0.16}^{0.17}$ & $-0.24_{0.24}^{0.24}$ & $-0.32_{0.22}^{0.18}$ & & $-0.26_{0.23}^{0.2}$ & $-0.36_{0.33}^{0.22}$ & $-0.34_{0.25}^{0.22}$\\
$f_{\rm esc}^{4\AA}$ & $0.99$ & $0.94_{0.02}^{0.02}$ & $0.95_{0.02}^{0.02}$ & - & & $0.88_{0.04}^{0.05}$ & $0.9_{0.04}^{0.08}$ & - & & $0.78_{0.11}^{0.12}$ & $0.81_{0.12}^{0.18}$ & -\\ \\
 & & \multicolumn{3}{c}{Source G} & & \multicolumn{3}{c}{Source H} & & \multicolumn{3}{c}{Source I} \\ \cline{3-5} \cline{7-9} \cline{11-13}
Prop. &  True & \igmz & \igm & No IGM & & \igmz & \igm & No IGM & & \igmz & \igm & No IGM \\ \hline
$\log V_{\rm exp}$ & $2.4$ & $2.25_{0.08}^{0.07}$ & $2.27_{0.14}^{0.18}$ & $2.4_{0.11}^{0.14}$ & & $2.3_{0.21}^{0.24}$ & $2.25_{0.45}^{0.27}$ & $2.43_{0.26}^{0.27}$ & & $2.01_{0.34}^{0.3}$ & $1.89_{0.3}^{0.27}$ & $1.75_{0.6}^{0.53}$\\
$\log N_{\rm H}$ & $18.4$ & $18.05_{0.31}^{0.27}$ & $18.21_{0.42}^{0.44}$ & $17.87_{0.43}^{0.27}$ & & $18.39_{0.7}^{0.48}$ & $18.3_{0.75}^{0.71}$ & $18.17_{0.93}^{0.55}$ & & $18.37_{0.85}^{0.52}$ & $18.41_{1.21}^{0.76}$ & $18.03_{0.97}^{0.48}$\\
$\log \tau_{\rm a}$ & $-0.6$ & $-2.05_{0.37}^{0.46}$ & $-2.1_{0.65}^{0.53}$ & $-2.14_{0.49}^{0.43}$ & & $-2.08_{0.3}^{0.3}$ & $-2.07_{0.42}^{0.39}$ & $-1.93_{0.87}^{0.4}$ & & $-2.03_{0.22}^{0.26}$ & $-2.11_{0.31}^{0.37}$ & $-2.11_{0.31}^{0.31}$\\
$\log EW_{\rm in }$ & $1.0$ & $1.0_{0.04}^{0.04}$ & $1.01_{0.03}^{0.04}$ & $0.92_{0.03}^{0.04}$ & & $1.06_{0.07}^{0.07}$ & $1.06_{0.07}^{0.09}$ & $0.98_{0.07}^{0.05}$ & & $1.32_{0.19}^{0.19}$ & $1.33_{0.2}^{0.2}$ & $1.2_{0.2}^{0.17}$\\
$\log W_{\rm in }$ & $-0.8$ & $-0.63_{0.11}^{0.1}$ & $-0.69_{0.1}^{0.12}$ & $-0.62_{0.09}^{0.09}$ & & $-0.56_{0.16}^{0.14}$ & $-0.65_{0.19}^{0.17}$ & $-0.67_{0.12}^{0.1}$ & & $-0.47_{0.21}^{0.21}$ & $-0.53_{0.19}^{0.17}$ & $-0.63_{0.23}^{0.19}$\\
$f_{\rm esc}^{4\AA}$ & $0.9$ & $0.82_{0.06}^{0.04}$ & $0.88_{0.04}^{0.06}$ & - & & $0.76_{0.1}^{0.1}$ & $0.81_{0.1}^{0.12}$ & - & & $0.67_{0.13}^{0.13}$ & $0.78_{0.14}^{0.16}$ & -\\ \\
\end{tabular}
\label{tab:quality_params}
\end{table*}
\endgroup

\begin{figure*} 
        \includegraphics[width=2.4in]{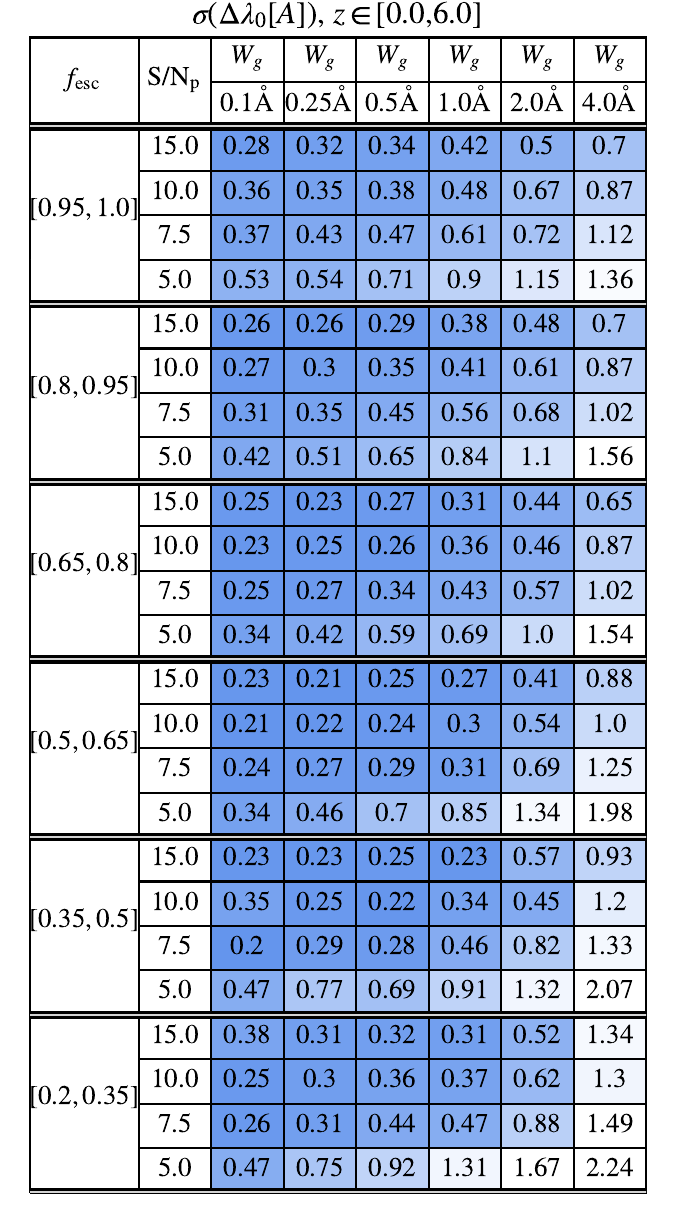}%
        \includegraphics[width=2.4in]{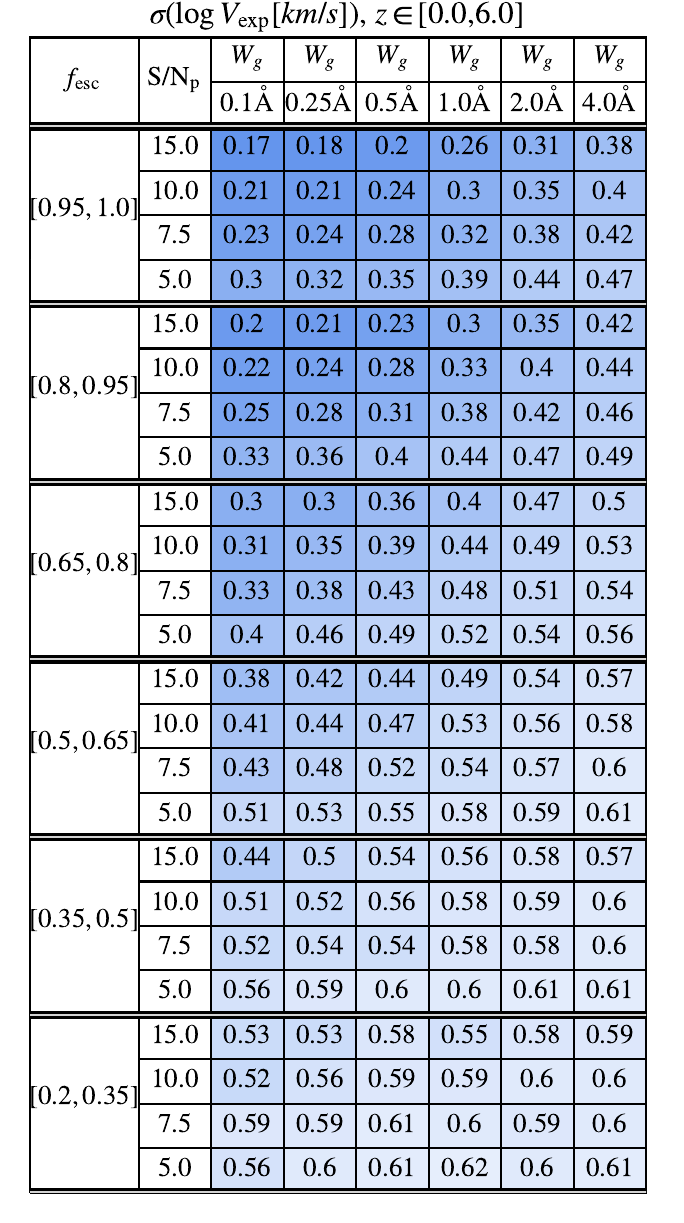}%
        \includegraphics[width=2.4in]{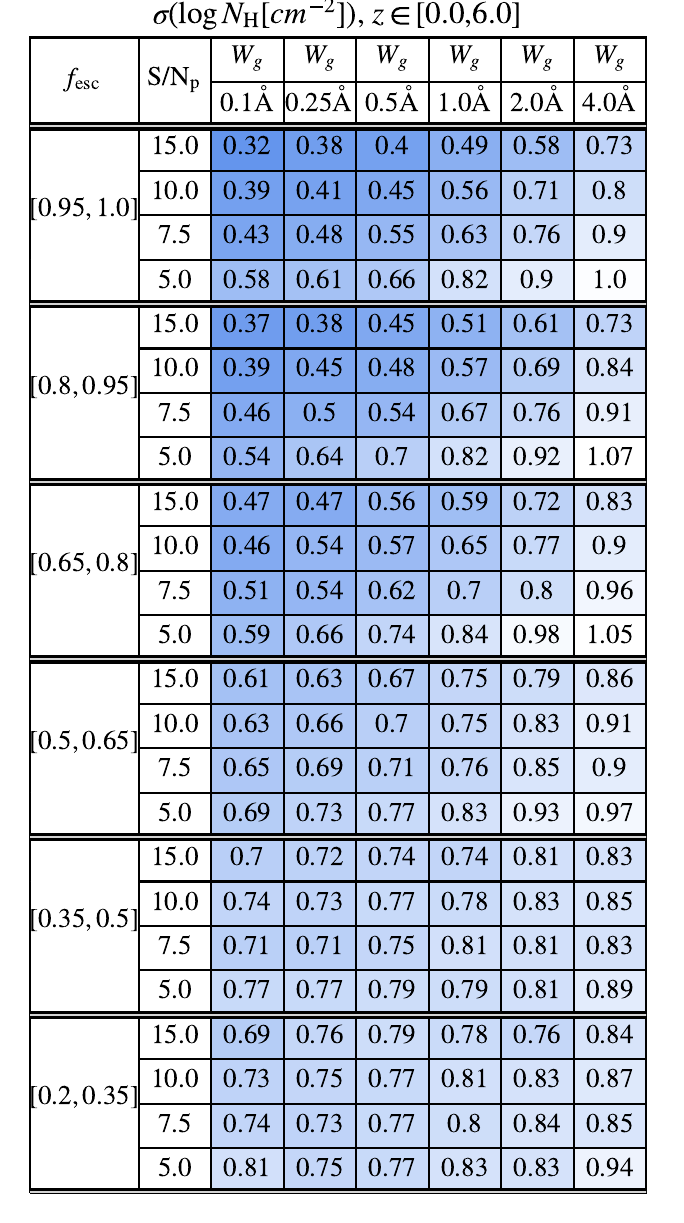}
        \includegraphics[width=2.4in]{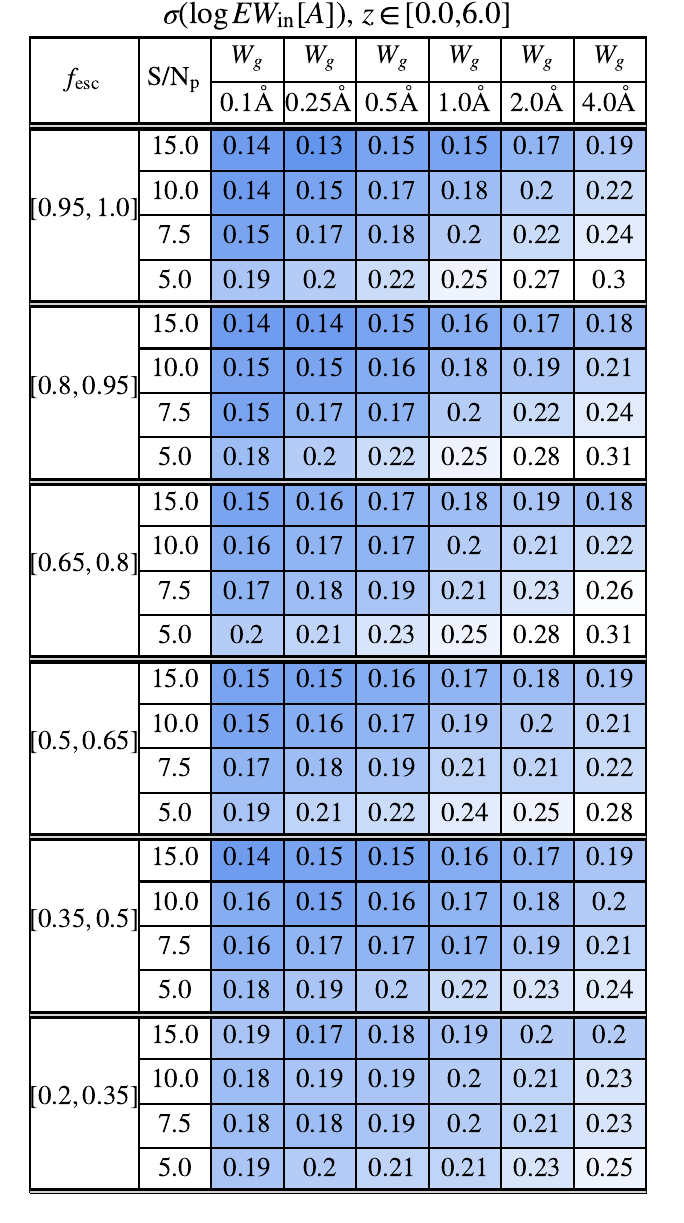}%
        \includegraphics[width=2.4in]{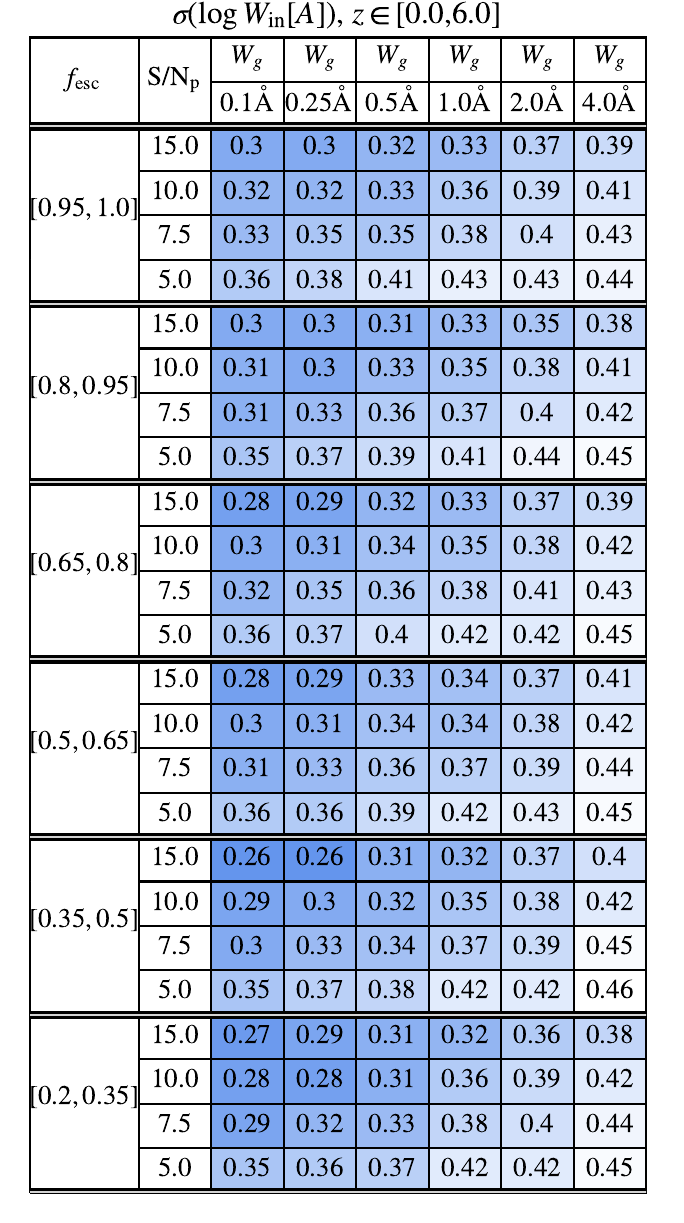}%
        \includegraphics[width=2.4in]{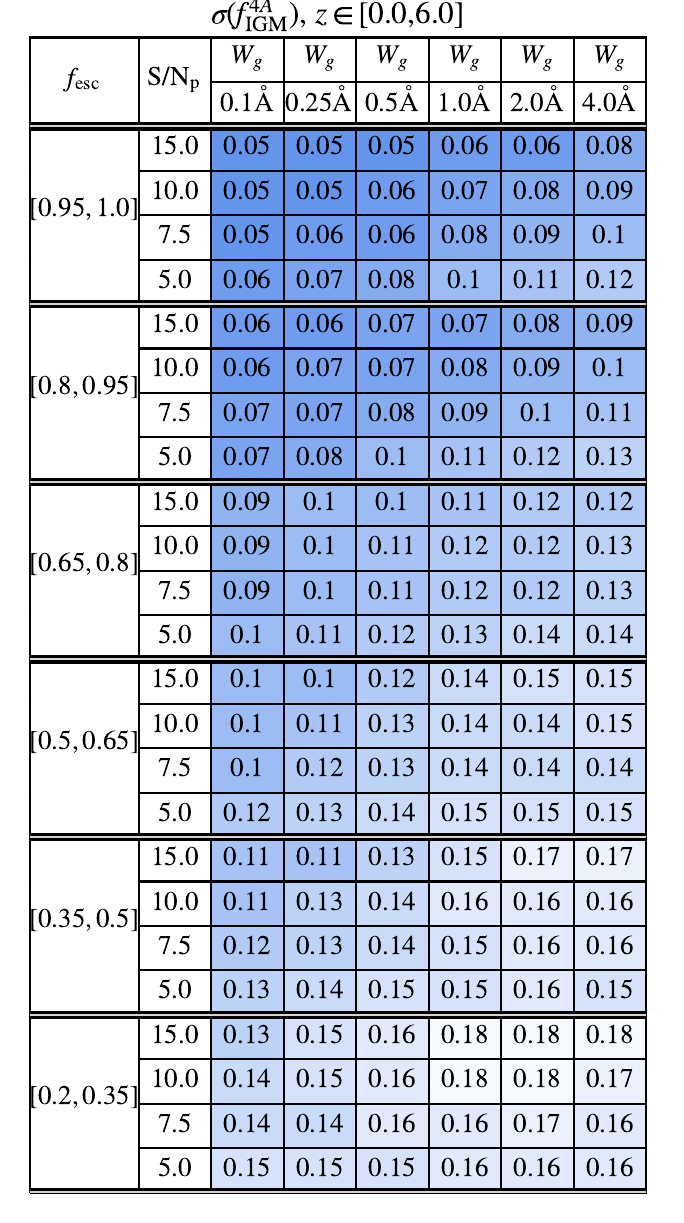}%
        \caption{ Accuracy measured through the standard deviation of the difference between the true output parameters and those predicted by \igmz . The top row shows the accuracy in rest frame wavelength of \lya, the outflow expansion velocity and the outflow neutral hydrogen column density from left to right. In the bottom row we show the accuracy for the intrinsic equivalent width and width before entering into the ISM and \fa from left to right. This is computed for 6 values of \wg and \dl and 4 of \sn, i.e., 24 mocks of 10000 \lya line profiles. These are split by their true \fa in 6 bins. The color of each cell is shown in darker for smaller values (better accuracy). }
        \label{fig:acuracy_IGM+z}
        \end{figure*}
\begin{figure*} 
        \includegraphics[width=2.3in]{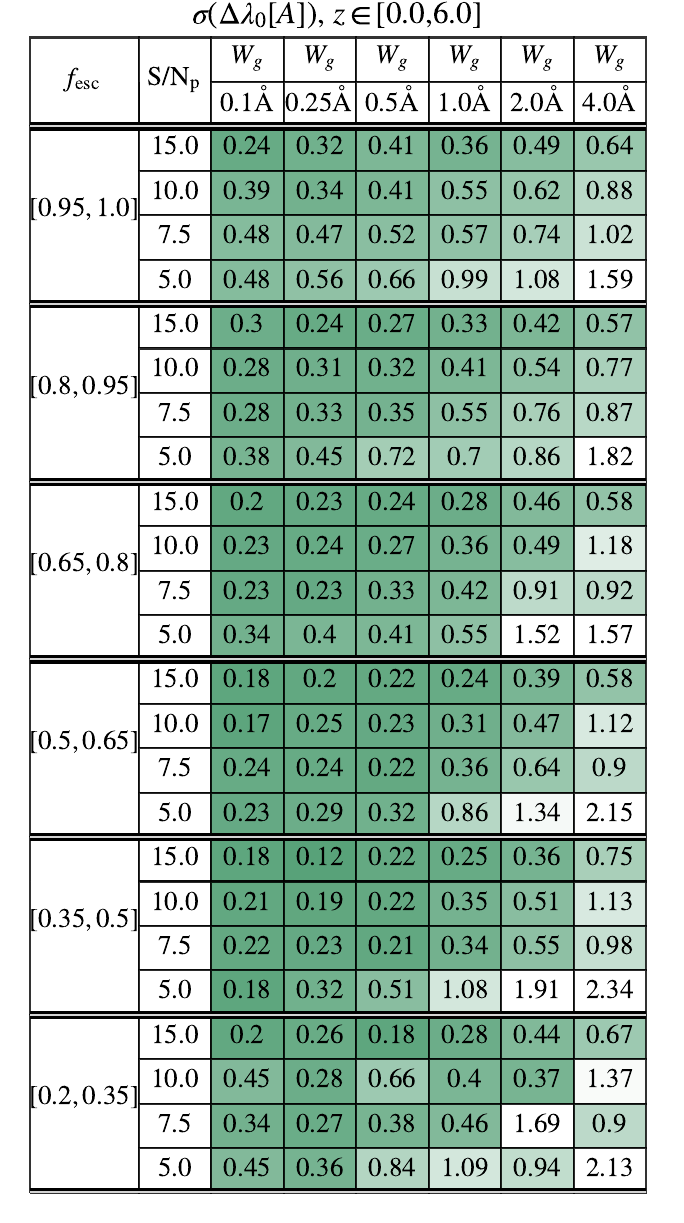}%
        \includegraphics[width=2.3in]{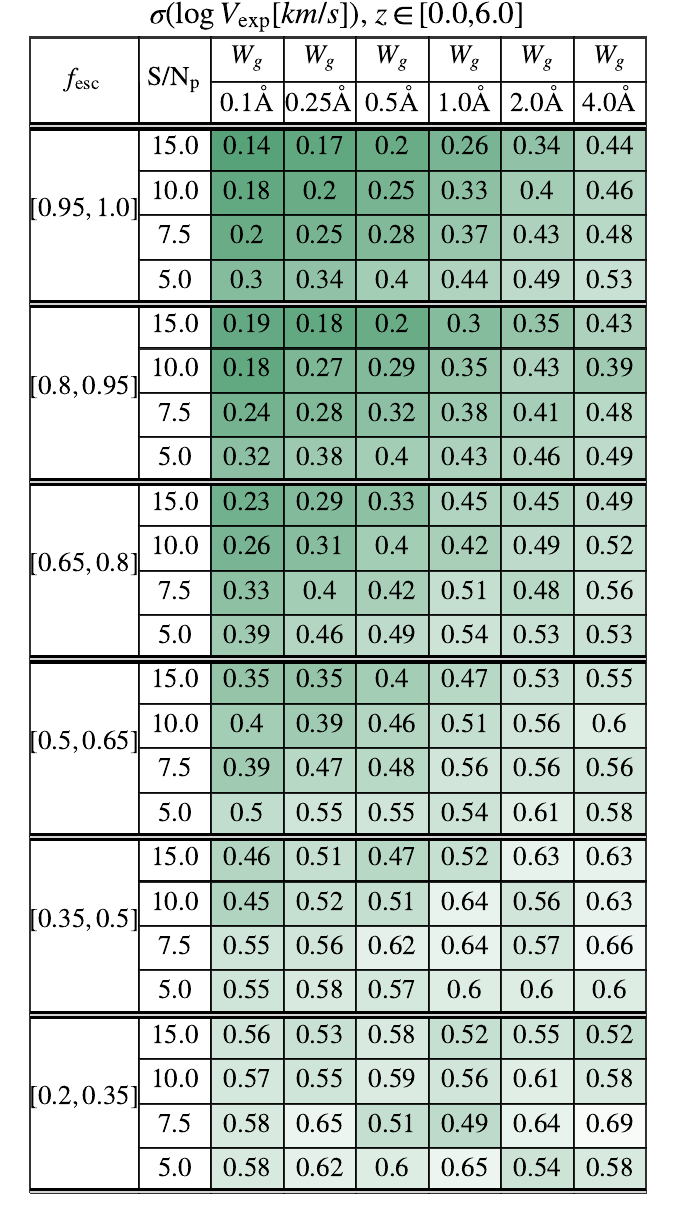}%
        \includegraphics[width=2.3in]{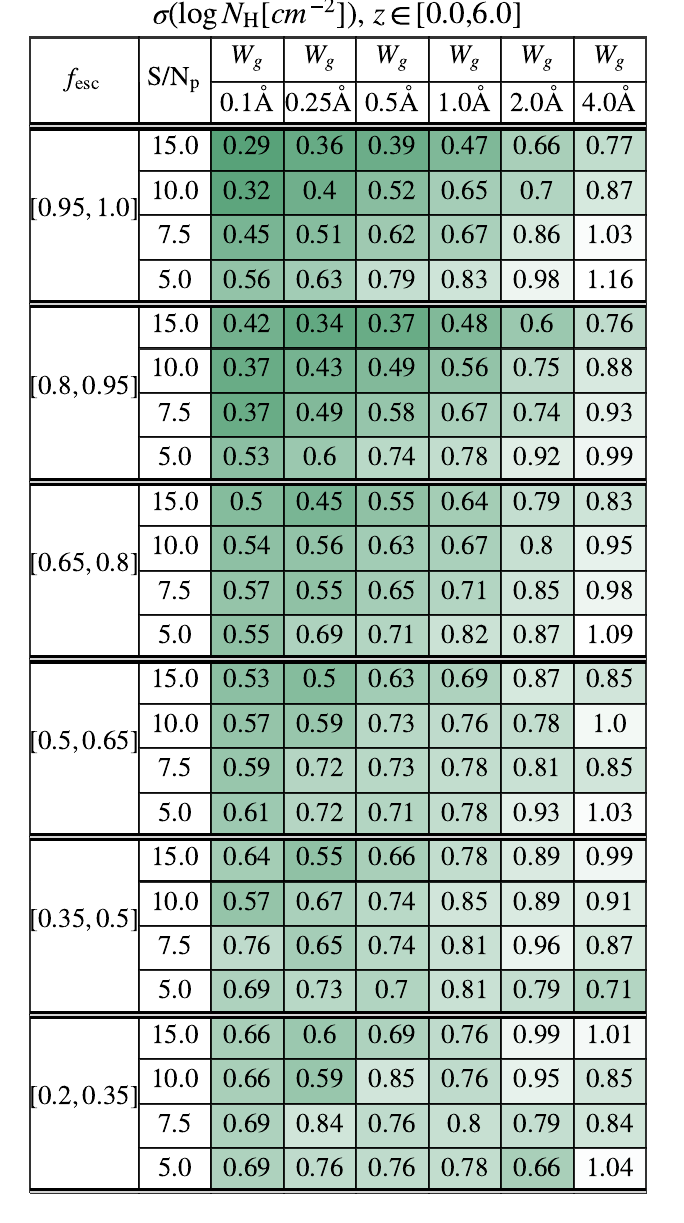}
        \includegraphics[width=2.3in]{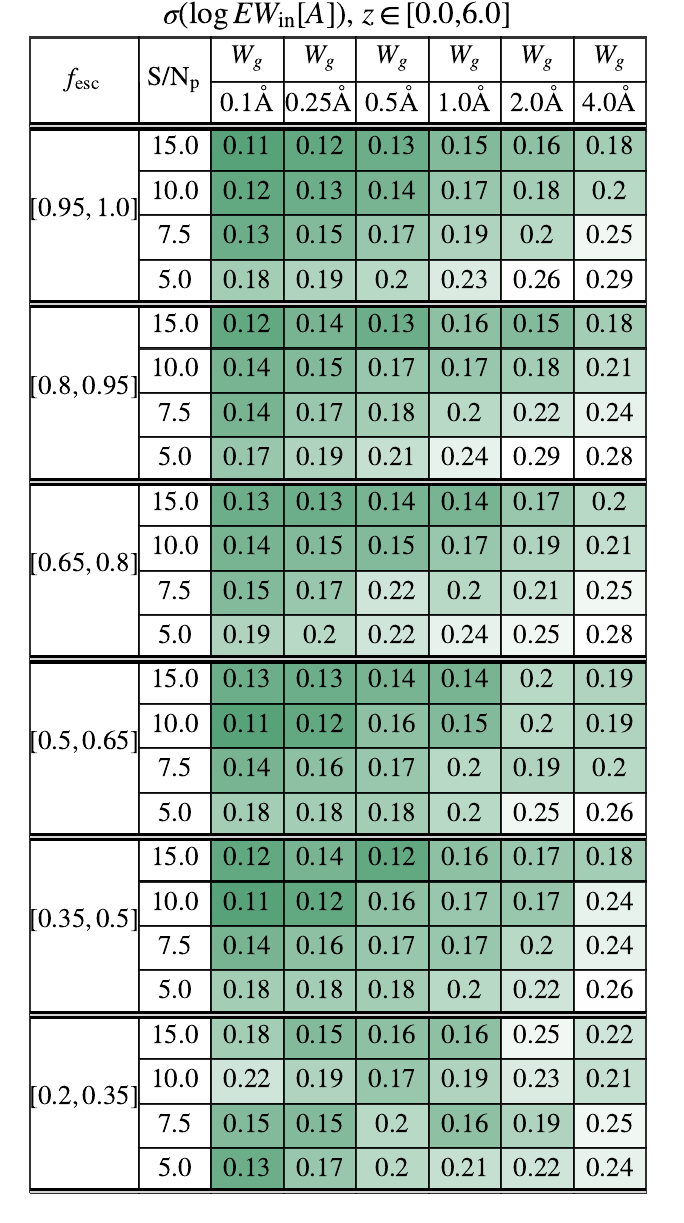}%
        \includegraphics[width=2.3in]{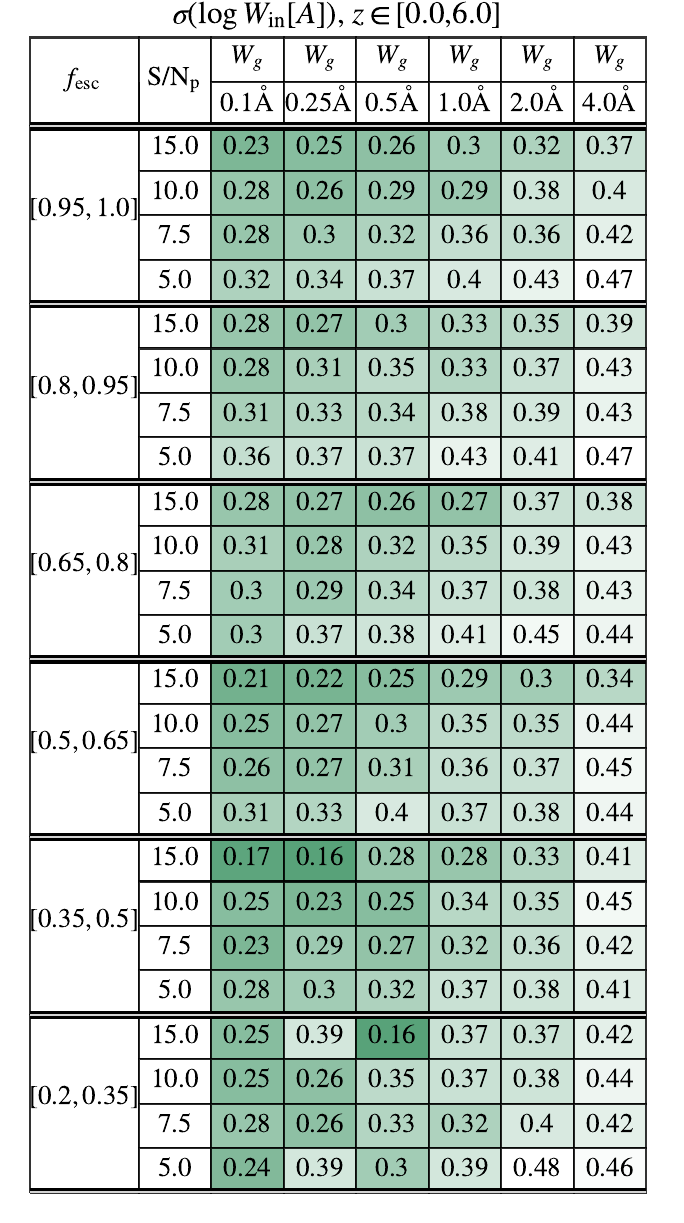}%
        \includegraphics[width=2.3in]{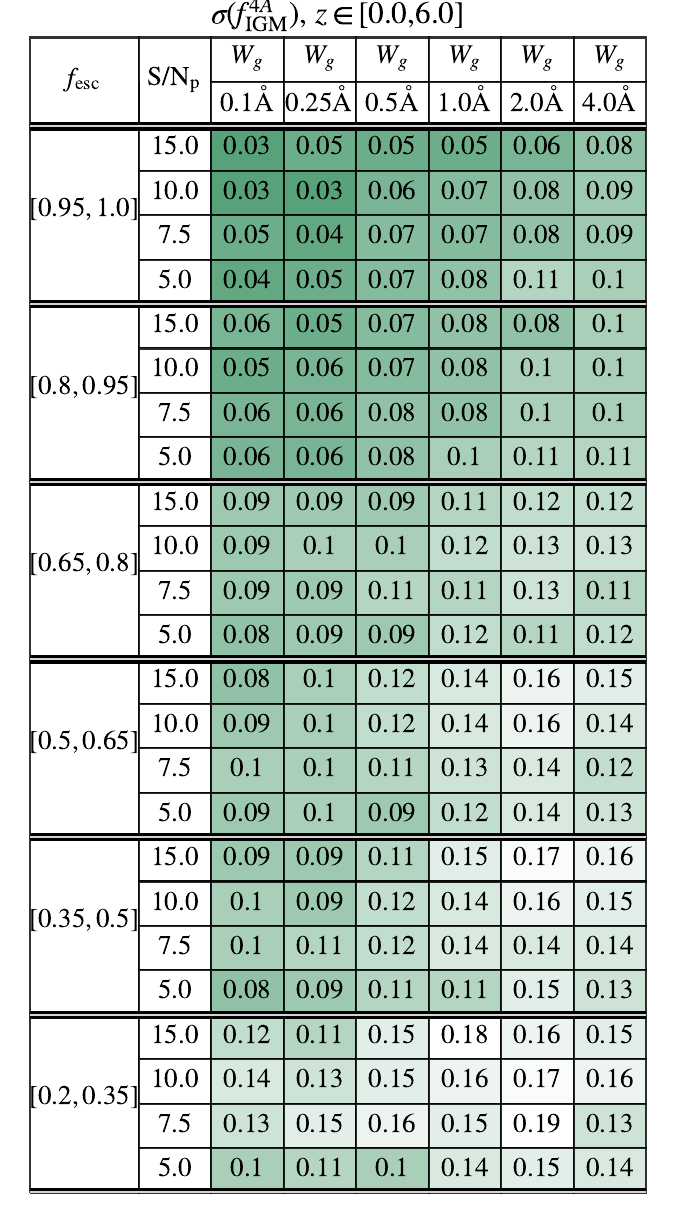}%
        \caption{ Same as Fig.~\ref{fig:acuracy_IGM+z} but for \igm . }
        \label{fig:acuracy_IGM-z}
        \end{figure*}
\begin{figure*} 
        \includegraphics[width=2.3in]{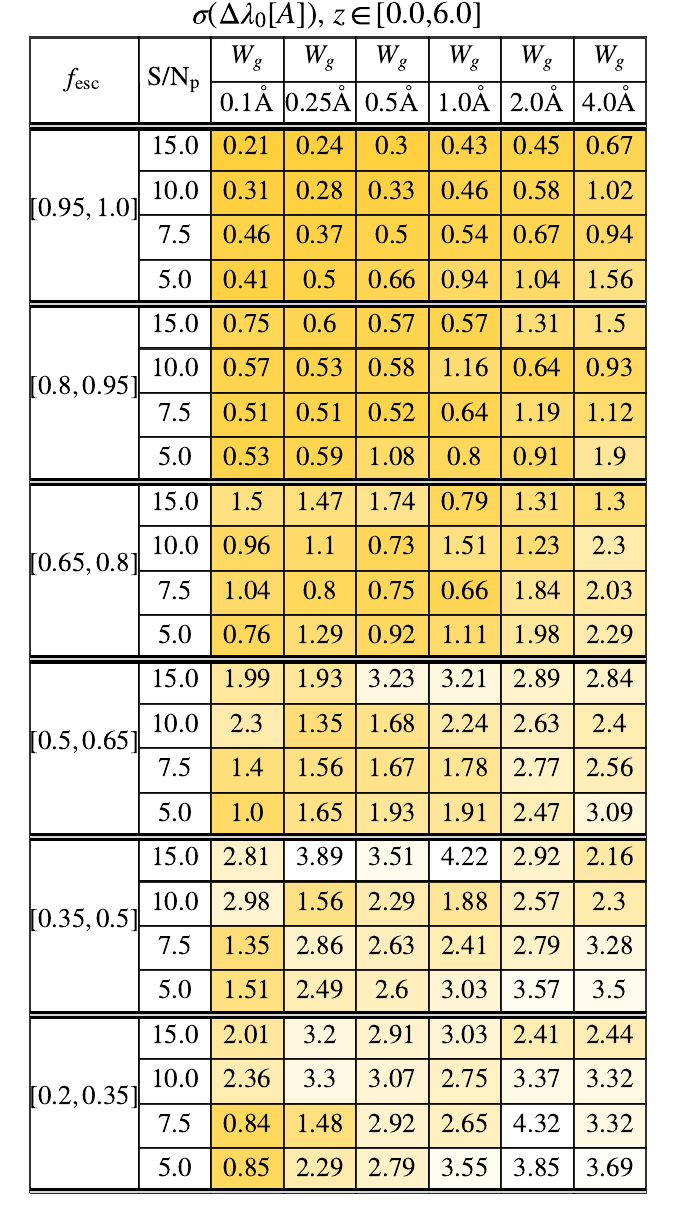}%
        \includegraphics[width=2.3in]{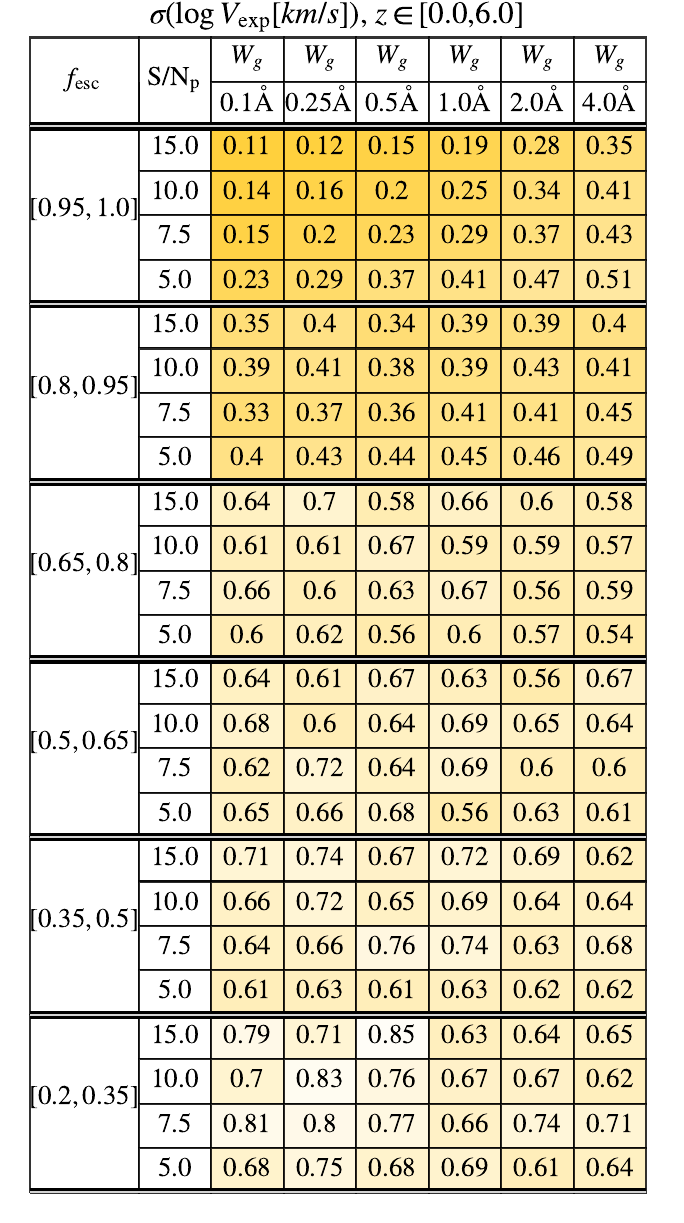}%
        \includegraphics[width=2.3in]{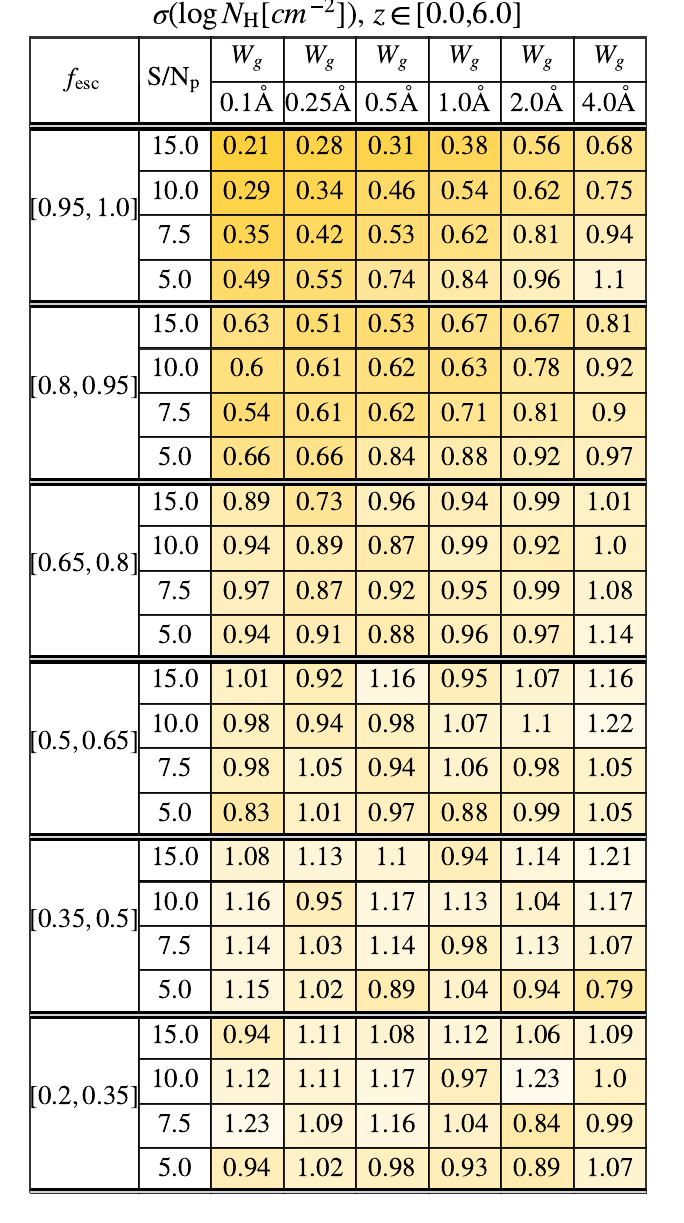}
        \includegraphics[width=2.3in]{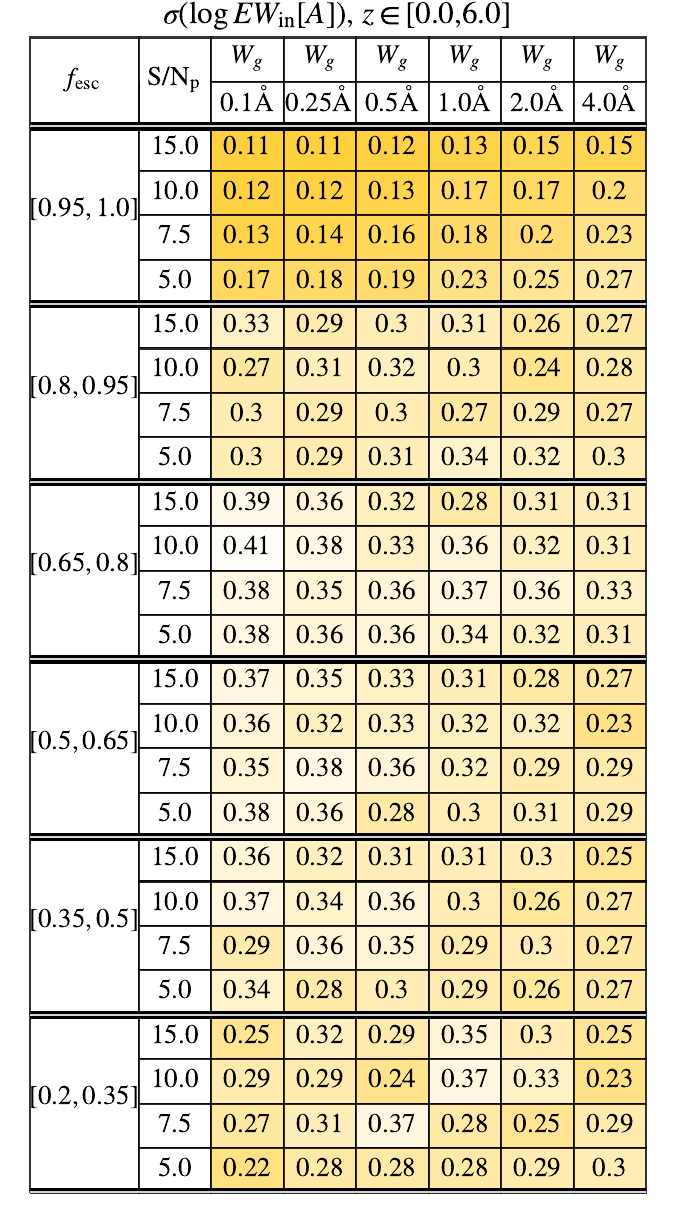}%
        \includegraphics[width=2.3in]{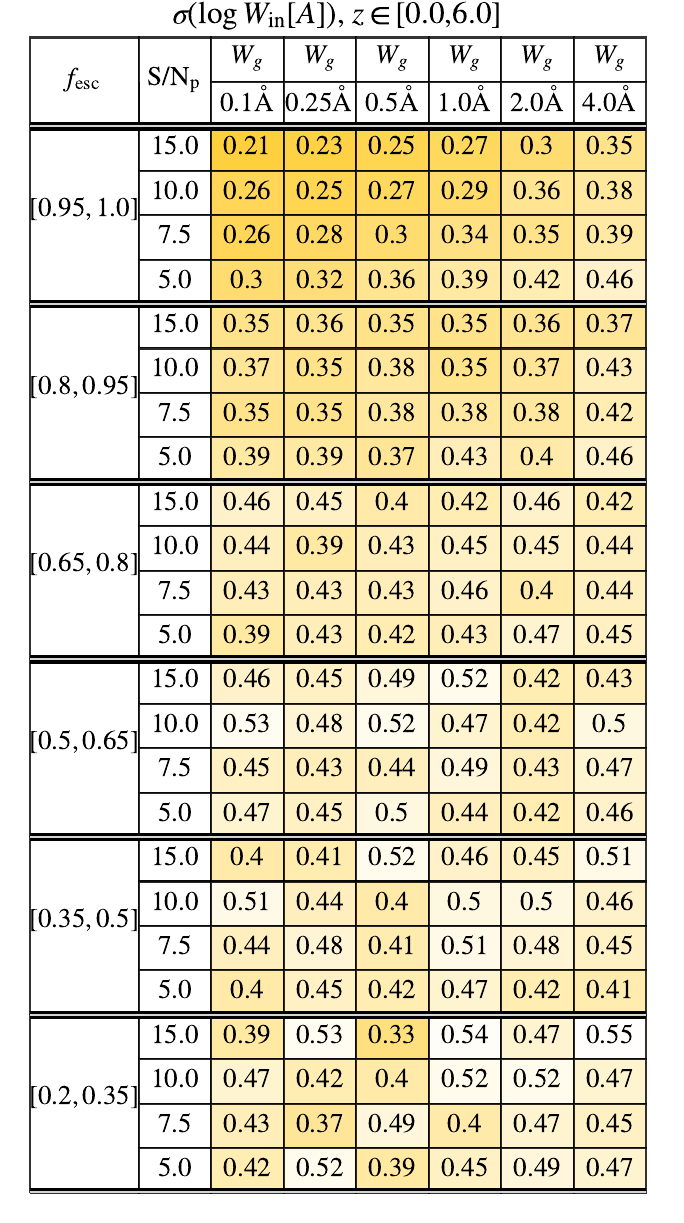}%
        \caption{ Same as Fig.~\ref{fig:acuracy_IGM+z} but for \noigm . }
        \label{fig:acuracy_NoIGM}
        \end{figure*}

As described in \zp, the accuracy of the ANN models on the outflow parameters and redshift depend on the quality of the observed spectrum. In general, spectrum with better quality will be better reconstructed by our ANN models.  

Individual cases of how the observed line profile quality changes the reconstructed line profile is shown in Fig.~\ref{fig:EXAMPLES_quality}. Each row shows the same intrinsic line and IGM absorption at different qualities (better to worse from left to right) and the true and predicted outflow parameters and IGM escape fraction are listed in Tab.\ref{tab:quality_params}. In general, for sources of the best quality ( {\it A}, {\it D} and {\it G}) the predicted parameters for \igmz and \igm are 1$\sigma$ compatible with the true values. However as the observed line profile becomes noisier and less resolved the accuracy in the parameters decreases. For example, in the middle row, for the best quality ({\it D}) the amplitude of the blue peak is reasonably well reconstructed and KS=0.04. Meanwhile, for the lowest quality ({\it F}), KS=0.09  and the positions and amplitude of the blue peak are slightly miss-predicted. 

In order to quantify the accuracy of \igmz or \igm and \noigm we computed 10000 line profiles with IGM  absorption between redshift 0 and 6 for 24 combinations of quality (4 for \sn and 6 for \wg). In Fig.~\ref{fig:acuracy_IGM+z}, \ref{fig:acuracy_IGM-z} and \ref{fig:acuracy_NoIGM} we show the accuracy for the output parameters of \igmz, \igm and \noigm, respectively. In general, we find that \igmz and \igm exhibit the same accuracy across \fa and spectral quality for all properties. Also, the accuracy of all the output parameters decreases with the line profile quality. Worse \lya line profiles (larger \wg and smaller \sn) are recovered with worse redshift, outflow parameters and \fa. This happens at every \fa interval. 

Focusing in the range \fa$\in[0.95,1.0]$, we find \noigm actually performs better than \igmz and \igm. For example, at \wg=0.1 and \sn=15.0, the accuracy of \noigm is better than that of \igmz and \igm for redshift, \vexp, \nh and \w (a $\sim 10\%$). This shows that when the \lya line profiles is mostly unabsorbed by the IGM the \noigm  works better than \igmz or \igm . This is due to the fact that \noigm is trained only with IGM free \lya line profiles. In comparison, \igmz or \igm are trained with a much more diverse \lya line profile population. This can be a consequence of the outflow parameters confusion discussed in \ref{sec:results_mock}. A small fraction of \lya line profiles with \fa$\sim 1.0$ are reconstructed as if they were more absorbed than what they actually are, lowering the accuracy in the outflow parameters and redshift. 

Meanwhile, \igmz and \igm outperform \noigm when \fa < 0.95. In general the accuracy for all the output parameters decreases with the true \fa. This is due to the fact that, at smaller values of \fa more information has been destroyed by the IGM. However, \noigm's accuracy drops fast with decreasing \fa, while they accuracy \igmz and \igm goes down slowly. 

For instance, focusing on \wg=0.1 and \sn=15.0, the accuracy in \vexp at  \fa$\in[0.95,1.0]$ for \noigm is 0.11\kms, while at \fa$\in[0.8,0.95]$ it drops to 0.35\kms, at \fa$\in[0.65,0.8]$ it drops to 0.64\kms. Meanwhile, for \igmz (\igm), the \vexp \fa$\in[0.95,1.0]$ is 0.17\kms(0.14\kms), while at \fa$\in[0.8,0.95]$ it is 0.20\kms(0.19\kms), at \fa$\in[0.65,0.8]$ it is 0.3\kms(0.23\kms). This same trend is also found at other spectral quality configurations for \nh, \ew and \ta (although \ta is not displayed).


We find that the accuracy in the accuracy in \lya wavelength  follows a different trend. While it is true that in \noigm it drops fast for decreasing \fa values, we find that for \igmz  and \igm it initially drops and then stabilizes at intermediate \fa values (\fa$\in[0.3,0.8]$) before rising again at \fa$\sim 0.3$.  focusing on \wg=0.1 and \sn=15.0,   \fa$\in[0.95,1.0]$ for \noigm is 0.21\AA{}, while at \fa$\in[0.8,0.95]$ it drops to 0.75\AA{}, at \fa$\in[0.65,0.8]$ it drops to 1.5\AA{}, and at \fa$\in[0.5,0.65]$ up to 2.0\AA. Meanwhile, for \igmz (\igm), the \lya wavelength rest frame accuracy at \fa$\in[0.95,1.0]$ is 0.28\AA{}(0.24\AA{}), while at \fa$\in[0.8,0.95]$ it is 0.26\AA{}(0.3\AA{}), at \fa$\in[0.65,0.8]$ it is 0.25\AA{}(0.20\AA{}), and at \fa$\in[0.5,0.65]$ up to 0.23\AA(0.18\AA{}). This is also found at other spectral quality configurations. While at lower values of \fa, more outflow information is erased by the IGM, it is also true that more IGM information is imprinted on the \lya line profile. The addition of this information, such us the position of a sudden flux drop (e.g. cases {\it S} and {\it T} of Fig.~\ref{fig:mock_example_successful_reconstruction}), enhances the accuracy redshift of the source.

Regarding the accuracy on \fa we find that both, \igmz and \igm, are able to give accurate prediction for \fa in a wide range of spectral quality and in a broad \fa regime. As for the outflow properties, we find that the \fa accuracy decreases for more absorbed \lya line profiles (smaller \fa). In the best case scenario considered (\wg=0.1 and \sn=15.0), the \fa accuracy of \igmz (\igm) is 0.05 (0.03) at \fa$\sim 1$, 0.06 (0.06) at  \fa$\sim 0.87$, 0.09 (0.09) at  \fa$\sim 0.72$, 0.12 (0.08) at  \fa$\sim 0.62$ and 0.14 (0.09) at  \fa$\sim 0.42$. Remarkable, \fa is relatively well estimated also for \lya line profiles with worse quality. The \fa uncertainty in the \igm is below 0.15, in general, even for \wg=4.0\AA and \sn=5.0 .

Comparing to the previous \zelda ANN presented in \zp (Their Fig.8),  \igmz , \igm and \noigm present the same or slightly better accuracy at the \fa $\in[0.95,1.0]$ regime. In \cite{gurung_lopez_2022} no IGM features were included, so the most direct comparison would be \noigm with the their model. For instance, the \noigm redshift accuracy for the best scenario in both works (\wg=0.1 and \sn=15.0) is 0.21\AA, while their was 0.26\AA{}. The \vexp \noigm accuracy is 0.11\kms while theirs was 0.14\kms. Meanwhile, \ta, \ew and \w exhibit the almost the same accuracy in \zp and here. \noigm is only a $\sim 10\%$ better for these output properties. This increase in accuracy is due to the different ANN configuration and the changes in the input. 

\section{ Accuracy in the shape of line profile reconstruction }\label{sssec:results_mock_parameters_line}

\begin{figure*} 
        \includegraphics[width=7.1in]{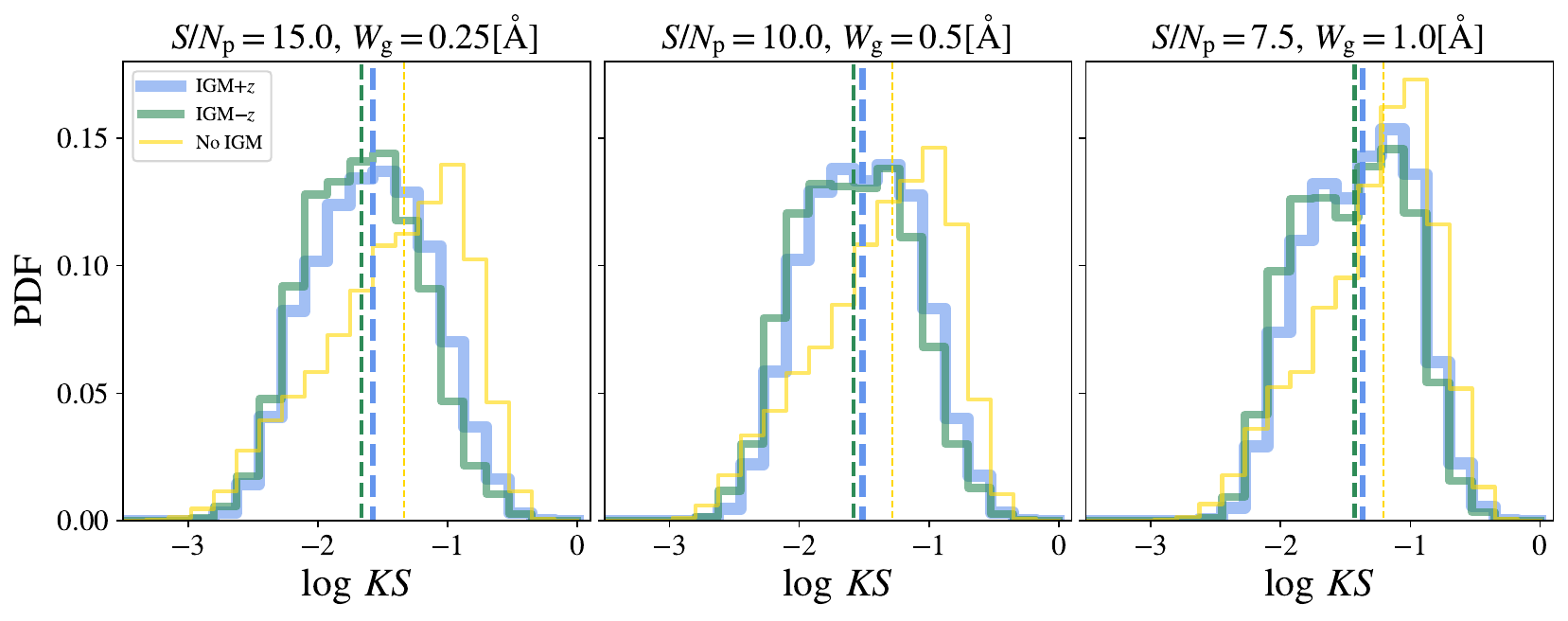}%
        \caption{ Kolmogórov-Smirnov estimator distribution comparison between models. \zelda's prediction using the \igmz, \igm and NoIGM models are displayed in blue, green and yellow, respectively. Each subpanel show the KS distribution at different observed line profile qualities. In particular, in the left panel \sn=15.0 and \wg=0.25\AA, in the middle panel \sn=10.0 and \wg=0.5\AA, in the right panel \sn=7.0 and \wg=1.0\AA. The vertical dashed lines mark the median of the KS distribution of the matching color. }
        \label{fig:KS_distributions}
        \end{figure*}
\begin{figure*} 
        \includegraphics[width=2.4in]{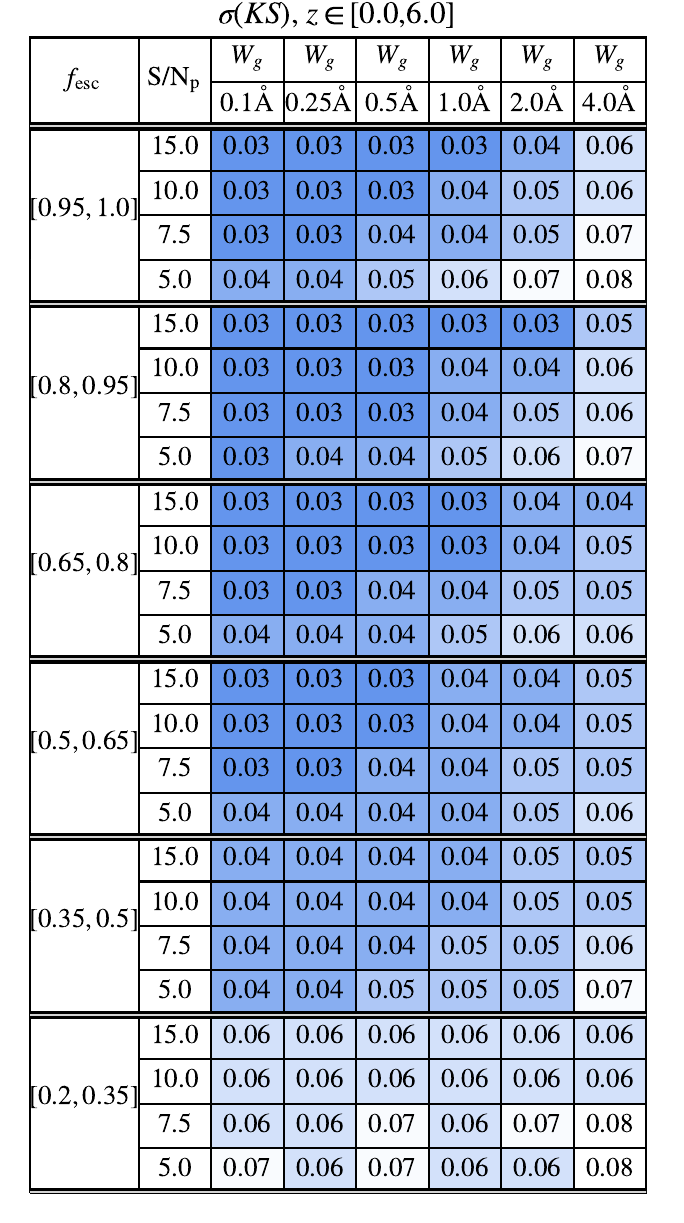}%
        \includegraphics[width=2.4in]{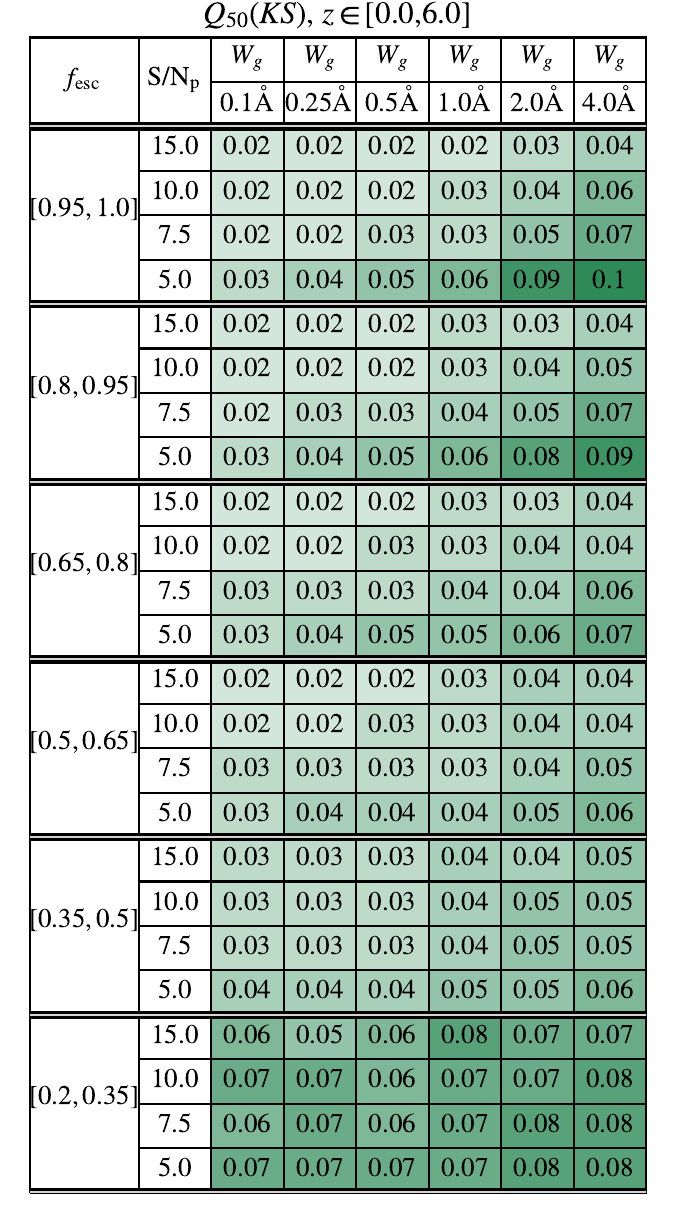}%
        \includegraphics[width=2.4in]{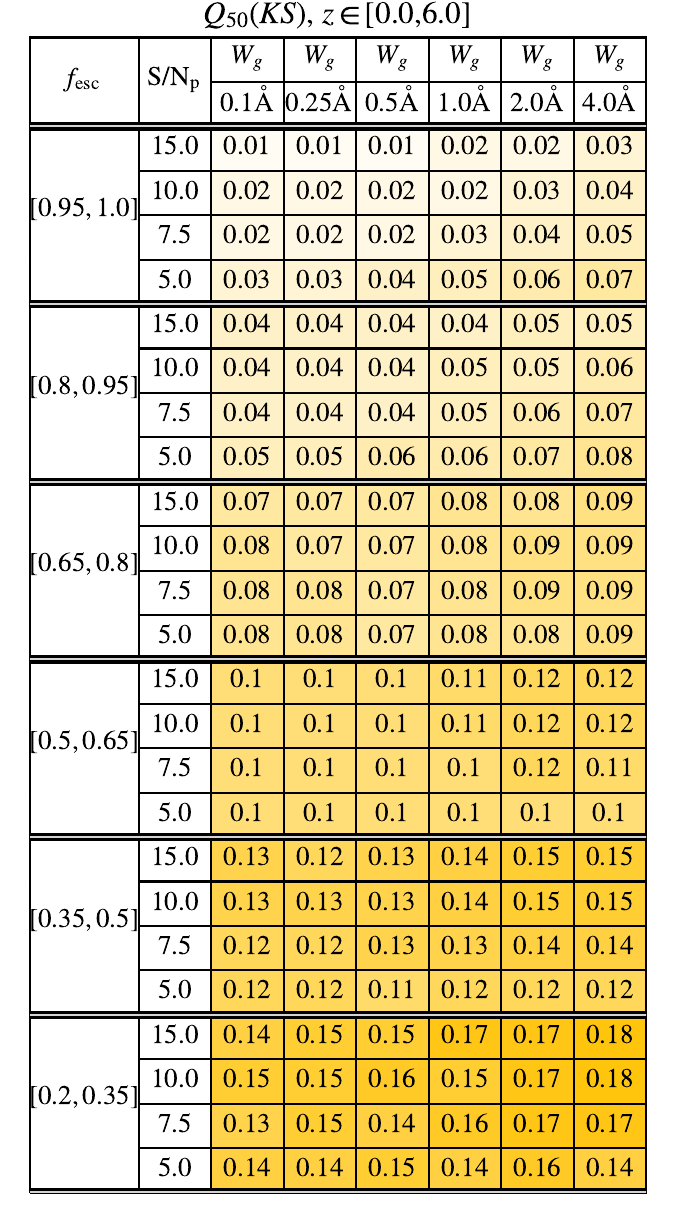}%
        \caption{ Median of the Kolmogórov-Smirnov distribution comparison between observed line profiles quality configurations and \zelda's models. The \igmz, \igm and NoIGM are shown from left to right. Lighter colors match smaller number and vice versa. }
        \label{fig:KS_tables}
        \end{figure*}
\begin{figure*} 
        \includegraphics[width=2.4in]{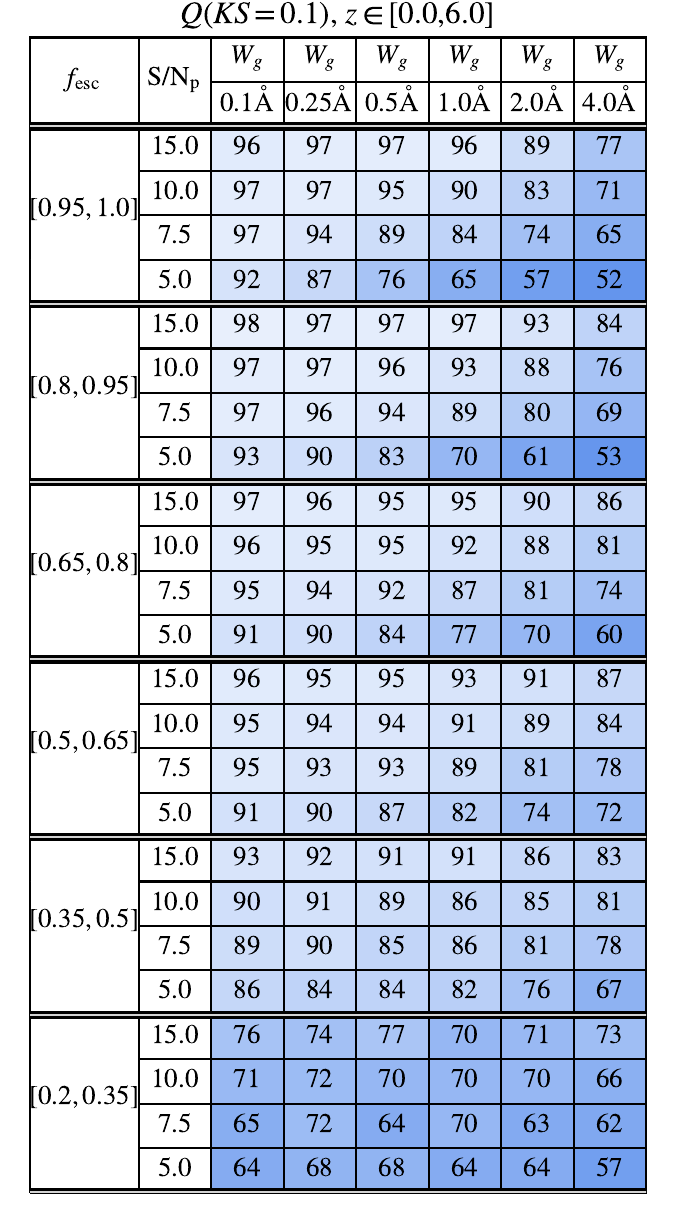}%
        \includegraphics[width=2.4in]{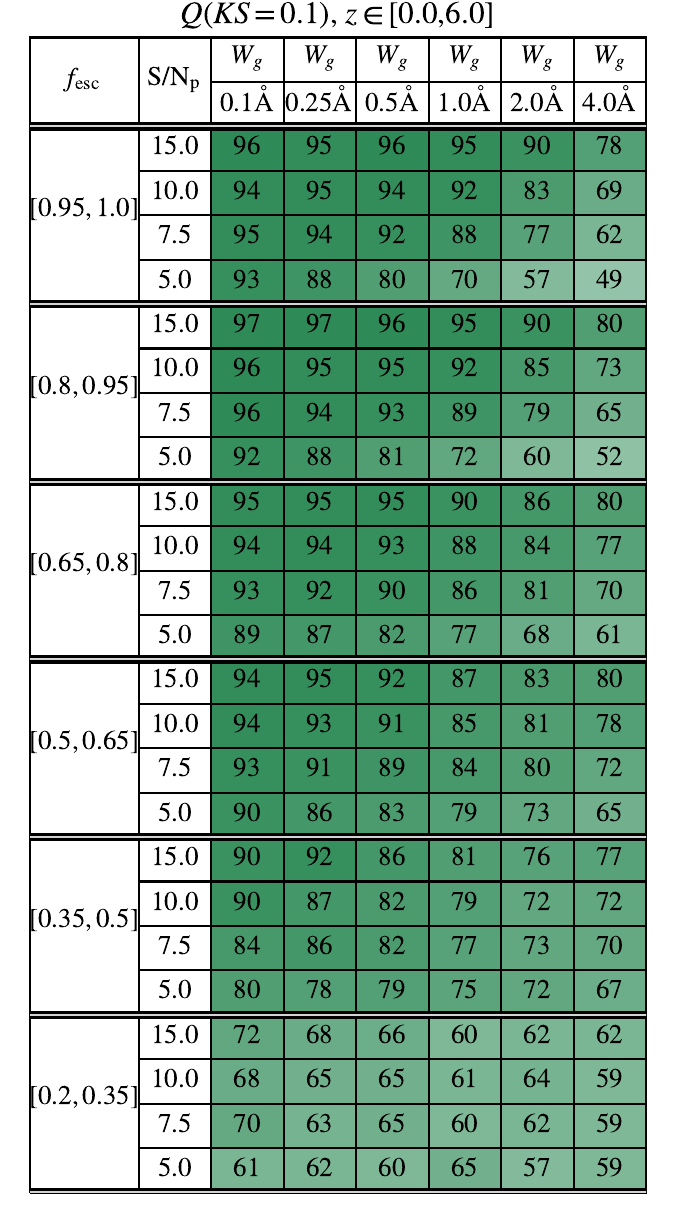}%
        \includegraphics[width=2.4in]{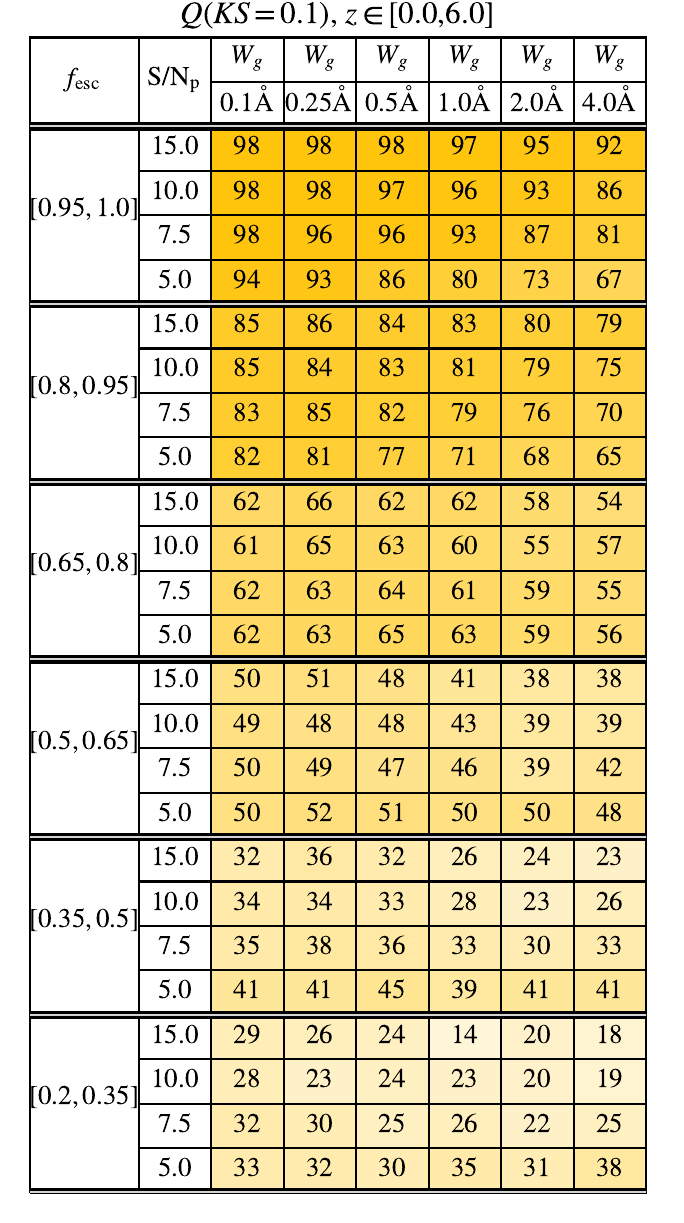}%
        \caption{ Percentile in the Kolmogórov-Smirnov distribution at which KS=0.1 for \igmz, \igm and NoIGM are shown from left to right.  Lighter colors match smaller number and vice versa. }
        \label{fig:KS_Q10}
        \end{figure*}

In this section we study the accuracy in the shape of the recovered intrinsic \lya line profile.  Fig.~\ref{fig:KS_distributions} shows the KS distribution for three mocks made in the same methodology as those for the parameter accuracy (Sect.~\ref{sssec:results_mock_parameters_parameters}). In particular, \sn=15.0 and \wg=0.25\AA, \sn=10.0 and \wg=0.5\AA and \sn=7.5 and \wg=1.0\AA are shown from left to right. The KS distribution of the predicted \lya line profiles by \igmz, \igm and \noigm are shown in blue, green and yellow, respectively. The horizontal dashed lines indicate the median of each distribution. In general, we find that \igmz and \igm outperform \noigm. While there is a significant difference between the accuracy of \igmz and \igm and \noigm, \igm performs slightly better than \igmz . \igmz and \igm show a median value around $10^{-1.5}$ while the median KS value of \noigm is around $10^{-1.2}$.

In Fig.~\ref{fig:KS_tables} we show the median of the KS distribution for the \lya line profile mocks shown in  Sect.~\ref{sssec:results_mock_parameters_parameters} as a function of \fa and for \igmz , \igm and \noigm , from left to right. As expected, the \lya line profile shape are better recover at better spectral quality configurations. As for the outflow parameters, we find that \noigm outperforms \igmz and \igm in the \fa $\in[0.95,1.0]$ regime. Also, the line profile accuracy of \noigm drops fast with \fa. In the best scenario (\wg=0.1 and \sn=15.0), at \fa $\in[0.95,1.0]$, the median KS is 0.01, at \fa $\in[0.8,0.95]$ is 0.04 and at \fa $\in[0.35,0.5]$ is 0.13. Meanwhile, \igmz and \igm \lya line profile reconstruction is better than that of \noigm for \fa<0.95. In particular, for \igmz and \igm the median KS is generally below 0.05 in all the quality configurations explored. Although for \wg=4.0\AA and small \fa values, the median KS goes up to 0.9.

In Sect.~\ref{sec:results_mock} we show that in some cases and outflow configuration after passing through the IGM might resembles another outflow configuration. This causes confusion when reconstructing the \lya lines (see examples in Fig.~\ref{fig:mock_example_unsuccessful_reconstruction}).  In order to quantify the fraction of cases in which our ANN models reconstruct properly the \lya we show in Fig.~\ref{fig:KS_Q10} the percentile at which KS=0.1. In general, reconstructed \lya line profiles with KS=0.1 are correctly recovered. Although in some cases \lya lines with KS>0.1 are properly recovered (and vice verse), the KS=0.1 threshold is in general valid. We find the same general trends for smaller KS thresholds like 0.08. 

As expected, the percentile at which KS=0.1 ($Q(KS=0.1)$) depends on the \lya line profile quality. The better the spectral quality, the higher the $Q(KS=0.1)$ is. As before, \noigm outperforms \igmz and \igm in the unabsorbed regime, being $Q(KS=0.1)$=98\%, 0.96\% and 0.96\% respectively at the best configuration. However, \igmz and \igm outperform \noigm when \fa<0.95. Remarkably $Q(KS=0.1)$ is greater than the 90\% for many quality configurations. Also, we find that for \sn>7.5 and \fa>0.5 the  $Q(KS=0.1)$>70\% typically, even at \wg=4.0\AA.

Notice that the outflow model confusion described in Sect.~\ref{sssec:results_mock_parameters_parameters} is not the only contributor to $Q(KS=0.1)$. The information destroyed by pixelization and noise affect $Q(KS=0.1)$. For instance, $Q(KS=0.1)$ for \noigm, in the unabsorbed regime, goes from 98\% in the best quality scenario down to 67\% at the worst spectral quality.  

\section{ Alternative models for \fa inference }\label{sec:disucssion_other_models}

\begin{figure*} 
        \includegraphics[width=3.6in]{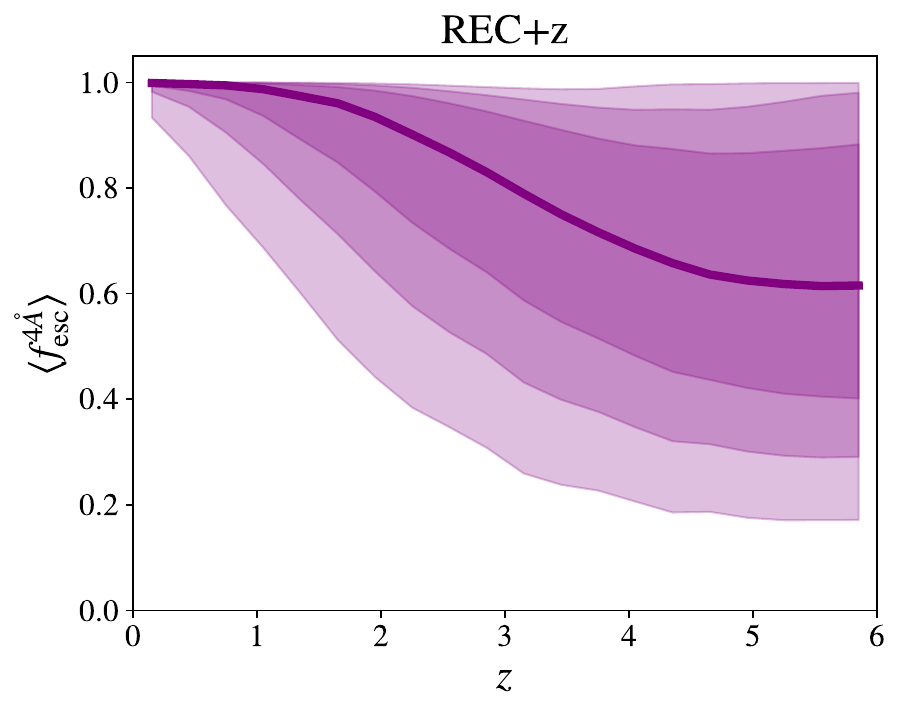}%
        \includegraphics[width=3.6in]{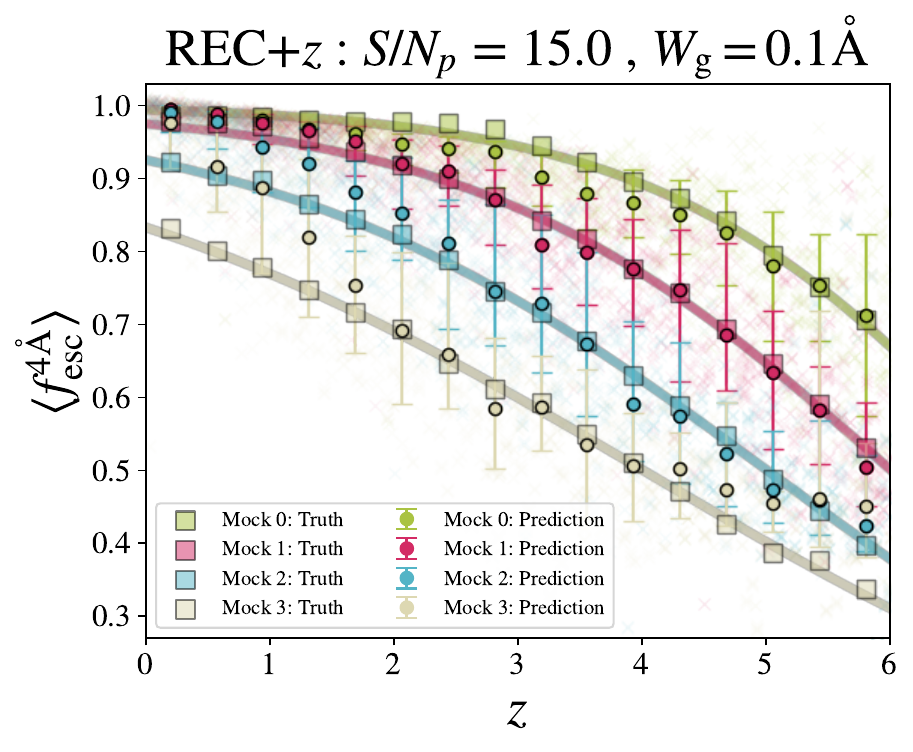}
        \caption{ {\bf Left} : \fa distribution as a function of redshift in the \recz model. The solid line shows the 50 percentile, while, from darker to lighter, the shadows show the 1$\sigma$, 2$\sigma$ and 3$\sigma$. {\bf Right} : Same as  as Fig.~\ref{fig:Mock_f_esc_IGM+z} but for  \recz at \sn=15.0 and \w=0.1\AA{}. }
        \label{fig:rec_model}
        \end{figure*}

In this section we briefly discuss another model, \recz. \recz uses the recalibrated IGM transmission curves (as \igm) at the source redshift includes it in the input (as \igmz). The redshift evolution in the \fa distribution in shown in the left panel of Fig.~\ref{fig:rec_model}. The \recz model shows a similar \fa distribution as \igmz (see Fig.~\ref{f_mean_training}), however, at $z<1.0$ the dispersion in \recz is smaller, as 2$\sigma$ of the sources have \fa>0.9. In comparison, in \igmz more than 2$\sigma$ of the sources exhibit \fa>0.8 at  $z<1.0$. The small variance at $z<1.0$ in this training set is due to the fact that the IGM absorption at the \lya wavelength is suppressed after the recalibration (see the right panel of Fig.~\ref{fig:IGM_T_mean}). 

The small scatter causes that \recz \fa measurements are biased towards high values at $z<1.0$. This is shown in the right panel of Fig.~\ref{fig:rec_model}. Focusing in the mock 4 (grey), \recz predicts a \mfa about 0.9 when the true \mfa is 0.8 at $z<1.0$. Similarly, this is also clear in mock 3 (blue).  The over prediction in \recz at $z<1.0$ is caused by the small \fa scatter of the training set. As in the training set there is no source with \fa$\sim 0.8$ at $z<1.0$,  \recz does not predicts \fa$\sim 0.8$ values at $z<1.0$ even if the source true \fa is 0.8. 

In the development of this work we tested several alternative models, such as \recz. We found that \igmz and \igm were the most unbiased models among those explored. In particular, a few models, just as \recz, were biased in the low redshift regime while they performed correctly at high redshift. For these reason, we display mainly \igmz and \igm in this work. Meanwhile, in the \zelda package, in addition to \igmz and \igm, the other tested models are also included.

\end{appendix}

\end{document}